\documentclass[11pt,a4paper]{article}

\usepackage{amssymb}
\usepackage{amsmath}
\usepackage{slashed}
\usepackage{dsfont}\usepackage{mathrsfs}
\usepackage{bbm}
\usepackage{microtype}
\usepackage{appendix}
\usepackage{youngtab}

\usepackage{xcolor}
\definecolor{darkblue}{rgb}{0.1,0.1,.7}
\definecolor{darkgreen}{rgb}{0,.5,0}
\usepackage[pdftex,breaklinks,colorlinks,linkcolor=darkblue,citecolor=darkgreen]{hyperref}

\newenvironment{acknowledgments}{\vspace{12pt}\begin{center}
\textbf{Acknowledgments}\end{center}\vspace{-12pt}}{}
\newcommand{\ack}[1]{\begin{samepage}\begin{acknowledgments} {#1} \end{acknowledgments}\end{samepage}}

\DeclareMathOperator{\tr}{tr}

\newcommand{\be}{\begin{equation}}
\newcommand{\ee}{\end{equation}}
\newcommand{\nn}{\nonumber}
\newcommand{\scr}{\scriptstyle}

\newcommand{\A}{{\cal A}}

\newcommand{\D}{{\cal D}}

\newcommand{\N}{{\cal N}}
\renewcommand{\P}{{\cal P}}

\renewcommand{\S}{{\cal S}}

\newcommand{\rmd}{{\rm d}}
\newcommand{\ts}{\textstyle}

\newcommand{\pr}{\partial}

\newcommand{\vep}{\varepsilon}
\newcommand{\vphi}{\varphi}

\newcommand \bbeta{\bar \beta}
\newcommand{\tlam}{{\tilde \lambda}}
\newcommand{\rg}{{\rm g}}

\begin{document}

\numberwithin{equation}{section}

\begin{titlepage}

\begin{flushright}
\small
DAMTP-2017-30\\
CERN-TH-2017-149\\
July 2017\\
\normalsize
\end{flushright}

\vspace{1cm}
\begin{center}

{ \LARGE Seeking Fixed Points in Multiple Coupling
Scalar Theories
\vspace{6pt}
in the $\vep$ Expansion}
\end{center}

\vspace{1cm}
\begin{center}

{
Hugh Osborn$^{a}$ and Andreas Stergiou$^{b}$}
\vskip 1cm

{$^a$Department of Applied Mathematics and Theoretical
Physics, Wilberforce Road,\\ Cambridge CB3 0WA, England\\
\vspace{3pt}
$^b$Theoretical Physics Department, CERN, Geneva, Switzerland}
\end{center}

\begin{abstract}

{Fixed points for scalar theories in $4-\varepsilon$, $6-\varepsilon$ and $3-\varepsilon$
dimensions are discussed. It is shown how a large range of known fixed points
for the four dimensional case can be obtained by using a general framework with two
couplings. The original maximal symmetry, $O(N)$, is broken to various subgroups,
both discrete and continuous. A similar discussion is applied to the six dimensional case.
Perturbative applications of the $a$-theorem are used to help classify potential
fixed points. At lowest order in the $\vep$-expansion it is shown that at
fixed points there is a lower bound for $a$  which is saturated at bifurcation points.}

\end{abstract}

\end{titlepage}
\pagenumbering{roman}
\newpage

\pagenumbering{arabic}

\setcounter{footnote}{0}

\section{Introduction}

In seeking conformal field theories, whether through finding zeros of $\beta$-functions or
applying bootstrap consistency conditions, it is natural to impose as much symmetry as
possible. This then restricts the number of couplings and correspondingly non zero
operator product coefficients so that the number of channels to be considered in a bootstrap
analysis is reduced. In the cardinal exemplar of the success of the numerical bootstrap,
the three dimensional Ising model with two relevant fields $\sigma, \epsilon$, there is a
${\mathbb Z}_2$ symmetry with $\sigma$ ${\mathbb Z}_2$ odd.

The imposition of symmetries might introduce a selection bias in finding non trivial CFTs.
On the other hand the fixed points of non symmetric theories might have an emergent symmetry. According to the notion of universality fixed points are determined
by the associated symmetry group.  To an extent it is often tacitly assumed that fixed points
without a high degree of symmetry may be neglected. Nevertheless for any potential classification of
CFTs in three or more dimensions it is necessary to determine all possible fixed points which
may have smaller symmetry groups than have been hitherto considered. Even with knowledge of the  fixed
points in quantum field theories in particular dimensions, an important question is how they are linked under
possible RG flows induced by perturbations by adding relevant operators.
Determining the spectrum of relevant operators is part of the data defining a CFT or can be found
by solving linear eigenvalue equations. What happens under RG flow, and whether new fixed points
are attained, is a non linear problem. For slightly relevant perturbations it is possible to use
conformal perturbation theory but this becomes difficult beyond lowest order. In a perturbative
context the $\vep$ expansion can be taken to higher order by determining $\beta$-functions
and anomalous dimensions in a time honoured fashion. This may be applied, as originally
suggested by Wilson and Fisher \cite{WilsonF}, in $4-\vep$ dimensions but can also be used in
$6-\vep$ and $3-\vep$ dimensions.

In two dimensions, at least for minimal models, there is a well understood framework of CFTs
which are linked by RG flows. In two dimensions conformal perturbation theory
can be used systematically, as originally proposed  by Zamolodchikov \cite{Zamolodchikov}
and extended subsequently \cite{Ludwig,Lassig,Lassig2,Gaberdiel,Poghossian,Ahn}.
A crucial organising principle is the $c$-theorem which requires that the Virasoro central charge $c$
for any unitary CFT at fixed points reached by RG flow arising from a relevant
perturbation has a lower value of $c$ than the original unperturbed CFT.
Minimal models, with $c<1$, then flow ultimately to the Ising model with $c=\frac12$.

There is no such simple picture in three or other dimensions of current interest.
Here we attempt to discuss potential fixed points and RG flows using the $\vep$-expansion
with various additional assumptions for simplicity. For determining critical exponents
in many different condensed matter systems with a wide range of symmetries
there is a large literature from the 1970's and later. A review discussing large number
such models which are applicable to critical phenomena for condensed matter systems is \cite{RGrev}.
The $\vep$-expansion is a version of conformal perturbation
theory starting from free field theories and which allows an extension beyond lowest order.
An assumption made here is that it is qualitatively correct
in determining possible CFTs in three dimensions starting from $4-\vep$ dimensions through essentially one
loop calculations. Of course in particular applications loop calculations to three or more
loops and sophisticated resummation methods may be used to obtain more accurate results
for critical exponents.

Results in the $\vep$-expansion depend critically on the number $\N$ of scalar fields. The
symmetry of the kinetic term is $O(\N)$. In $4-\vep$ dimensions there is a fixed point with $O(N)$, $N=\N $
symmetry. This is stable  for $N \lesssim 4$. For $N>4$ there are relevant perturbations which lead to a variety of fixed points.
Whether there are fixed points which cannot be reached by simple RG flows from free fields
is not yet clear. A related question is how many couplings are necessary in order to find particular
fixed points. For the Ising model it is sufficient to consider just $\N=1$
and the $\phi^4$ perturbation
with a single coupling. In $6-\vep$ dimensions if $\N=N+1$ an $O(N)$ renormalisable theory
is formed by fields forming an $N$-vector and a singlet. There are then fixed points with
$O(N)$ symmetry if $N$ is large enough.

In $\vep$-expansions the starting point is a potential of the form
\be
V(\phi;g) =  {\ts \sum_I }\,g^I P_I(\phi) \, ,
\ee
where $\{ P_I(\phi)\} $  are polynomials of degree
$4$, or $3$ or $6$, depending on the dimension. The symmetry group $H_g$ for $V$ is determined
by
\be
V(\rg \phi;g) = V(\phi ; g) \quad \mbox{for all} \quad \rg \in H_g \subset O(N) \, .
\label{symV}
\ee
For each polynomial $P_I(\phi)$ there is a corresponding symmetry group $H_I$ defined as in \eqref{symV}.
For generic couplings  $H_g \supseteq H \simeq \cap \, H_I$, but $H_g$  may be larger than $H$.
Under RG flow the symmetry is in general $\cap\, H_g$ for all $g$, but the symmetry may be enhanced
at particular points in the space of couplings.  It is also important to note that the larger symmetry defined
by
\be
V(\rg \phi;g) = V(\phi ; g\, \rg ) \, , \ \ (g\,\rg)^I = g^J \rg_J{}^I \, ,\quad \mbox{for all} \quad \rg \in G_g \subset O(N) \, ,
\ee
defines a physically equivalent theory if $ (g\,\rg)^I \ne g^I$.
At a fixed point with $g=g_*$ if $G_{g_*}/H_{g_*}$ is discrete then
there are equivalent fixed points with identical critical exponents. If  $G_{g_*}/H_{g_*}$  is continuous there are
submanifolds of equivalent fixed points which contain apparently exactly marginal operators.
In general $G_g \supset G$ where $G \subset O(N)$ is the symmetry associated to
the set of polynomials $\{P_I\}$ such that for $\rg \in G$,  $P_I(\rg \phi) = \rg_I{}^J P_J(\phi)$.
Of course if  $\{P_I\}$ includes all $\binom{N+p}{p}$ polynomials of degree $p$ then $G\simeq O(N)$,
this is still true if $\{P_I\}$ forms a representation space for a faithful irreducible representation of $O(N)$.

Determining all possible polynomials invariant under subgroups of $O(N)$,
discrete or continuous, is very non trivial task. For $N=2,3$
the results are simple but general results for quartic polynomials when $N=4$ \cite{Michel2,Brezin2} or $N=6$ are much more
involved \cite{Hatch,Hatch1,Hatch2}. An important result due to Michel \cite{Michel3} in $4-\vep$ dimensions is that stable fixed points are unique
(an essentially identical proof is described in \cite{VicariZ}).
This then  implies at such a fixed point  $G_{g_*}\simeq H_{g_*}$. This does not hold if there is a submanifold
of continuously connected equivalent fixed points but otherwise provides constraints on the possible existence
of stable fixed points for particular breakings of $O(N)$, or in $6-\vep$ dimensions.

RG flow for multiple couplings, which is determined by solving
\be
\frac{\rmd}{\rmd t} g^I = - \beta^I(g) \, ,
\ee
is potentially very non trivial but is constrained by the $a$-theorem in
four dimensions and its possible extensions in other cases. In a strong version of the $a$-theorem there
exists a function $a(g)$ which decreases monotonically under RG flow until the flow reaches
fixed points $g^I= g_*{\!}^I $ where the $\beta$-functions vanish. For a stable fixed point $a(g)$ has a local minimum
and the stability matrix $\big [ M_I{}^J\big ]$ defined by
\be
 M_I{}^J = \pr_I \beta^J(g) \big |_{g=g_*}  \, ,
\ee
is positive definite.
 In four dimensions an  $a$-function satisfying the requirements for a strong $a$-theorem can be
 constructed directly from the one and two loop $\beta$-functions \cite{WallaceG}, the same is true in six
dimensions \cite{Jack6d} and, as will be shown later, in three dimensions. In four \cite{Weyl}
and six \cite{Stergiou6d} there are arguments that these results, expressing the $\beta$-functions
as a gradient flow, can be extended to arbitrary perturbative order. This imposes non trivial
constraints on the higher loop $\beta$-functions, even for purely scalar theories.
Using minimal subtraction the expressions for $a(g)$
can be easily extended to $4-\vep$ and $6-\vep$ dimensions and so the existence of an $a$-function
is relevant for fixed points accessible through an $\vep$-expansion. There are however no
comparable arguments for theories starting from three dimensions. In four dimensions at a fixed
point $a$, determined by the trace anomaly for a curved background or the three point function
for the energy momentum tensor on flat space, is necessarily positive.
However in a perturbative discussion in $4-\vep$ dimensions the
results for $a(g)$ away from fixed points are not necessarily bounded below. Under RG flow $a(g)$
decreases but it may be that no fixed point is attained and the assumption that the couplings
are ${\rm O}(\vep)$ breaks down. In particular $a(g)$ could become negative in situations
where the potential $V(\phi;g)$ is also unbounded below and there is no stable ground state.
These conclusions hold even more strongly in perturbative discussions in $6-\vep$ dimensions,
since cubic potentials of course are never bounded below.

In our discussions in $6-\vep$ and $4-\vep$ dimensions we consider perturbative expressions for
$a(g)$, neglecting the free field contributions which are irrelevant here and also allowing for
convenience an arbitrary overall normalisation. The most stable fixed point $g_*$ is then the one at
which $a(g_*)$ is the  least local minimum. In the framework of the $\vep$-expansion we are able to show that
$a(g_*)$ has a lower bound, of order $\vep^2, \, \vep^3$ in $6-\vep, 4-\vep$ dimensions.
This bound is attained when two fixed points collide  and then disappear from the set of fixed points for
real couplings, so that there is a saddle-node bifurcation. If the initial couplings are small but
are such that $a(g)$ is less than the bound then the RG flow cannot reach perturbative fixed points
calculable in the $\vep$-expansion.  The existence of $a(g)$ nevertheless avoids the possibility of
more complicated bifurcations \cite{Gukov}.

In this paper we discuss a range of fixed points present in scalar field theories when
the number of fields $\N>1$ and there are two or more couplings. For simplicity we first
consider in sections \ref{twoFlavSix} and \ref{twoFlavFour} the cases when $\N=2$ in $6-\vep$ and $4-\vep$ dimensions.
For an arbitrary renormalisable theory a complete analysis of potential fixed points is
possible. In the four dimensional case there is just the $O(2)$ symmetric point and
two decoupled Ising fixed points. In both cases the couplings transform under $O(2)$
and potential fixed points are related to the different irreducible components.

For general $N$ there are many more possibilities of finding fixed points where the
initial symmetry is broken by slightly relevant perturbations. In section
\ref{redSymSix} we consider
perturbations of the $O(N)$ symmetric theory in six dimensions.
The unperturbed theory has a critical point in $6-\vep$ dimensions so long as
$N>N_{\rm crit}$ \cite{Fei,Fei3loop}.  With one additional
coupling for a relevant operator there is a breaking $O(N) \to O(m)\times O(n)$, $m+n=N$.
There is then a new fixed point corresponding to an $O(m)$ invariant theory with $n$
decoupled free fields so long as $m >N_{\rm crit}$, or equivalently $m\leftrightarrow n$.
We also consider breakings $O(N) \to O(m)\times O(n)$, $m\,n=N$ involving additional
fields which form symmetric traceless tensor under $O(m)$.

In section \ref{redSymFour} we analyse breakings of the $O(N)$ symmetric theory in $4-\vep$ dimensions due to perturbations
formed by a symmetric traceless $4$-tensor. At lowest order in the $\vep$-expansion this
becomes relevant for $N>4$. We consider a general framework in which the RG flow may be restricted
to a two dimensional space of couplings. In the simplest case to three loops there are three
coefficients $a,b,c$, but overall normalisation is arbitrary and there are bounds which constrain
the allowed values. This formalism covers a range of theories considered in the
literature, including ones with hypercubic and hypertetrahedral symmetry, and also cases
where there is an unbroken $O(m)$ symmetry, $N=n\, m$. It is shown how the $a$-function
has a lower bound at fixed points which is attained at bifurcation points where new
fixed points emerge. The  general formalism is shown to give  identical results to three
loops with previous calculations for particular models.

We also discuss in section \ref{threeD} scalar theories in $3-\vep$ dimensions with a $\phi^6$ interaction using results
for $\beta$-functions for a general renormalisable potential at two and four loops. Such
theories have an $O(N)$ invariant fixed point but, by determining the anomalous
dimensions of scalar operators with dimensions less than three, there is no
apparent large $N$ theory valid for dimensions away from three (the $O(N)$ theories
in four and six dimensions at large $N$ can be related to an $O(N)$ model for any $d$
with an interaction $\sigma \, \phi^2$). The $O(N)$ fixed point is unstable against perturbations
for any $N$ and we consider the simple case leading to a fixed point with hypercubic symmetry.

In section \ref{multCoupFour} we extend the framework of section \ref{redSymFour} to multiple couplings corresponding to
several slightly relevant operators formed by symmetric traceless 4-tensors. The algebraic
conditions necessary for a closed RG flow are generalised to this case and it is
shown how the bound on the $a$-function remains valid to lowest order in $\vep$ and
that this bound is realised at the bifurcation point where there is an exactly marginal
operator. An example with three couplings, in which there are five inequivalent fixed points,
is considered. A  theory with arbitrarily many couplings with a closed RG flow is also
considered.

Various details are contained in five appendices. In appendix \ref{bifSix} we analyse the bifurcation
point for the $O(N)$ symmetric theory in $6-\vep$ dimensions. Appendix \ref{altfp} describes
a different  $4-\vep$ fixed point than the ones considered in section \ref{redSymFour}.
In appendix \ref{boundsTP} we obtain
bounds on the coefficients $a,b$ which appear in the general discussion in $4-\vep$ dimensions.
The bounds are saturated in the case of hypercubic symmetry. Appendix
\ref{pertsHT} considers perturbations
of the theory with hypertetrahedral symmetry. The analysis shows in general the presence of
relevant operators. Finally in appendix \ref{centralCharge} we show how the   perturbative expression for $C_T$,
the coefficient of the energy momentum tensor two point function, at the fixed point can be constrained
by large $N$ results. This allows $C_T$ to be found for arbitrary theories in the $\vep$ expansion
in $4-\vep$ dimensions to order $\vep^3$.

In general the different sections are more or less independent and may be read, if desired,
separately. Detailed results for many cases have already appeared in the literature but
are here presented in a hopefully unified framework.

\section{Two Flavour Scalar Theory in
  \texorpdfstring{$\mathbf{6\boldsymbol{-}\boldsymbol{\vep}}$}{6-epsilon} Dimensions}
\label{twoFlavSix}

A general two component $\phi^3$ theory is described by a potential
\be
V(\phi) =
\tfrac16 \, \lambda_1 \, \phi_1{\!}^3 + \tfrac16 \, \lambda_2 \, \phi_2{\!}^3
+ \tfrac12 \, g_1 \, \phi_1 \, \phi_2{\!}^2 + \tfrac12 \, g_2 \, \phi_1{\!}^2 \phi_2 \, ,
\label{Vphi}
\ee
depending on couplings $G_I = (\lambda_1, \lambda_2 ,g_1 ,g_2)$. Results
for $\beta$-functions for such renormalisable $\phi^3$ theories, with fields $\phi_i$,
in $6-\vep$ dimensions are succinctly given in  terms of $\beta_V(\phi)$
where, taking $V\to (4\pi)^{\frac32} V$,
\be
\beta_V(\phi) =  \vep \, V(\phi) + V_i(\phi)\big (- \tfrac 12 \, \vep \, \delta_{ij}
+ \gamma_{ij} \big  ) \phi_j + \bbeta_V(\phi) \, , \quad V_i = \pr_i V \,
\label{betaV3}
\ee
for $\gamma_{ij}$ the symmetric  anomalous dimension matrix and $\bbeta_V$ corresponding  to the
contributions of one particle irreducible graphs.
Taking $V(\phi) = \tfrac16 \, \lambda_{ijk} \phi_i \phi_j \phi_k$ then
$\beta_V(\phi) = \tfrac16\, \beta_{ijk} \phi_i\phi_j \phi_k$. At one loop
\be
\bbeta_V{\!}^{(1)} =- \tfrac16 \, V_{ij}V_{jk} V_{ki} \, , \qquad
\gamma_{ij}{\!}^{(1)} \phi_i \phi_j = \tfrac{1}{12} \, V_{ij} V_{ij} \, ,
\label{betaV31}
\ee
and at two loops\footnote{These expressions can be obtained from results in
\cite{Mckane0,Fei3loop,Grinstein6d,Jack6d}.}
\begin{align}
\bbeta_V{\!}^{(2)} = {}&
 - \tfrac18 \, V_{ij}V_{ik} V_{lm} \lambda_{jln}\lambda_{kmn}
+ \tfrac{7}{144} \, V_{ij} V_{jk} (\lambda^2)_{kl} V_{li} - \tfrac{1}{12} \,
V_{il}V_{jm} V_{kn} \lambda_{ijk} \lambda_{lmn} \, , \nn \\
\gamma_{ij}{\!}^{(2)} \phi_i \phi_j = {}&  \tfrac{1}{18}  \big ( V_{ij} V_{kl} \lambda_{ikm} \lambda_{jlm} -
\tfrac{11}{24}\, V_{ij} (\lambda^2)_{jk} V_{ik} \big ) \, , \quad (\lambda^2)_{ij} =
\lambda_{ikl} \lambda_{jkl}\, , \ \ \lambda_{ijk} = V_{ijk} \, .
\label{betaV32}
\end{align}
and also at three loops
\begin{align}
\bbeta_V{\!}^{(3)} = {}& \tfrac16 \Big (- V_{ij}V_{kl}V_{mn} \big ( \lambda_{ikp}\lambda_{jmq}\lambda_{lnr}\lambda_{pqr}
+\tfrac{1}{24} (23 - 24\, \zeta_3) \,  \lambda_{ipq}\lambda_{kpr}\lambda_{mqr}\lambda_{jln}  \big )
\nn \\
\noalign{\vskip -2pt}
&{}  - V_{ij}V_{kl} \big ( \tfrac98 \, \delta_{ik} \lambda_{jrp} \lambda_{lrq}
+ \tfrac{47}{288} \, (\lambda^2)_{ik} \delta_{jp}\delta_{lq} + \tfrac{47}{144} \, \delta_{ik} (\lambda^2)_{jp} \delta_{lq}
+  \tfrac{3}{16}\, \lambda_{ikp} \lambda_{jlq}  \big )
  \lambda_{psm}\lambda_{qsn} V_{mn} \nn \\
 &{}+ V_{ij}V_{jk}V_{kl} \big ( \tfrac{23}{96}\, \lambda_{imn}\lambda_{mpr}\lambda_{nqr} \lambda_{lpq}
 - \tfrac{19}{108} \, \lambda_{imp}\lambda_{lmq} (\lambda^2)_{pq} + \tfrac{11}{576}\,(\lambda^2)_{im} (\lambda^2)_{ml} \big ) \nn \\
 &{}+V_{ij}V_{ik}V_{mn} \big (  \tfrac{11}{576}\, (\lambda^2)_{jm} (\lambda^2)_{kn} + \tfrac{5}{9} \, \lambda_{jpq}\lambda_{kpn}(\lambda^2)_{qm}
 +\tfrac34\, \lambda_{jpq}\lambda_{krn}\lambda_{qsm}\lambda_{prs} \big )
 \nn \\
 &{}+ V_{ij} V_{kl} V_{mn}\big ( \tfrac{15}{16}\,  \lambda_{jpr}\lambda_{irm} \lambda_{kps}\lambda_{lsn}
 + \tfrac{11}{72} \, \delta_{jk} \lambda_{ipm}\lambda_{lqn} (\lambda^2)_{pq}
 +\tfrac{11}{48}\,  \lambda_{ikp}\lambda_{jln} (\lambda^2)_{pm} \big )
 \nn \\
&{}  + V_{ij}V_{kl} V_{mn} \big ((1-3\,\zeta_3)\,   \lambda_{jpq}\lambda_{kpr} \lambda_{lnq}\lambda_{imr}
 - \tfrac{1}{16} (29-24\,\zeta_3)\, \delta_{jk}
 \lambda_{ipr} \lambda_{lqs} \lambda_{pqm} \lambda_{rsn} \big )\Big ) \, , \nn \\
\gamma_{ij}{\!}^{(3)}& \phi_i \phi_j =
\tfrac{1}{108} \,    V_{ij}V_{kl} \, c_{ijkl} \, ,\nn \\
 c_{ijkl} ={}&  \tfrac{71}{16}\, \lambda_{imn}\lambda_{jmp} \lambda_{nqk}\lambda_{pql}
+ \tfrac{7}{8} \,   \lambda_{inp}\lambda_{jlm} \lambda_{knq}\lambda_{mpq}
- \tfrac{103}{96} \, \lambda_{ikm} \lambda_{jln}  (\lambda^2)_{mn}  \nn \\
&{} - 2 \,  \lambda_{kmn} \lambda_{jlm}  (\lambda^2)_{in}  -
\tfrac{121}{48} \, \delta_{jl}\, \lambda_{imn} \lambda_{pq k} \lambda_{mpr}\lambda_{nqr}
+ \tfrac{23}{576}\,  (\lambda^2)_{ik}  (\lambda^2)_{jl}  \nn \\
\noalign{\vskip -2pt}
&{} -
\tfrac{13}{288}\,  \delta_{jl} \, (\lambda^2)_{im}  (\lambda^2)_{mk}
+ \tfrac{103}{72}\,  \delta_{jl} \, \lambda_{imp} \lambda_{knp}  (\lambda^2)_{mn}
+\tfrac92 ( \tfrac74 - \zeta_3)  \lambda_{imn}\lambda_{jpq} \lambda_{mqk}\lambda_{npl}   \, .
\label{betaV33}
\end{align}
Including ${\rm O}(\phi^2)$ contributions in $V(\phi)$ then results for \eqref{betaV3} immediately
determine the associated anomalous dimensions for such operators.

For the two component theory determined by \eqref{Vphi} the maximal symmetry
is determined by the transformations
\be
\phi_{\theta,1} = \cos \theta \, \phi_1 + \sin \theta \, \phi_2 \, , \qquad
\phi_{\theta,2} =  - \sin \theta \, \phi_1 + \cos \theta \, \phi_2 \, .
\label{varp}
\ee
Then
\be
V_\theta(\phi) = V (\phi_\theta) \, ,
\ee
with $V_\theta$ expressed in terms of transformed couplings
\be
G_{\theta,I} = R_{IJ}(\theta) \,  G_J \, ,
\label{GMG}
\ee
where
\be
\big [ R_{IJ} (\theta) \big ] = \begin{pmatrix}  {\scr \cos^3 \theta}  & {\scr  - \sin^3 \theta} &
{\scr 3 \, \cos \theta \, \sin^2 \theta} &  {\scr -  3 \, \cos^2 \theta \, \sin \theta} \\
{\scr \sin^3 \theta} &  {\scr \cos^3 \theta} & {\scr 3 \, \cos^2 \theta \,  \sin \theta} &
{\scr 3 \, \cos \theta \, \sin^2 \theta} \\
{\scr \cos \theta  \, \sin^2  \theta} & {\scr -  \cos^2 \theta \, \sin\theta} &
{\scr \cos^3 \theta  - 2\, \cos  \theta \, \sin^2 \theta}
& {\scr - \sin^3 \theta + 2\, \cos^2 \theta \, \sin \theta} \\
 {\scr \cos^2 \theta  \, \sin  \theta} & {\scr \cos \theta \, \sin^2 \theta} &
 {\scr \sin^3 \theta  - 2\,  \cos^2 \theta \sin  \theta} &
 {\scr \cos^3 \theta - 2\, \cos \theta \, \sin^2 \theta }
 \end{pmatrix}  \, ,
 \label{M6}
 \ee
defines an equivalent theory, related by a field redefinition.  The $SO(2)$
transformations of the couplings given by \eqref{GMG} and \eqref{M6} may be
extended to $O(2)$ by
including reflections generated by $\phi_1 \leftrightarrow \phi_2$
which give an equivalence under $\lambda_1 \leftrightarrow \lambda_2, \,
g_1 \leftrightarrow g_2$.

The couplings may be decomposed in terms of real two-dimensional
$SO(2)$ irreducible representations
\be
v_{-3} = \begin{pmatrix} \lambda_1 - 3 g_1 \\ \lambda_2 - 3 g_2 \end{pmatrix} \, ,
\qquad
v_1 = \begin{pmatrix} \lambda_1 +g_1 \\ \lambda_2+ g_2  \end{pmatrix}  \, ,
\ee
where under $G_I \to G_{\theta,I} $
\be
v_n \to R_n (\theta) v_n \, , \qquad  R_n (\theta) =
\begin{pmatrix} \cos n\theta & - \sin n \theta \\
 \sin  n \theta &  \cos n\theta  \end{pmatrix} \, .
 \label{fix3}
 \ee
The four couplings can be reduced to three invariants
\begin{align}
I_1 = {}& (\lambda_1-3g_1)^2 + (\lambda_2-3g_2)^2 \, , \qquad
I_2 = (\lambda_1  + g_1)^2 + (\lambda_2 + g_2)^2 \, , \nn \\
I_3 + i \, I_4 ={}& \big (  (\lambda_1  + g_1) + i (\lambda_2 + g_2) \big )^3
\big ( (\lambda_1-3g_1) + i (\lambda_2-3g_2) \big ) \, ,
\end{align}
where $I_3{\!}^2 + I_4{\!}^2 = I_1 \, I_2{\!}^3$. Clearly $I_4$ is odd under
$\lambda_1 \leftrightarrow \lambda_2, \, g_1 \leftrightarrow g_2$.
The invariants determine the couplings up to an $SO(2)$ equivalence.

From \eqref{fix4} invariant subspaces under $SO(2)$ are obtained by imposing
either $v_1$ or $v_{-3}$ to be zero giving
the conditions
\be
{\rm a}) \ \
\lambda_1 = - g_1 \, , \ \lambda_2 = - g_2  \quad \mbox{and} \quad
{\rm b}) \ \  \lambda_1 =  3 g_1 \, , \ \lambda_2 = 3 g_2 \, .
\label{fix}
 \ee
Case a) corresponds to $\lambda_{ijk}$ in \eqref{Vphi} being symmetric traceless
and case b) to taking $\lambda_{ijk} =3\,  \delta_{(ij} v_{k)}$ with $v_i = (g_1,g_2)$.
With the restrictions in \eqref{fix} the only invariants are $I_1$ or $I_2$ respectively.
If $\lambda_2 = g_2 = 0$ and $g_1=-\lambda_1$  then
the $O(2)$ symmetry is reduced to ${\mathbb Z}_3$; in this case
$V(\phi) = \tfrac{1}{12}\lambda_1 \big ( (\phi_1 + i \phi_2)^3 + (\phi_1 - i \phi_2)^3\big )$.
If $g_{1,\theta}=g_{2,\theta}= 0 $ for some $\theta$, so that the theory is equivalent to
two decoupled theories, then it is necessary that $I_1 = I_2$, or
\be
g_1{\!}^2 + g_2{\!}^2 = \lambda_1 g_1 +  \lambda_2 g_2 \, .
\label{decouple6}
\ee

The symmetric $2\times 2$ anomalous dimension matrix for this theory
is $\gamma(G)$ and transforms as
\be
\gamma(G_\theta)  = R_1(\theta)  \gamma(G)  \, R_1(\theta)^T \, .
\label{grot}
\ee
Hence $\tr \gamma, \, \det \gamma$, and the eigenvalues of $\gamma$, are
scalars under the equivalence transformations of the couplings and are
functions of the invariants $I_{1,2,3}$.
For the $\beta$-function
\be
\beta_I ( G_\theta) = R_{IJ} (\theta ) \, \beta_J (G)\, .
\label{Brot}
\ee
This ensures that for a fixed point when $ \beta_I(G_*)=0$
then also  $\beta_I(G_{*\, \theta} )=0$.

For the theory defined by \eqref{Vphi} the $\beta$-functions to one loop are
given from \eqref{betaV31}:
\begin{align}
\hskip - 1cm
\beta_{\lambda_1}{\!}^{(1)}
={}& - \tfrac12 \, \vep \, \lambda_1 - \lambda_1{\!}^3 - g_1{\!}^3
- 3 (\lambda_1 + g_1 )  g_2{\!}^2 \nn \\
&{} + \tfrac14 \big ( \lambda_1 ( \lambda_1{\!}^2 + g_1{\!}^2 + 2 g_2{\!}^2 )
+ g_2( \lambda_1 g_2 + \lambda_2 g_1 + 2 g_1 g_2 ) \big )  \, , \nn \\
\hskip - 1cm
\beta_{\lambda_2}{\!}^{(1)}
={}& - \tfrac12 \, \vep \, \lambda_2 - \lambda_2{\!}^3 - g_2{\!}^3
- 3 (\lambda_2 + g_2 )  g_1{\!}^2 \nn \\
&{} + \tfrac14 \big ( \lambda_2 ( \lambda_2{\!}^2 + 2 g_1{\!}^2 + g_2{\!}^2 )
+ g_1( \lambda_1 g_2 + \lambda_2 g_1 + 2 g_1 g_2) \big )  \, , \nn \\
\hskip - 1cm
 \beta_{g_1}{\!}^{(1)}
={}& - \tfrac12 \, \vep \, g_1  - (\lambda_1 + g_1 ) ( g_1{\!}^2 +  g_2{\!}^2 )
- g_1 (\lambda_2 + g_2 )^2 \nn \cr
&{} + \tfrac{1}{12} \big ( g_1
( \lambda_1{\!}^2 + 2  \lambda_2{\!}^2 + 5 g_1{\!}^2 + 4 g_2{\!}^2 )
+ (\lambda_2 + 2 g_2 )( \lambda_1 g_2 + \lambda_2 g_1 + 2 g_1 g_2 ) \big )  \, , \nn \\
\hskip - 1cm
 \beta_{g_2}{\!}^{(1)}
={}& - \tfrac12 \, \vep \, g_2  - (\lambda_2 + g_2 ) ( g_1{\!}^2 +  g_2{\!}^2 )
- g_2 (\lambda_1 + g_1 )^2 \nn \\
&{} + \tfrac{1}{12} \big ( g_2
( 2 \lambda_1{\!}^2 + \lambda_2{\!}^2 + 4 g_1{\!}^2 + 5 g_2{\!}^2 )
+ (\lambda_1 + 2 g_1 )( \lambda_1 g_2 + \lambda_2 g_1 + 2 g_1 g_2 ) \big ) \, .
\end{align}
The positive contributions in the second line of each expression
arise from the  $2\times 2$  anomalous dimension matrix which
at lowest order is, from \eqref{betaV31},
\be
\gamma^{(1)} = \frac{1}{12} \begin{pmatrix} \lambda_1{\!}^2 + g_1{\!}^2 + 2 \, g_2{\!}^2 &
\lambda_1 g_2 + \lambda_2 g_1 + 2\, g_1 g_2 \\
\lambda_1 g_2 + \lambda_2 g_1 + 2\, g_1 g_2 &  \lambda_2{\!}^2 + 2\, g_1{\!}^2 +  g_2{\!}^2
\end{pmatrix}  \, .
\ee

Finding zeros of the $\beta$-functions giving fixed points requires relations
between the couplings.
Requiring that the couplings satisfy a) in \eqref{fix} the perturbative
expansion of the $\beta$-function to all orders
takes the form\footnote{For case b) $SO(2)$ invariance restricts
$\beta_I(G)  \big |_{\lambda_1=3g_1, \lambda_2=3g_2} =  G_I \,
f_1(g_1{\!}^2 + g_2{\!}^2  )  + G'{\!}_I \, f_2( g_1{\!}^2 + g_2{\!}^2  ) $
where $G'{\!}_I =g_1(g_1{\!}^2 - 3g_2{\!}^2)(1,0,-1,0) +g_2(g_2{\!}^2 - 3g_1{\!}^2)(0,1,0,-1)$.
The perturbative results to three loops give
$f_1(x) = - \tfrac12 \vep -\tfrac{35}{6} x - \tfrac{6241}{108}x^2 -( \tfrac{7287775}{7776} + \tfrac{1061}{3}
\zeta_3)x^3$
and $f_2(x) = -3 - \tfrac{763}{18} x - (\tfrac{982045}{1296} +
310 \zeta_3)x^2$.}
\be
\beta_I(G)  \big |_{g_1=-\lambda_1, g_2 = - \lambda_2} =  G_I \,
f\big ( \lambda_1{\!}^2 + \lambda_2{\!}^2 \big ) \,
\ee
with
\be
f(x) = - \tfrac12 \, \vep +  \sum_{n\ge 1} b_n \, x^n \, , \qquad
b_1 = \tfrac12 \, , \quad b_2 = - \tfrac{83}{36} \, , \quad
b_3 = - \tfrac{10627}{2592} - \zeta (3) \, .
\ee
The fixed point in this case then arises from the zeros of $f(x)$
\be
x_* = \vep + \tfrac{83}{18} \, \vep^2
+ \big ( \tfrac{21913}{432} + 2\, \zeta_3 \big ) \vep^3   + {\rm O}(\vep^4) \, .
\ee
Other fixed points require complex couplings and so are not considered here.

Following from \eqref{grot}  the perturbative expansion the anomalous
dimension matrix then takes the form
\be
\gamma (G) \big |_{g_1=-\lambda_1, g_2 = - \lambda_2} = \sum_{n\ge 1} a_n \,
(\lambda_1{\!}^2 + \lambda_2{\!}^2 )^n \, \mathds{1} \, ,
\ee
with
\be
a_1 = \tfrac16 \, , \qquad a_2= - \tfrac{11}{108} \, , \qquad
a_3 = \tfrac{5537}{7776} - \tfrac13 \, \zeta_3 \, .
\ee
To three loops
\be
\gamma(G_*) = \big ( \tfrac16 \, \vep + \tfrac23 \, \vep^2 + \tfrac{443}{54} \, \vep^3
+ {\rm O}(\vep^4) \big ) \, \mathds{1} \, .
\ee

As a consequence of \eqref{Brot} at a fixed point satisfying a) in  \eqref{fix}
by considering the derivative of \eqref{Brot} at $\theta=0$
\be
H_I \, \pr_I \beta_J(G) \big |_{G=G_*} = 0 \, ,
\label{zero}
\ee
which gives in this case
\be
\big [ H_I \big ]=
\big  ( \lambda_{2 *} \  {-  \lambda_{1 *} }\ { - \lambda_{2 *}} \  \lambda_{1 *} \big )  \, .
\label{zero3}
\ee
As a consequence $\det [ \pr_I \beta_J(G)] \big |_{G=G_*}  = 0 $ so that the matrix
$[ \pr_I \beta_J(G)]|_{G=G_*} $ has a zero eigenvalue. At lowest order
\begin{align}
[ \pr_I \beta_J(G)^{(1)}] \big |_{g_1=-\lambda_1, g_2 = - \lambda_2}
= {}& - \tfrac12 \, \vep \, {\mathds 1} \nn \\
&{}- \begin{pmatrix}  {\scr 2\, \lambda_1{\!}^2 + \frac94 \, \lambda_2{\!}^2}  & {\scr  - \frac14 \, \lambda_1 \lambda_2} &
{\scr \frac76\, \lambda_1{\!}^2 + \frac{11}{12} \, \lambda_2{\!}^2} &  {\scr \frac14 \, \lambda_1 \lambda_2 } \\
{\scr  -\frac14 \, \lambda_1 \lambda_2} &  {\scr \frac94\, \lambda_1{\!}^2 + 2 \, \lambda_2{\!}^2}
& {\scr \frac14 \, \lambda_1 \lambda_2  } &
{\scr \frac{11}{12}\, \lambda_1{\!}^2 + \frac{7}{6} \, \lambda_2{\!}^2} \\
{\scr \frac72\, \lambda_1{\!}^2 + \frac{11}{4} \, \lambda_2{\!}^2 } & {\scr \frac34 \, \lambda_1 \lambda_2 } &
{\scr  - \frac13\, \lambda_1{\!}^2 + \frac{5}{12} \, \lambda_2{\!}^2}
& {\scr - \frac34 \, \lambda_1 \lambda_2 } \\
 {\scr \frac34 \, \lambda_1 \lambda_2 } & {\scr \frac{11}{4}\, \lambda_1{\!}^2 + \frac{7}{2} \, \lambda_2{\!}^2} &
 {\scr - \frac34 \, \lambda_1 \lambda_2 } &
 {\scr \frac{5}{12}\, \lambda_1{\!}^2 - \frac{1}{3} \, \lambda_2{\!}^2}
 \end{pmatrix}  \, .
\end{align}
The eigenvectors at the fixed point are unchanged by higher order corrections to
$\beta$ although the eigenvalues are modified. Scaling operators with definite scale dimension
are determined by the left eigenvectors of $[ \pr_I \beta_J(G)]|_{G=G_*} $.
There is always a zero mode given by \eqref{zero3}.
The other three left eigenvectors and corresponding eigenvalues to three loops are given by
\begin{align}
\big ( \lambda_{1*} \  {\lambda_{2*}} \ {- \lambda_{1*}} \
{-  \lambda_{2*}} \big ) \, & , \qquad
\vep - \tfrac{83}{18}\, \vep^2 - \big ( \tfrac{38183}{648} + 4 \zeta_3 \big ) \vep^3\, , \nn\\
\big ( {1} \ {0} \ { \tfrac13} \ {  0}  \big )   ,
 \big (  {0} \  {1} \ {0} \ {\tfrac13} \big ) \, & ,
\quad - \tfrac{11}{3} \,
\vep  - \tfrac{158}{9}\, \vep^2 - \big ( \tfrac{17380}{9} + 16 \zeta_3 \big ) \vep^3  \, .
 \end{align}
There are two degenerate relevant operators.

An alternative reduction to that in \eqref{fix} arises if we impose
\begin{align}
\big ( (\lambda_1 + g_1 ) - i (\lambda_2 +g_2)\big )^3 = {}&
C \big ( (\lambda_1 - 3g_1 ) + i( \lambda_2 - 3 g_2) \big ) \, , \nn \\
\big ( (\lambda_1 + g_1 ) - i (\lambda_2 +g_2)\big )^2 \big ( (\lambda_1 - 3g_1 ) - i( \lambda_2 - 3 g_2) \big )= {}&
C' \big ( (\lambda_1 + g_1 ) + i( \lambda_2 + g_2) \big ) \, ,
\label{CCp}
\end{align}
with $C,C'$ real invariants such that
$I_2{\!}^3 = C^2 I_1 \, , \ I_3 = C \, I_1 = C' \, I_2  \, , \ I_4 = 0$. Subject to \eqref{CCp}
\be
\begin{pmatrix} \beta_{\lambda_1} + \beta_{g_1} \\ \beta_{\lambda_2} + \beta_{g_2} \end{pmatrix}
= F  \begin{pmatrix} \lambda_1 + g_1 \\ \lambda_2 + g_2 \end{pmatrix} \, , \qquad
\begin{pmatrix} \beta_{\lambda_1} -3 \,  \beta_{g_1} \\ \beta_{\lambda_2} -3\,  \beta_{g_2} \end{pmatrix}
= F'  \begin{pmatrix} \lambda_1 -3\, g_1 \\ \lambda_2 - 3\,  g_2 \end{pmatrix} \, ,
\ee
with $F,F'$ functions of invariants. To find  fixed points it is then sufficient to require just
$F=F'=0$ where $F,F'$ depend on two independent variables.
For a perturbative expansion valid away from a fixed point
it is necessary that $C,C'$ are expressible as
a series expansion in the couplings. This can be achieved in \eqref{CCp} by requiring
\be
\lambda_1 \lambda_2 = g_1 g_2 \, , \qquad \lambda_1 g_1 = g_2{\!}^2 \, ,
\qquad \lambda_2 g_2 = g_1{\!}^2 \, , \qquad C= C'=I_1 = I_2 \, .
\label{const}
\ee
In this case $F=F'$ and
\be
\beta_I ( G) = G_I \, F(I_1 ) \, , \qquad
\gamma(G) = \begin{pmatrix} \lambda_1 +g_1 \\ \lambda_2 +g_2 \end{pmatrix}
 \begin{pmatrix} \lambda_1 +g_1 & \lambda_2 +g_2 \end{pmatrix} \,
G (I_1 ) \, .
\ee
As \eqref{const} implies \eqref{decouple6} this corresponds to a
decoupled free field theory combined with a single field $\phi^3$
theory. Note that since $\det \gamma = 0$ in this case there is always
a zero anomalous dimension.
To three loops,
\begin{align}
F (x) = {}& - \tfrac12 \, \vep - \tfrac34 \, x
- \tfrac{125}{144}\, x^2 - \big (\tfrac{33085}{20736} + \tfrac58 \zeta_3 \big ) x^3 \, ,
\nn \\
G ( x ) = {}& \tfrac{1}{12} +\tfrac{13}{432}\, x + \big (\tfrac{5195}{62208} -
\tfrac{1}{24} \zeta_3 \big ) x^2 \, .
\end{align}
Of course $ F(x) $ has no perturbative zeros for real couplings for $\vep>0$.

\section{Two Flavour Scalar Theory in
  \texorpdfstring{$\mathbf{4\boldsymbol{-}\boldsymbol{\vep}}$}{4-epsilon}
Dimensions}
\label{twoFlavFour}

The structure of fixed points can be fully analysed for just two
scalars \cite{ZiaW2}.
 A general two component $\phi^4$ theory is determined by the potential
\be
V(\phi) =  \tfrac{1}{24} \, \lambda_{ijkl} \phi_i \phi_j \phi_k \phi_l =
\tfrac{1}{24} \, \lambda_1 \, \phi_1{\!}^4 + \tfrac{1}{24} \, \lambda_2 \, \phi_2{\!}^4
+ \tfrac14 \, \lambda_0 \,   \phi_1{\!}^2\, \phi_2{}^2
+ \tfrac16 \, g_1 \, \phi_1{\!}^3 \, \phi_2 + \tfrac16 \, g_2 \, \phi_1 \phi_2{\!}^3 \, ,
\label{Vphi4}
\ee
with couplings $G_I=( \lambda_1, \lambda_2, \lambda_0,g_1,g_2)$. The transformations of the
couplings $G_I\to G_{\theta,I}$, generating equivalent theories as in \eqref{GMG},  induced by
\eqref{varp}, can be decomposed into a singlet and two real two dimensional
vectors.
 The transformations of the couplings may be reduced to irreducible components
 as
 \begin{align}
 I_1 =
 \lambda_1 + \lambda_2 + 2 \lambda_0 \, , \qquad
v_4= \begin{pmatrix} \tfrac14 ( \lambda_1 + \lambda_2 - 6\lambda_0) \\ g_1 - g_2 \end{pmatrix} \, , \qquad
v_2 =
\begin{pmatrix} \tfrac12(\lambda_1-\lambda_2)   \\  g_1 + g_2 \end{pmatrix} \, ,
\label{fix5}
 \end{align}
 with $v_2,v_4$ transforming as in \eqref{fix3}.
 The invariants in this case in addition to $I_1$ are then
 \begin{align}
 I_2 = {}&  \tfrac{1}{16} (\lambda_1 + \lambda_2 - 6 \lambda_0)^2 + (g_1-g_2)^2 \, , \nn \\
I_3 = {}&   \tfrac{1}{4} (\lambda_1 - \lambda_2 )^2 + (g_1 + g_2)^2 \, , \nn \\
I_4+i \, I_5 ={}& \big ( \tfrac12(\lambda_1-\lambda_2) - i (g_1+g_2)\big )^2
\big ( \tfrac14 (\lambda_1 + \lambda_2 - 6 \lambda_0) + i (g_1-g_2) \big ) \, ,
\label{Ifour}
\end{align}
where $I_4{\!}^2 + I_5{\!}^2 = I_2 \, I_3{\!}^2$.
$I_5$ is odd under
$\lambda_1 \leftrightarrow \lambda_2, \, g_1 \leftrightarrow g_2$.

 From \eqref{fix5} invariant subspaces are also given in this case by setting
 two of $I_1, \, v_2, \, v_4$ to zero or
 \be
 \lambda_1 = \lambda_2= - \lambda_0  \, , \ g_1=-g_2 \, ,\quad
 \lambda_1 = - \lambda_2\, , \lambda_0 = 0 \, , \ g_1=g_2 \, , \quad
 \lambda_1 = \lambda_2= 3 \lambda_0  \, , \ g_1=g_2 = 0 \, ,
\label{space1}
\ee
corresponding to $\lambda_{ijkl}$ being symmetric traceless, $\lambda_{ijkl} = \delta_{(ij} \, s_{kl)}, \,
s_{kk}=0$, $s_{11}=-s_{22}=\lambda_1, \, s_{12} = 2 g_1$,
and $\lambda_{ijkl} = 3 \lambda_0 \, \delta_{(ij} \, \delta_{kl)}$.
For the last case $V(\phi) =  \tfrac18 \lambda_0 \,
 (\phi_1{\!}^2 + \phi_2{\!}^2)^2$ describing an $O(2)$ invariant theory.
Setting $\lambda_1=\lambda_2 = -\lambda_0, \, g_1=g_2 =0$ gives a ${\mathbb Z}_4$ invariant
potential. For a decoupled theory we require
$g_{\theta,1}= g_{\theta,2} = \lambda_{\theta,0}=0$
for some $\theta$. This gives the constraint
$I_1{\!}^2  = 16\, I_2, \, I_1 I_3 = 4\, I_4$ or
\begin{align}
2\lambda_0{\!}^2 - (\lambda_1+\lambda_2) \lambda_0 + (g_1-g_2)^2 = {}& 0 \, ,\nn \\
(g_1+g_2)^2 ( 2\lambda_0 - \lambda_1-\lambda_2) + 2 (\lambda_1-\lambda_2)
(g_1{\!}^2 - g_2{\!}^2 ) - \lambda_0(\lambda_1 - \lambda_2)^2 = {}& 0 \, .
\label{decouple}
\end{align}

 For arbitrary quartic potentials $V(\phi)$ as in \eqref{Vphi4} the corresponding
$\beta$-function can be written as in \eqref{betaV3}, with now $\vep = 4-d$,
where  to three loops from \cite{Analogs}
\begin{align}
\bbeta_V = {}& \tfrac12 \, V_{ij} V_{ij} - \tfrac12\, V_{ij}V_{ikl} V_{jkl} +
\tfrac14 \, \lambda_{ijmn} \, \lambda_{kl mn} V_{ik}V_{jl}
- \tfrac{3}{16}\, \lambda_{iklm} \, \lambda_{jklm} V_{in} V_{jn} \nn \\
&{} + \lambda_{ijkl}\big  ( 2\, V_{im} V_{kln} V_{lmn} - \tfrac14 \, V_{mn} V_{ijm } V_{kln} \big ) \nn \\
&{} - \tfrac18 \, V_{ikl} V_{jkl} V_{imn} V_{jmn}
+ \tfrac12 \zeta_3 \, V_{ijk} V_{ilm} V_{jln} V_{kmn}  \, ,
\nn \\
&  \gamma_{ij}{}^{(2)} =  \tfrac{1}{12} \,  \lambda_{iklm} \lambda_{jklm} \, , \qquad
\gamma_{ij}{}^{(3)} = -\tfrac{1}{16} \,  \lambda_{iklm} \lambda_{jknp} \lambda_{lmnp} \, .
\label{BVfour}
\end{align}
The lowest order contributions to the $\beta$-functions are then easily determined,
taking now $G_I \to (4\pi)^2 \, G_I$,
\begin{align}
\beta_{\lambda_1}{\!}^{(1)}
={}& - \vep \, \lambda_1 + 3 (\lambda_1{\!}^2 + \lambda_0{\!}^2  +  2 \, g_1{\!}^2 ) \, , \nn \\
\beta_{\lambda_2}{\!}^{(1)}
=  {}& - \vep \, \lambda_2 + 3 (\lambda_2{\!}^2 + \lambda_0{\!}^2  +  2 \, g_2{\!}^2 ) \, , \nn \\
\beta_{\lambda_0}{\!}^{(1)}
= {}& - \vep \, \lambda_0 + (\lambda_1 + \lambda_2) \lambda_0 + 4\, \lambda_0{\!}^2  +  2 \, g_1{\!}^2
+  2 \, g_2{\!}^2 + 2\, g_1 g_2 \, , \nn \\
 \beta_{g_1}{\!}^{(1)}
={}& - \vep \, g_1 + 3 (2\, \lambda_0 g_1 + \lambda_0 g_2  +  \lambda_1 g_1) \, , \nn \\
 \beta_{g_2}{\!}^{(1)}
={}& - \vep \, g_2 + 3 (  \lambda_0 g_1 +  2 \, \lambda_0 g_2    +
\lambda_2 g_2) \, .
\end{align}
The anomalous dimension matrix is first non zero at two loops
\be
\gamma^{(2)} = \frac{1}{12} \begin{pmatrix} \lambda_1{\!}^2 + 3 \lambda_0{\!}^2 +
3 g_1{\!}^2 +  g_2{\!}^2 &
3 \lambda_0 ( g_1+g_2) + \lambda_1 g_2 + \lambda_2 g_1  \\
3 \lambda_0 ( g_1+g_2) + \lambda_1 g_2 + \lambda_2 g_1  &
\lambda_2{\!}^2 +3 \lambda_0{\!}^2 +  g_1{\!}^2 +  3 g_2{\!}^2
\end{pmatrix}  \, .
\ee

Finding  fixed points with multiple couplings is a non trivial exercise.
Analysing first the zeros for real couplings determined by the one loop
$\beta$-functions imposes various constraints. These constraints
reduce the representation content of the couplings to a single invariant so that
higher order contributions to the $\beta$-functions are determined by a function
of a single variable. There are three cases
\begin{align}
{\rm a})& \quad \lambda_{1} = \lambda_{2} \, , \ \ g_{1}= - g_{2} \, , \ \ 2 \, g_{1}{\!}^2 =
\lambda_0 ( \lambda_{1} - \lambda_0) \, , \quad  \
\Rightarrow  \ \beta_I(G) = G_I \, f_a(\lambda_1 + \lambda_0)  \, , \nn \\
{\rm b}) & \quad   \lambda_0{\!}^2 = \lambda_{1}\lambda_{2} = g_{1}g_{2} \, , \  \
g_1{\!}^2 = \lambda_{1}\lambda_0\, , \ \  g_ {2}{\!}^2 = \lambda_{2}\lambda_0\, ,  \ \
\lambda_0 \, g_{1} = \lambda_{1} g_{2} \, , \ \  \lambda_0 \, g_{2} = \lambda_{2} g_{1} \, , \nn \\
& \quad \ \Rightarrow \big ( \tfrac12 (\lambda_1-\lambda_2) + i (g_1+g_2) \big )^2 =
I_1 \, \big ( \tfrac14 ( \lambda_1 + \lambda_2 - 6 \lambda_0) + i (g_1 - g_2) \big ) \, , \nn
\\ & \quad  \
\Rightarrow  \ \beta_I(G) = G_I \, f_b(\lambda_1 + \lambda_2 + 2 \lambda_0)  \, , \nn \\
{\rm c}) & \quad   \lambda_{1} = \lambda_{2} = 3 \, \lambda_0 \, , \ \   g_{1} = g_{2} = 0 \, ,
\quad
\Rightarrow \   \beta_I(G) = G_I \, f_c(\lambda_0)  \, .
\label{fix4}
\end{align}
In case a) $I_2 = \frac{1}{16} \, I_1{\!}^2, \, I_3=I_4=I_5=0$ while in case b)
$4 I_2 =  I_3 = \frac{1}{4} \, I_1{\!}^2, \, I_4=\frac{1}{16} \, I_1{\!}^3, \, I_5=0$
and in case c) only $I_1$ is non zero.

From perturbative results to three loops the functions $f_{a,b,c}$ in \eqref{fix4}
and their zeros as an expansion in $\vep$ to ${\rm O}(\vep^3)$  are given by
\begin{align}
f_a(x) = f_b(x) ={}&  - \vep +3 x - \tfrac{17}{3}\, x^2 + \big (\tfrac{145}{8}  + 12 \zeta_3 \big ) x^3\
\, , \nn \\
&  x_* = \tfrac13\, \vep + \tfrac{17}{81} \, \vep^2 + \big ( \tfrac{709}{17496} - \tfrac{4}{27}
\zeta_3\big )  \, \vep^3
\, ,  \nn \\
f_c(x) = {}& -\vep + 10 x  - 60 x^2 + \big (617 + 384 \zeta_3 \big ) x^3\, ,  \nn \\
& x_* = \tfrac{1}{10} \, \vep + \tfrac{3}{50} \, \vep^2
+ \big ( \tfrac{103}{10000} -\tfrac{24}{625} \zeta_3\big )  \, \vep^3
\, .
\end{align}
The zeros determine  fixed points.
 Case ${\rm a})$ is equivalent to two decoupled Ising models, within the $\vep$-expansion,
 corresponding to taking  $\lambda_{0*}=g_{1*} = g_{2*}=0, \,  \lambda_{1*} = \tfrac13 \vep +{\rm O}(\vep^2)$
 and also to the theory obtained by taking $\lambda_{0*}= \lambda_{1*}, \, g_{1*} = g_{2*}=0, \,
 \lambda_{1*} =\tfrac16 \vep +{\rm O}(\vep^2)$.
 Case ${\rm b})$ is described by an
 Ising model and a free field theory, it is equivalent to taking  $\lambda_{2*}= \lambda_{0*}=g_{1*} = g_{2*}=0$
 when $\phi_2$ has no interaction. It is easy to check that these cases satisfy \eqref{decouple}.
 Case ${\rm c})$ is the $O(2)$ symmetric Heisenberg fixed point. Although the original
 theory defined by \eqref{Vphi4} has in general just a reflection ${\mathbb Z}_2$ symmetry, the fixed points all have at
least a ${\mathbb Z}_2 \times {\mathbb Z}_2 $ reflection symmetry.

 For case a) the left eigenvectors and eigenvalues of $[ \pr_I \beta_J(G)]|_{G=G_*}$
always include a zero mode given by
\be
\big ( {- 4 g_{1*}} \ {-4 g_{1*}} \ 4 g_{1*} \  \lambda_{1*}- 3 \lambda_{0*} \
{- \lambda_{1*}} + 3 \lambda_{0*} \big )  \, ,
\ee
and  to ${\rm O}(\vep^3)$ the results for the others are given by
\begin{align}
&  \big ( \lambda_{1*} \  \lambda_{1*}  \  \lambda_{0*} \  g_{1*} \  {-  g_{1*}} \big )  ,
\big (  \lambda_{1*}-\lambda_{0*} \  \lambda_{0*} -  \lambda_{1*} \  0 \ g_{1*} \ g_{1*}   \big ) ,  \ \
   \vep - \tfrac{17}{27}\, \vep^2 +
\big ( \tfrac{1603}{2916} + \tfrac{8}{9} \zeta_3 \big ) \vep^3  , \nn \\
 & \big ( { - 2 \lambda_{0*}} \  2 \lambda_{0*} \  0 \   \lambda_{1*} \
 \lambda_{1*} \big )   , \hskip 2.1cm
\tfrac{1}{54} \, \vep^2 + \tfrac{109}{5832} \, \vep^3 \, , \nn \\
& \big (  \lambda_{0*} \  \lambda_{0*} \ \tfrac13 (\lambda_{1*}- 2 \lambda_{0*})  \
{- g_{1*}} \ g_{1*} \big )  , \quad  - \tfrac13 \, \vep + \tfrac{19}{81} \, \vep^2
+ \big ( \tfrac{937}{8748} - \tfrac{8}{27} \zeta_3 \big ) \vep^3 \, .
\label{Ising}
   \end{align}The eigenvectors lie within the invariant subspaces in \eqref{space1}.
There is one slightly relevant operator. For case b) the corresponding results are
\begin{align}
&  \big (  {-4 g_{1*}}   \ 4g_{2*}   \  2(g_{1*} - g_{2*}) \ \lambda_{1*} - 3 \lambda_{0*} \
 - \lambda_{2*} + 3 \lambda_{0*}  \big )  ,
\quad  \ 0 \nn \\
&  \big ( {-4 g_{2*}} \  4g_{1*}  \   { -  2}(g_{1*} - g_{2*}) \  {-  \lambda_{2*}} + 3 \lambda_{0*} \
 \lambda_{1*} - 3 \lambda_{0*} \big ),
\quad  -  \vep
 + \tfrac{1}{108}\, \vep^2 + \tfrac{109}{11664} \, \vep^3  ,
\nn \\
& \big (  \lambda_{1*}  \  \lambda_{2*} \  \lambda_{0*} \  g_{1*} \ g_{2*}   \big )  ,
 \hskip 2.8 cm  \vep
  - \tfrac{17}{27}\, \vep^2 + \big ( \tfrac{1603}{2916} + \tfrac{8}{9} \zeta_3 \big ) \vep^3  ,
\nn \\
& \big ( \lambda_{2*}  \ \lambda_{1*} \  \lambda_{0*} \  {- g_{2*}} \ { - g_{1*} } \big )   ,
\hskip 2cm -  \vep   ,
\nn \\
 & \big (  6 \lambda_{0*} \  6 \lambda_{0*} \ \lambda_{1*} + \lambda_{2*} - 4 \lambda_{0*}  \
{- 3}( g_{1*} - g_{2*} ) \  3( g_{1*} - g_{2*} ) \big )  , \nn \\
& \hskip 6.25cm   - \tfrac23 \, \vep
+ \tfrac{19}{162} \, \vep^2+ \big ( \tfrac{937}{17496} - \tfrac{4}{27} \zeta_3 \big ) \vep^3 \, .
   \end{align}

 For the $O(2)$ symmetric case c) the left eigenvectors and
associated eigenvalues are
 \begin{align}
&  \big (  1 \  1  \  {-1} \  0 \  0 \big ) ,
\big ( 0 \  0 \  0 \ 1 \  {-1} \big )   ,   &  \hskip - 1cm &
  \tfrac15  \vep - \tfrac{1}{5}\, \vep^2 +
\big ( \tfrac{29}{2500} + \tfrac{192}{625} \zeta_3 \big ) \vep^3  , \nn \\
 & \big (  1 \  {-1} \  0 \   0 \ 0 \big )  ,
 \big ( 0 \ 0 \ 0 \ 1 \ 1 \big ) ,  &  \hskip - 1cm &
\tfrac45 \, \vep - \tfrac{1}{2} \, \vep^2 +  \big ( \tfrac{1777}{5000}  +
\tfrac{408}{625}\zeta_3 \big ) \, \vep^3 \, , \nn \\
& \big (  3 \  3 \ 1 \ 0  \ 0  \big )  ,  & \hskip - 1cm  &
\vep  -  \tfrac35 \, \vep^2
+ \big ( \tfrac{257}{500} + \tfrac{96}{125} \zeta_3 \big ) \vep^3 \, .
   \end{align}
All perturbations then correspond to irrelevant operators.

Assuming the constraints on the couplings in \eqref{fix4} we have to all orders
\begin{align}
\gamma \big |_{\lambda_1=\lambda_2, g_1 = - g_2, 2g_1{}^2
= \lambda_0(\lambda_1-\lambda_0)}
= {}& g_a ( \lambda_1 + \lambda_0 ) \, {\mathds 1} \, , \nn \\
\gamma \big |_{\lambda^2 = \lambda_{1}\lambda_{2} = g_{1}g_{2}, \,
g_1{\!}^2 = \lambda_{1}\lambda_0, \,  g_ {2}{\!}^2 = \lambda_{2}\lambda_0,  \,
\lambda_0 \, g_{1} = \lambda_{1} g_{2} , \,  \lambda_0 \, g_{2} = \lambda_{2} g_{1}}
= {}& g_b ( \lambda_1 + \lambda_2 + 2\lambda_0 ) \,
\left ( \begin{smallmatrix} \lambda_1 + \lambda_0 & g_1+g_2 \\ g_1+g_2  & \lambda_2 + \lambda_0
\end{smallmatrix} \right )\, , \nn \\
\gamma \big |_{\lambda_1=\lambda_2 = 3 \lambda_0, \, g_1=g_2=0}
= {}& g_c (\lambda_0 ) \, {\mathds 1} \, ,
\end{align}
with
\be
x\, g_b(x) = g_a(x) \, .
\ee
To three loops
\be
g_a(x) = \tfrac{1}{12} \, x^2 - \tfrac{1}{16} \, x^3 \, , \qquad
g_c(x) = x^2 - \tfrac{5}{2} \, x^3 \
\ee
It is easy to see, subject to the necessary conditions for case b), that
$\det \left ( \begin{smallmatrix} \lambda_1 + \lambda_0 & g_1+g_2 \\ g_1+g_2  & \lambda_2 + \lambda_0
\end{smallmatrix} \right ) = 0$ so that the eigenvalues are $\lambda_1+\lambda_2 + 2\lambda_0 ,
\, 0$.
The non zero anomalous dimensions at the fixed points to three loops are then
\be
\tfrac{1}{108} \big ( \vep^2 + \tfrac{109}{108} \, \vep^3 \big ) \, , \qquad
\tfrac{1}{100} \big ( \vep^2 + \tfrac{19}{20} \, \vep^3 \big ) \, ,
\ee
for the Ising model case and the $O(2)$ theory respectively.

The constraints in \eqref{fix4} reduce the RG flow to one dimensional
trajectories up to an $O(2)$ equivalence. Relaxing the restrictions such
as imposing just $v_2=0$ the RG flow is restricted to a two dimensional
surface since
\begin{align}
\beta_I(G) 
= {}& G_I \,  a(\lambda_1 + \lambda_0, J )
+ G^0{\!}_I \, b(\lambda_1+ \lambda_0,J) \, , \qquad
\gamma(G) 
=  c ( \lambda_1 + \lambda_0,J  ) \, {\mathds 1} \, , \nn \\
G_I = {}&  ( \lambda_1 , \, \lambda_1, \, \lambda_0 , \, g_1, \, - g_1 ) \, , \quad
G^0{\!}_I  =  ( 3, \, 3, \, 1,\, 0 , \, 0) \, , \qquad
J =  \lambda_0(\lambda_1-\lambda_0) - 2 \, g_1{\!}^2 \, .
\end{align}
In terms of \eqref{fix4} $f_a(x) = a(x,0), \ f_c(x)= a(4x,2x^2) + 2x \,  b(4x,2x^2) $.
From perturbation theory to three loops
\begin{align}
a(x,y)= {}& - \vep + 3 x - \tfrac13 ( 17 x^2 + 2y) + \tfrac18  \, x \big ((145 + 96\, \zeta_3) x^2 -168\, y \big ) \, , \nn \\
b(x,y) = {}& -y + 4 \, x \, y   -
\tfrac18  \big ((89 + 96\, \zeta_3) x^2  + 38\, y \big )y \, , \nn \\
c(x,y) = {}& \tfrac{1}{12} ( x^2-2y) - \tfrac{1}{16} \, x ( x^2 - 3y) \, .
\end{align}
Fixed points require $a(x,y)=b(x,y)=0$ unless $G_I \propto G^0{\!}_I $ requiring
$\lambda_1 = 3 \lambda_0, \, g_1=0$ so that $\lambda_1+\lambda_0=4\lambda_0 , J = 2\lambda_0{\!}^2$
and $c(4x, 2x^2) = x^2 -  \tfrac52\, x^3$ corresponding to
the $O(2)$ invariant theory. Perturbatively $b(x,y)=0$ requires $y=0$.
On this surface the RG flow links the fixed points corresponding to case a) and
case c) in \eqref{fix4}.

\section{Scalar Theories with Reduced Symetry in
\texorpdfstring{$\mathbf{6\boldsymbol{-}\boldsymbol{\vep}}$}{6-epsilon} Dimensions}
\label{redSymSix}

For $\N$ scalars $\phi_i$  the kinetic term has $O(\N)$ symmetry.
 However for a renormalisable theory in $6-\vep$ dimensions with a cubic
 potential there is of course no $O(\N)$ invariant theory.  For $\N=N+1$ and
  decomposing  $\phi_i = (\sigma, \vphi_i)$ there is an $O(N)$ invariant  theory
with $\sigma=\phi_0$ a singlet and $\vphi_i$ transforming as a vector.
As analysed in detail in \cite{Fei3loop} there is a fixed point
with $O(N)$ symmetry in $6-\vep$ dimensions. Here we consider perturbations
which break the $O(N)$ symmetry but maintain the reflection symmetry
under $\vphi_i \to - \vphi_i$ based on the potential with three couplings
\be
V(\phi) = \tfrac{1}{6} \, \lambda_{ijk} \phi_i \phi_j \phi_k =
\tfrac12 \, g \, \sigma \, \vphi_i \vphi_i
+ \tfrac12 \, h \, \sigma \, d_{ij} \vphi_i \vphi_j
+ \tfrac16 \lambda \, \sigma^3 \, ,
\label{VsixN}
\ee
where $d_{ij}=d_{ji}$ is assumed to satisfy, for a convenient choice of
normalisation,
\be
d_{ii} = 0 \, , \qquad d_{ik} \, d_{kj} = \delta_{ij} + b \, d_{ij} \, .
\label{drel}
\ee
For $h=0$ this is just the $O(N)$ invariant theory. Theories in which the
tensors $d_{ij}$ are related by $O(N)$ transformations are equivalent.
 Invariance under reflections $\sigma \to - \sigma$ ensures that
$(\lambda,g,h) \simeq (-\lambda,-g,-h)$ defining equivalent theories.

To lowest order the $\beta$-function in general is given by, after rescaling
$\lambda_{ijk} \to (4 \pi)^{\frac32} \lambda_{ijk}$,
\be
4 \, \beta_{ijk} = \frac{\pr}{\pr \lambda_{ijk}} A \, , \qquad
A = - \vep \, \lambda_{ijk} \lambda_{ijk}
-  \lambda_{ijk} \lambda_{ilm}\lambda_{jln} \lambda_{kmn}
+\tfrac14\,  \lambda_{ikl} \lambda_{jkl}\lambda_{imn} \lambda_{jmn} \, .
\label{A3gen}
 \ee
 Under RG flow to the IR $A$ decreases. However since the lowest
 order result for $A$ in \eqref{A3gen} has no global minimum, the couplings may become
 large under RG flow and the approximation loses its validity. At any local minimum
of $A$ at which $\beta_{ijk}=0$ with $\lambda_{* ijk} = {\rm O}(\vep)$ \cite{VicariZ}
 \be
 A_* = - \tfrac12 \, \vep \, \lambda_{* \hskip 0.5pt ijk}  \lambda_{* \hskip 0.5pt ijk}
 = - 6 \, \vep \, \tr ( \gamma_*) \, ,
\label{locA}
\ee
with $\gamma_*$ the lowest order contribution to the anomalous dimension at the
critical point.

For the three coupling theory in \eqref{VsixN} the general form \eqref{betaV31} determines
\begin{align}
\! \beta_\lambda
= {}& - \tfrac12 \, \vep \, \lambda   - \tfrac34 \, \lambda^3
 - N \big (  g^3- \tfrac14  \, (g^2 +h^2 ) \lambda
+ 3 \, g h^2  + b \, h^3 \big ) \, , \nn\\
\! \beta_g = {}& - \tfrac12 \, \vep \, g + \tfrac{1}{12} (N-8)  \, g^3
- ( g^2 +h^2) \lambda + \tfrac{1}{12}  \, \lambda^2  g + \tfrac{1}{12}(N-24) \, gh^2
 - \tfrac23 \, b \, h^3 \, , \nn \\
\! \beta_h = {}& - \tfrac12 \, \vep \, h + \tfrac{1}{12} (N-24)  \, g^2 h
- 2\, g \lambda \, h + \tfrac{1}{12}    \,  \lambda^2 h  -2 \, b \,  gh^2
- b \, \lambda h^2 +\big ( \tfrac{1}{12}(N-8)  - \tfrac23 \, b^2 \big )  h^3 \, ,
\label{blgh}
\end{align}
and also the anomalous dimensions for $\sigma, \, \vphi$ become
\begin{align}
\gamma_\sigma = {}& \tfrac{1}{12}\big ( N (g^2 + h^2) + \lambda^2 \big ) \, , \nn \\
\gamma_{\vphi \, ij} = {}& \tfrac16 (g^2 + h^2 ) \, \delta_{ij} +
\tfrac16 ( 2\, gh + b \, h^2 ) \, d_{ij} \, .
\label{gamsp}
\end{align}
Of course for $h=0$ these reduce to standard results for the $O(N)$ symmetric
theory.

For $h=0$ the fixed points for were analysed by Fei {\it et al} \cite{Fei3loop}.
In the $\vep$ expansion possible fixed points are in general determined by the lowest
order one loop contributions to the $\beta$-functions, higher  orders give an
expansion in powers of $\vep$ for the positions of the fixed points.
At large $N$ there are three inequivalent  $O(N)$ fixed points,
obtained by the vanishing of $\beta_\lambda, \, \beta_h$ in \eqref{blgh}, with non zero
real $(\lambda_*,g_*) \simeq (-\lambda_*,-g_*)$,  given by, to first order
in $\vep$,
\be
g_*{\!}^2  \approx \frac{6}{N} \, \vep \, , \quad
\lambda_*{\!}^2  \approx \frac{6^3}{N} \, \vep \, , \quad g_* \lambda_* > 0 \, ,
\quad \mbox{and} \quad g_*{\!}^2  \approx \frac{5}{N} \, \vep \, , \quad
\lambda_*{\!}^2  \approx  \vep .
\label{fixN}
\ee

Defining $x=\lambda/g$  and then writing the one loop $\beta$-functions as
\begin{align}
\beta_\lambda = {}& - \tfrac12 \, \vep \, \lambda + g^3 f_\lambda(x) \, , \qquad
f_\lambda(x)= - \tfrac14 (3\, x^3 +4  N - N \, x) \, , \nn \\
\beta_g = {}& - \tfrac12 \, \vep \, g + g^3 f_g(x) \, , \qquad
f_g(x) = \tfrac{1}{12}(N-8 - 12\, x + x^2) \, ,
\label{fgf}
\end{align}
the fixed point equations require
\be
f(x) = xf_g(x) - f_\lambda(x) =\tfrac16 \big( 5 x^3 - 6x^2 -(N+4) x + 6N\big )  = 0 \, .
\label{f6d}
\ee
For $N>0$ this cubic equation has coincident roots, so that
$f(x)=f'(x)=0$,  at
\be
\big (x_{{\rm crit}}, \,  N_{{\rm crit}} \big ) =
( 8.7453 , \, 1038.27 ) \, , \qquad
 \big (x'{\!}_{{\rm crit}}, \,  N'{\!}_{{\rm crit}} \big ) =
( 1.10339 , \, 1.02145 ) \, .
\label{bifurc}
\ee
For $N'{\!}_{{\rm crit}}<N<N_{{\rm crit}}$ there
is only one real solution  of $f(x)=0$ with $x<0$  for any $N>0$.
For $N>N_{\rm crit}$ there  are two roots for
$x>0$, which merge at $x= x_{{\rm crit}} $ when
$(\lambda^2, N g^2) = ( 0.4581, 6.218)\vep$, and one root with $x<0$.
For real couplings it is necessary that at the fixed point $f(x_*)=0$ then
$f_g(x_*)>0$ or $(x_*-6)^2 >44-N$.
For large $N$ the roots are $6, \, \pm \sqrt{\frac{N}{5}}$
which  correspond to the three inequivalent fixed points given by \eqref{fixN}.
For $N=1$ $f(x)$ has roots at $x=\pm 1, \, \tfrac65$ but
there is just one inequivalent fixed point with real couplings when
$\lambda_* = - g_*$ and $\lambda_*{\!}^2 = \vep$. The roots at $x=1,\tfrac65$ correspond
to the fixed points which merge when $N\to N'{\!}_{{\rm crit}} $ and give $g^2<0$, and
so both $\lambda,g$ are imaginary. In contrast to four dimensions, as a consequence of the
reflection symmetry $(\lambda,g)\sim -(\lambda,g)$, the stable fixed point is not unique.

A discussion of the $\vep,\vep^2$ corrections to the bifurcation point in \eqref{bifurc}
is given in appendix \ref{bifSix}.

The stability matrix in this case is
\be
M = \begin{pmatrix} \pr_\lambda \beta_\lambda & \pr_g
\beta_\lambda \\ \pr_\lambda \beta_g & \pr_g \beta_g \end{pmatrix}
\bigg |_{\lambda=\lambda_*,g=g_*,h=0} \ ,
\ee
and, with the $\beta$-functions given by \eqref{blgh} for $h=0$.
\be
\det M = - \vep \, g_*{\!}^2 f'(x_*) \, , \quad \tr M =
\vep - g_*{\!}^2 f'(x_*) \, , \quad  x_* = \lambda_* /g_* \, .
f(x_*)= 0 \, ,
\ee
This determines the eigenvalues of $M$ to lowest order in $\vep$
\be
\kappa_0 = \vep\, , \qquad \kappa_1 = - \lambda_* f'(x_*) \, .
\label{eigen6}
\ee
For $(\lambda_*,g_*)$  to be a stable fixed point the eigenvalues  must be
positive which requires $ f'(x_*) <0$. At the roots for large $N$
$f'(x*) \approx - \frac16 \, N, \, \frac13 \, N$.
When $f'(x_*) = 0$ there is
a marginal operator. For any $x_*$ the corresponding $g_*$ is determined to
this order by $g_*{\!}^2 = \tfrac12\, \vep/f_g(x_*)$.

From \eqref{A3gen} for $h=0$
\be
A = - \vep ( 3N\, g^2 + \lambda^2)
- 2 N\, g^3 (  g + 2 \, \lambda ) - \lambda^4 \
+ \tfrac14 ( N \, g^2 + \lambda^2)^2 \, ,
\ee
which satisfies
\be
\frac{\pr}{\pr g} A = 12N \, \beta_g \, , \qquad \frac{\pr}{\pr \lambda} A = 4 \,
\beta_\lambda \, ,
\ee
and at a fixed point
\be
A_* = - \tfrac12 ( 3 N g_*{\!}^2 + \lambda_*{\!}^2 )\vep \, ,
\ee
in accord with \eqref{locA}. This gives
\be
A_* = - 3 \, \frac{3N + x^2}{N-8-12 \, x + x^2} \, \vep^2 \bigg |_{f(x)=0} \, ,
\ee
and we may then obtain
\be
A_* \ge A_{*{\rm min}} = - 3 \,
\frac{3 \, N_{{\rm crit}} + x_{{\rm crit}}{}^2}
{  N_{{\rm crit}}  -8-12 \, x_{{\rm crit}}  + x_{{\rm crit}}{}^2} \,
\vep^2 \approx - 9.557\, \vep^2 \, .
\ee
The minimum is attained when the fixed points merge.
At the large $N$ fixed points in \eqref{fixN}
\be
A_*  \approx - 9 \, \vep^2 \, , \ - 8 \, \vep^2  \, ,
\ee
with $A_*$ lower at the stable fixed point.

The $h=0$ and $N=1$  results here are related to the discussion in
section \ref{redSymSix} by taking $\lambda_1 = \lambda, \, g_1 = g$ and $\lambda_2=g_2=0$.
The fixed point obtained with $\lambda_* = - g_*$ has a
${\mathbb Z}_3$ symmetry as described there. For $\lambda= g$  the condition
\eqref{decouple6} is satisfied and the theory becomes the sum of two
decoupled $N=0$ theories, with $\lambda_1=\lambda_2 = \lambda/\sqrt 2, \,
g_1=g_2 =0$.

However from \eqref{blgh} the large $N$  fixed point is unstable for
non zero $h$ since
\be
\pr_h \beta_h \big |_{g=g_*, \lambda=\lambda_*,h=0} = - \tfrac43 \, g_*{\!}^2
- g_* \lambda_* <0 \quad \mbox{if} \quad N > N_{{\rm crit} }  \, ,
\ee
so that the coupling $h$ corresponds to a relevant operator. To find fixed
points which are reached for non zero $h$ we consider the linear
transformation
\be
g= \frac{1}{1+\alpha^2} \, ( g' + \alpha^2\, h') \, , \qquad h= \frac{\alpha}{1+\alpha^2} (g' - h') \, , \qquad
\alpha^2 - 1 = \alpha\,  b \, .
\label{gab}
\ee
and the $\beta$-functions become
\begin{align}
\beta_\lambda
= {}& - \tfrac12 \, \vep \, \lambda   - \tfrac34 \, \lambda^3
 - m \big (  g'{\hskip 0.5pt}^3 - \tfrac14  \, \lambda \, g'{\hskip 0.5pt}^2 \big  )
 -  n \big ( h' \hskip 0.5pt{}^3 -  \tfrac14\,  \lambda \, h' \hskip 0.5pt{}^2  \big )
\, , \nn\\
\beta_{g'} = {}& - \tfrac12 \, \vep \, g'
+ \tfrac{1}{12} (m-8)  \, g'{\hskip 0.5pt}^3
-\lambda\, g' \hskip 0.5pt{}^2 + \tfrac{1}{12} \,\big (n  \,
h' \hskip 0.5pt{}^2 + \lambda^2\big ) g'\, , \nn \\
\beta_{h'} = {}& - \tfrac12 \, \vep \, h'
+ \tfrac{1}{12}\big  (m  \, g'{\hskip 0.5pt}^2 + \lambda^2 \big ) \, h' -  \lambda\,
h' \hskip 0.5pt{}^2
+ \tfrac{1}{12} \, \big (n  - 8 \big )  h' \hskip 0.5pt{}^3\, ,
\label{blg}
\end{align}
for
\be
m = \frac{1}{1+\alpha^2} \, N \, , \qquad n = \frac{\alpha^2}{1+\alpha^2} \, N \, .
\label{Np}
\ee
For $h'=0$ the $\beta$-functions are just those of the unperturbed $O(m)$ theory, for
$m$ an integer, and so have the same fixed points, while  $n$ fields become free.
So long as $m  \, g'{\hskip 0.5pt}^2 + \lambda^2 > \vep$ the $h'$ coupling is
irrelevant so under RG flow the couplings are attracted to the sub manifold
corresponding to $h'=0$.
A similar reduction arises for $g'=0$ taking $  m \leftrightarrow n$. There
are also fixed points for $g',h'$ non zero. For $x=g'/\lambda, \, y = h' /\lambda$
the vanishing of the lowest order $\beta$-functions in \eqref{blg} requires
\be
m ( x^3 - \tfrac16\, x^2) + n ( y^3 - \tfrac16 \, y^2) = \tfrac23 \, x^2 + x -\tfrac56
= \tfrac23 \, y^2 + y -\tfrac56 \, ,
\label{xy3}
\ee
with then, for any solution of \eqref{xy3}, $\lambda$ at the fixed point
is given by
\be
- \tfrac12 \, \vep = \lambda_*{\!}^2 \big ( m ( x^3 - \tfrac14 \, x^2) +  n ( y^3 - \tfrac14 \, y^2 )
+ \tfrac34 \big ) \, .
\ee
Consistency of \eqref{xy3} requires $x=y$ or $x+y= - \tfrac32$.
In the first case then from \eqref{gab} $h=0$ and $g'=h'=g$ and
the theory reduces to the unbroken $O(N)$ theory and has the same
fixed points as in \eqref{fixN} for large $N$.
For the second  case if $m  \gg n $ then we may take
$y\approx - \tfrac32$ and
$m \, x^2 \approx 5 (1 - \tfrac92\, n )$. With $n $ taking integer
values this does not lead to real solutions. The two cases coincide for $x=y = - \tfrac34$
which requires $N=\tfrac29$.

To determine the possible values of $\alpha$ which are given in terms of $b$ in \eqref{gab}
it is necessary to find the various solutions of \eqref{drel}.
To this end we consider eigenvectors satisfying $d_{ij} \, v_j
= \lambda \, v_i$ where there are two possible eigenvalues
$\lambda_{\pm} = \tfrac12 \big ( b \pm  \sqrt{b^2 + 4} \big )$. Assuming $r$ $\lambda_+$
and $N-r$ $\lambda_-$ the trace condition gives
\be
r(N-r) \, b^2 = (N-2r)^2  \, , \quad r=1,\dots, N-1 \, ,
\ee
giving, for $b=(N-2r)/\sqrt{r(N-r)}$,
\be
\lambda_+ = \sqrt{\frac{N-r}{r}} \, , \ \ \lambda_- = - \sqrt{\frac{r}{N-r}} \, .
\ee
Other solutions are obtained for $r\to N-r$.
The solutions for $\alpha$ are then
\be
\alpha = \lambda_{\pm} \quad \Rightarrow \quad m = r , \, N-r \, .
\ee
The relevant perturbations corresponding to the coupling $h$
then break the symmetry $O(N) \to O(m) \times O(n)$
with $m+n=N$, depending on the solutions chosen for \eqref{drel}.
The RG flow  leads to a non trivial CFT with $O(m)$ or $O(n)$
symmetry, assuming $m$ or $n$ are large enough, while the
remaining fields become free. Assuming \eqref{gamsp} and \eqref{gab} then
\begin{align}
\gamma_\sigma = {}& \tfrac{1}{12}\big ( m \,  g'{\hskip 0.5pt}^2  +
n \,  h'{\hskip 0.5pt}^2  + \lambda^2 \big ) \, , \nn \\
\gamma_{\vphi \, ij} = {}& \tfrac{1}{6} \big (
g'{\hskip 0.5pt}^2  \,  \delta_{+\, ij} + h'{\hskip 0.5pt}^2  \, \delta_{- \, ij} \big )  \, ,
\label{gamsp2}
\end{align}
for $\delta_{\pm}$ projection operators given by
\be
\delta_{+\, ij} = \frac{1}{1+\alpha^2} \big ( \delta_{ij} + \alpha \, d_{ij}  \big ) \, , \qquad
\delta_{-\,  ij} = \frac{1}{1+\alpha^2} \big ( \alpha^2 \,  \delta_{ij} - \alpha \, d_{ij}  \big ) \, .
\label{project}
\ee
The
eigenvalues of $\gamma_{\vphi\, ij}$ are $\tfrac16 \,  g'{\hskip 0.5pt}^2, \,
\tfrac16 \,  h'{\hskip 0.5pt}^2 $ according to
whether the eigenvalue of $d_{ij}$ is $\alpha, - 1/\alpha$.

The results for the $O(m) \times O(n)$ theory can be extended to higher loops. Thus
the two loop $\beta$-functions are given by applying \eqref{betaV32}
\begin{align}
\beta_\lambda{\!}^{(2)}
= {}& -\tfrac{1}{144}  \Big ( 125 \,  \lambda^5
 + m\, g'{\hskip 0.5pt}^2 \big (  24 \, g'{\hskip 0.5pt}^3 + 322\, \lambda \,
g'{\hskip 0.5pt}^2 + 60 \, g' \lambda^2 -  31 \, \lambda^3\big  ) \nn \\
\noalign{\vskip -4pt}
& \hskip 2.8cm{}
 + n \, h'\hskip 0.5pt{}^2 \big ( 24 \, h' \hskip 0.5pt{}^3 +  322\,
\lambda \, h' \hskip 0.5pt{}^2  + 60 \, h' \lambda^2 - 31 \, \lambda^3 \big ) \Big )
\, , \nn\\
\beta_{g'}{\!}^{(2)} = {}&
- \tfrac{1}{432}\, g' \Big (   (86m+536)  \, g'{\hskip 0.5pt}^4 - 12(11m -30) g'{\hskip 0.5pt}^3 \lambda
+  (11m +628) g'{\hskip 0.5pt}^2 \lambda^2 + 24\,  g' \lambda^3  \nn \\
\noalign{\vskip -4pt}
& \hskip 1.8cm{}- 13\, \lambda^4 -  n  \, h' \hskip 0.5pt{}^2 \big (   2 \, h' \hskip 0.5pt{}^2 + 48\,
h'  \lambda - 11\,  \lambda^2 + 20 \, g' \hskip 0.5pt{}^2 - 108\, g' h' + 84\, g'\lambda \big ) \Big ) \, , \nn \\
\beta_{h'}{\!}^{(2)} = {}& - \tfrac{1}{432}\, h' \Big (   (86 n+536)  \, h'{\hskip 0.5pt}^4 -
12(11n -30) h'{\hskip 0.5pt}^3 \lambda
+  (11n +628) h'{\hskip 0.5pt}^2 \lambda^2
+ 24\,  h' \lambda^3 - 13\, \lambda^4 \nn \\
\noalign{\vskip -4pt}
& \hskip 1.8cm{}- m \,  g' \hskip 0.5pt{}^2 \big (   2 \, g' \hskip 0.5pt{}^2 + 48\,
g'  \lambda - 11\,  \lambda^2 + 20 \, h' \hskip 0.5pt{}^2 - 108\, g' h' + 84\, h'\lambda \big ) \Big ) \, ,
\label{blg2}
\end{align}
and for the two loop anomalous dimensions
\begin{align}
\gamma_\sigma{\!}^{(2)} = {}& \tfrac{1}{432}\Big ( m \, g'{\hskip 0.5pt}^2
( 2  g'{\hskip 0.5pt}^2 - 11 \lambda^2  + 48 \lambda g' )
+ n \,  h'{\hskip 0.5pt}^2 ( 2  h'{\hskip 0.5pt}^2 - 11 \lambda^2
+ 48 \lambda h' ) + 13 \lambda^4 \Big ) \, , \nn \\
\gamma_{\vphi \, ij}{\!}^{(2)} = {}& - \tfrac{1}{432} \Big (
g'{\hskip 0.5pt}^2\big ( (11 m -26) g'{\hskip 0.5pt}^2 + 11  n \, h'{\hskip 0.5pt}^2
- 48\, g' \lambda + 11\, \lambda^2 \big )  \,
\delta_{+\, ij} \nn \\
\noalign{\vskip -4pt}
& \hskip 1.3cm {}
+ h'{\hskip 0.5pt}^2 \big ( 11 m\,  g'{\hskip 0.5pt}^2 + (11 n
- 26 ) h'{\hskip 0.5pt}^2 - 48 \, h' \lambda + 11\, \lambda^2 \big )  \, \delta_{- \, ij} \Big )
\, .
\label{gamsp3}
\end{align}

\subsection{\texorpdfstring{$\boldsymbol{O(m)\times O(n)}$}{O(m) x O(n)} Theories in
\texorpdfstring{$\boldsymbol{6-\vep}$}{6-epsilon} Dimensions}

By letting $\phi_i = ( \sigma, \Phi_{ar}, \rho_{ab} )$, $a,b =1, \dots m, \, r = 1, \dots n$ and
where  $\rho_{ab} = \rho_{ba}, \, \rho_{aa}=0$  are
$\frac12(m-1)(m+2)$ additional scalars, the $O(N)$ symmetric
theory can be broken to $O(m) \times O(n) /{\mathbb Z}_2$, $N= m \, n$,
by extending the potential to
\be
V = \tfrac12 \, g_\sigma \, \sigma \, \Phi^2 +   \tfrac12 \, g_\rho \, \rho_{ab} \Phi_{ar} \Phi_{br}
+ \tfrac16 \, \lambda_\sigma \, \sigma^3  + \tfrac16 \, \lambda_\rho  \tr(\rho^3 ) +
\tfrac12 \,  \tlam \, \sigma \tr(\rho^2) \, .
\label{Vsix}
\ee
In general there are then five couplings.
For $m=2$ $\tr( \rho^3)=0$ so that the $\lambda_\rho$ coupling is irrelevant. By considering
reflections of $\sigma$ and $\rho$ it is clear that there are equivalence relations
\be
(g_\sigma, g_\rho , \lambda_\sigma , \lambda_\rho, \tlam) \simeq
(- g_\sigma, g_\rho , - \lambda_\sigma , \lambda_\rho, - \tlam) \simeq
(g_\sigma, - g_\rho , \lambda_\sigma , - \lambda_\rho,\tlam) \, .
\label{g6equiv}
\ee
The restricted theory with $g_\sigma, \lambda_\sigma$ and $\tlam$ set to zero has been discussed
in \cite{Herbut} and the general theory in \cite{Gracey2}. For this case there is a tractable large
$n$ limit valid for arbitrary $d$ which links perturbative results in six and four dimensions.

The potential \eqref{Vsix}  defines a renormalisable theory in $d=6-\vep$ dimensions and the $\beta$-functions have the form
\begin{align}
\beta_{g_\sigma} = {}& \bbeta_{g_\sigma}
+ ( -  \tfrac12 \, \vep  +  \gamma_\sigma + 2 \, \gamma_\Phi ) g_\sigma \, , \qquad
\beta_{g_\rho} =  \bbeta_{g_\rho}  +  ( -  \tfrac12 \, \vep + \gamma_\rho + 2 \, \gamma_\Phi ) g_\rho \, , \nn \\
\beta_{\lambda_\sigma} = {}& \bbeta_{\lambda_\sigma}
+  \big ( -\tfrac12 \, \vep  + 3 \, \gamma_\sigma  \big )  \lambda_\sigma \, , \hskip 1.35cm
\beta_{\lambda_\rho} =  \bbeta_{\lambda_\rho} +
\big ( -  \tfrac12 \, \vep + 3 \, \gamma_\rho \big )  \lambda_\rho \, , \nn \\
\beta_{\tlam} = {}& \bbeta_{\tlam} + ( -  \tfrac12 \, \vep +  \gamma_\sigma  + 2\, \gamma_\rho)  \tlam \, ,
\end{align}
As a consequence of \eqref{g6equiv} the $\beta$-functions satisfy the identities
\begin{align}\label{bequiv}
\beta_{g_\sigma,\lambda_\sigma,\tlam}(g_\sigma,g_\rho,\lambda_\sigma,\lambda_\rho,\tlam) = {}&
- \beta_{g_\sigma,\lambda_\sigma,\tlam}(-g_\sigma,g_\rho,-\lambda_\sigma,\lambda_\rho,-\tlam) \, , \nn \\
\beta_{g_\rho,\lambda_\rho}(g_\sigma,g_\rho,\lambda_\sigma,\lambda_\rho,\tlam) = {}&
 \beta_{g_\rho,\lambda_\rho}(-g_\sigma,g_\rho,-\lambda_\sigma,\lambda_\rho,-\tlam) \, , \nn \\
 \beta_{g_\sigma,\lambda_\sigma,\tlam}(g_\sigma,g_\rho,\lambda_\sigma,\lambda_\rho,\tlam) = {}&
 \beta_{g_\sigma,\lambda_\sigma,\tlam}(g_\sigma,-g_\rho,\lambda_\sigma,-\lambda_\rho,\tlam) \, , \nn \\
\beta_{g_\rho,\lambda_\rho}(g_\sigma,g_\rho,\lambda_\sigma,\lambda_\rho,\tlam) = {}&
-  \beta_{g_\rho,\lambda_\rho}(g_\sigma,-g_\rho,\lambda_\sigma,-\lambda_\rho,\tlam) \, .
 \end{align}

At one loop, using $\pr_i = ( \pr_\sigma, \pr_{\Phi,ar} , \pr_{\rho,ab})$,
where $\pr_{\rho,ab} \, \rho_{cd} = \tfrac12( \delta_{ac} \, \delta_{bd} +
\delta_{ad} \, \delta_{bc} ) - \tfrac{1}{m} \, \delta_{ab} \, \delta_{cd}$
and $\pr_i \pr_i = \pr_\sigma{\!}^2 +  \pr_{\Phi,ar}  \pr_{\Phi,ar} +
\pr_{\rho,ab}\pr_{\rho,ab}$ then from \eqref{betaV3}
\begin{align}
\bbeta_{g_\sigma}{\!}^{(1)}  = {}& -  g_\sigma{\!}^3 -
g_\sigma{\!}^2 \lambda_\sigma
- \tfrac{1}{2m}(m-1)(m+2) \,\big  (g_\sigma + \tlam \big )   g_\rho{\!}^2  \nn \\
\bbeta_{g_\rho}{\!}^{(1)} = {}&  -  g_\sigma{\!}^2 g _\rho - 2 \, g_\sigma g_\rho\, \tlam
-  \tfrac{1}{2m} (m-2) \, g_\rho{\!}^3   - \tfrac{1}{4m}(m-2)(m+4)  g_\rho{\!}^2 \lambda_\rho \nn \\
\bbeta_{\lambda_\sigma}{\!}^{(1)}  = {}&  - N\, g_\sigma{\!}^3 -
\lambda_\sigma{\!}^3 - \tfrac12(m-1)(m+2) \tlam{}^3  \, , \nn \\
\bbeta_{\lambda_\rho}{\!}^{(1)}  = {}&  - n\, g_\rho{\!}^3 -
\tfrac{1}{8m}(m^2+4m-24)\, \lambda_\rho{\!}^3 - 3\, \lambda_\rho \,\tlam{}^2  \, , \nn \\
\bbeta_{\tlam}{\!}^{(1)}  = {}& - n\, g_\sigma \, g_\rho{\!}^2 -
\lambda_\sigma \,\tlam{}^2  -  \tlam{}^3 - \tfrac{1}{4m} (m-2)(m+4)  \lambda_\rho{\!}^2 \tlam \, ,
\end{align}
with anomalous dimensions
\begin{align}
\gamma_\Phi{\!}^{(1)} ={}& \tfrac{1}{12} \big ( 2\, g_\sigma{\!}^2 +  \tfrac{1}{m}(m-1)(m+2) \, g_\rho{\!}^2 \big ) \, , \nn \\
\gamma_\sigma{\!}^{(1)} ={}& \tfrac{1}{12} \big ( N\, g_\sigma{\!}^2 + \lambda_\sigma{\!}^2 +
\tfrac{1}{2}(m-1)(m+2) \, \tlam{}^2 \big ) \, , \nn \\
\gamma_\rho{\!}^{(1)}  ={}& \tfrac{1}{12} \big ( n\, g_\rho {\!}^2 +  \tfrac{1}{4m}(m-2)(m+4) \, \lambda_\rho{\!}^2
+ 2 \, \tlam{}^2 \big ) \, .
\end{align}
At two loops the results for the $\beta$-functions are rather non trivial but can be obtained from \eqref{betaV3}
\begingroup
\allowdisplaybreaks
\begin{align}
\bbeta_{g_\sigma}{\!}^{(2)}  = {}& -  \tfrac{1}{72}(11N+98) \, g_\sigma{\!}^5  + \tfrac{1}{36}(7N - 38) \,
g_\sigma{\!}^4 \lambda_\sigma  - \tfrac{101}{72} \, g_\sigma{\!}^3 \lambda_\sigma{\!}^2
- \tfrac{1}{18} \,    g_\sigma{\!}^2 \lambda_\sigma{\!}^3 \nn \\
&{} +\tfrac{1}{2}(m-1)(m+2) \big ( \tfrac{7}{72}\, g_\sigma{\!}^3   \tlam{}^2
- \tfrac14 \, g_\sigma{\!}^2   \tlam{}^3 + \tfrac{7}{36}\, g_\sigma{\!}^2 \lambda_\sigma  \tlam{}^2 \big ) \nn \\
&{}- \tfrac{1}{2m}(m-1)(m+2) \big (  \tfrac{49}{18} \, g_\sigma{\!}^3
 + \tfrac{37}{18} \, g_\sigma{\!}^2  \tlam  +  \tfrac{5}{9} \, g_\sigma{\!}^2 \tlam
+ \tfrac32 \, g_\sigma \tlam \lambda_\sigma
 + \tfrac{101}{36} \, g_\sigma  \tlam{}^2  +  \tfrac14 \, \tlam{}^2 \lambda_\sigma
 - \tfrac{5}{36} \,   \tlam{}^3 \big ) g_\rho{\!}^2 \nn \\
 &{}+ \tfrac{1}{144m^2}(m-1)(m+2) \big ( 2(7N - m^2 -10m+20) \, \tlam
 - (11 N - 5m^2  +49 m -98) \, g_\sigma \big ) g_\rho{\!}^4
 \nn \\
  &{}- \tfrac{1}{16m^2}(m-1)(m-2)(m+2)(m+4) \big ( g_\sigma g_\rho{\!}^3\lambda_\rho
  + 3 \, g_\rho{\!}^3 \tlam \lambda_\rho + \tfrac19 \, g_\rho{\!}^2 \tlam \lambda_\rho{\!}^2
 - \tfrac{7}{36}  \, g_\sigma g_\rho{\!}^2 \lambda_\rho{\!}^2  \big ) \, , \nn \\
\noalign{\vskip  4 pt}
\bbeta_{g_\rho}{\!}^{(2)} = &{}
- \tfrac12 \, g_\sigma{\!}^3 g_\rho\,  \lambda_\sigma -  \tfrac32 \, g_\sigma{\!}^2 g_\rho \, \tlam \lambda_\sigma
- \tfrac12 \, g_\sigma  g_\rho \,  \tlam{}^2 \lambda_\sigma
+  \tfrac{7}{72}\, g_\sigma{\!}^2 g _\rho\,  \lambda_\sigma{\!}^2 +
\tfrac{7}{36}\, g_\sigma g _\rho\,   \tlam  \lambda_\sigma{\!}^2
\nn \\
&{} -  \tfrac{1}{2m} (m-2)(m+4)\big ( \tfrac{5}{18} \, g_\sigma{\!}^2  g_\rho{\!}^2  \lambda_\rho
+  \tfrac{9}{4} \, g_\sigma  g_\rho{\!}^2  \, \tlam \lambda_\rho
+  \tfrac{13}{72} \,  g_\rho{\!}^2 \, \tlam{}^2   \lambda_\rho
+ \tfrac{11}{72} \, g_\sigma  g_\rho \, \tlam \lambda_\rho{\!}^2 \big )
\nn \\
&{} + \tfrac{1}{288m^2}(m-2)(m+4) \big ( (14N + 7m^2 - 38m+76) g_\rho{\!}^4 \lambda_\rho
- \tfrac12 (18m^2 +101m -418)  g_\rho{\!}^3 \lambda_\rho {\!}^2 \nn \\
\noalign{\vskip -2pt}
& \hskip 9.5cm
{} +  \tfrac14 (5 m^2 -8 m  +104)  g_\rho{\!}^2 \lambda_\rho {\!}^3 \big )
\nn \\
&{}- \tfrac{1}{144m^2}\big ( N ( m-2)( 9m  + 29 )  + 22 m^3 + 149 m^2 - 196 m + 196
\big ) g_\rho{\!}^5 \nn \\
&{}-  \tfrac{1}{36m }( 18 N + 2  m^2 +49  m - 98 ) g_\sigma{\!}^2  g_\rho{\!}^3
 +  \tfrac{7}{72}(N-14)\, g_\sigma{\!}^4 g _\rho \nn \\
 &{}+ \tfrac{1}{36m} ( 7 N - 11 m^2 -47 m + 94)  g_\sigma g _\rho{\!}^3 \, \tlam
 +   \tfrac{1}{36}(7N- 40)\, g_\sigma{\!}^3 g _\rho \,  \tlam  \nn \\
 &{}- \tfrac{1}{72m} (18 m^2 + 47 m - 94) \, g _\rho{\!}^3 \, \tlam{}^2
+  \tfrac{1}{144} (7 m^2 + 7 m - 446) \, g_\sigma{\!}^2 g _\rho \, \tlam{}^2 \nn \\
\noalign{\vskip -2pt}
& \hskip 9cm
+  \tfrac{1}{72} (7 m^2 + 7 m -  22) \, g_\sigma g _\rho \, \tlam{}^3 \, , \nn \\
\bbeta_{\lambda_\sigma}{\!}^{(2)}  = {}&  - \tfrac16 N\, g_\sigma{\!}^5
- \tfrac{9}{4}N \,  g_\sigma{\!}^4 \lambda_\sigma - \tfrac{3}{4}N \,  g_\sigma{\!}^3 \lambda_\sigma{\!}^2
+ \tfrac{7}{24}N \,  g_\sigma{\!}^2 \lambda_\sigma{\!}^3
-  \tfrac{23}{24}  \, \lambda_\sigma{\!}^5 \nn \\
&{} - n\,  \tfrac12(m-1)(m+2) \big ( \tfrac16  \,   g_\sigma{\!}^3 g_\rho{\!}^2
+ \tfrac94 \,  g_\sigma{\!}^2 g_\rho{\!}^2 \tlam + \tfrac34 \,  g_\sigma g_\rho{\!}^2 \tlam{}^2
-  \tfrac{7}{24} \,   g_\rho{\!}^2 \tlam{}^3 \big )  \nn \\
&{}- \tfrac12(m-1)(m+2) \big ( \tfrac16\, \tlam{}^5 + \tfrac94 \, \tlam{}^4 \lambda_\sigma
+ \tfrac34 \, \tlam{}^3 \lambda_\sigma{\!}^2 -\tfrac{7}{24} \, \tlam{}^2 \lambda_\sigma{\!}^3 \big )  \nn \\
&{}- \tfrac{23}{192m}(m-1)(m-2)(m+2) (m+4) \, \tlam{}^3 \lambda_\rho{\!}^2 \, , \nn \\
\noalign{\vskip  4 pt}
\bbeta_{\lambda_\rho}{\!}^{(2)}  = {}&  - n \big ( \tfrac16\,  g_\rho{\!}^3  g_\sigma{\!}^2 +
 \tfrac92 \,  g_\rho{\!}^3  g_\sigma \tlam + \tfrac34 \,  g_\rho{\!}^3  \tlam{}^2
 + \tfrac32 \,  g_\rho{\!}^2  g_\sigma \tlam \lambda_\rho -
 \tfrac{7}{12}  \,  g_\rho{\!}^2  \tlam{}^2 \lambda_\rho \big ) \nn \\
 &{}+ \tfrac{7}{24} N \, g_\sigma{\!}^2  \tlam{}^2 \lambda_\rho - \tfrac52\, \tlam{}^3 \lambda_\rho \lambda_\sigma
 +  \tfrac{7}{24} \, \tlam{}^2 \lambda_\rho \lambda_\sigma{\!}^2 \nn \\
 &{}+ n \big ( \tfrac{1}{24m} ( 7m^2 - 2m +4) \, g_\rho{\!}^5 - \tfrac{3}{16m} ( m^2  +6 m  -24) \, g_\rho{\!}^4 \lambda_\rho\nn \\
 \noalign{\vskip -1pt}
 & \hskip 2.5cm {} - \tfrac{3}{32m} ( m^2 + 4m -24) \,  g_\rho{\!}^3 \lambda_\rho{\!}^2
+ \tfrac{7}{192m} ( m^2  +4 m  - 24) \, g_\rho{\!}^2 \lambda_\rho{}^3  \big ) \nn \\
&{} +  \tfrac{1}{48}(7m^2+ 7m- 282 )\, \tlam{}^4 \lambda_\rho  - \tfrac{1}{96m} ( 87m^2 +304 m -1736)
\,\tlam{}^2   \lambda_\rho{\!}^3  \nn \\
&{}- \tfrac{1}{384m^2} (m^4 + 33 m^3 - 108 m^2 -1160 m  + 3840 )  \lambda_\rho{\!}^5  \, , \nn \\
\noalign{\vskip 4pt}
\bbeta_{\tlam}{\!}^{(2)}  = {}& \tfrac{1}{72} N \big ( -18 \, g_\sigma{\!}^3 \tlam{}^2   +
7 \, g_\sigma{\!}^2 \tlam{}^3 + 14 \,  g_\sigma{\!}^2 \tlam{}^2  \lambda_\sigma \big ) \nn \\
&{}  + n \big ( -\tfrac16\,  g_\sigma{\!}^3 - 3 \,  g_\sigma{\!}^2 \tlam -
\tfrac34 \,  g_\sigma \tlam{}^2 + \tfrac{7}{36} \,  \tlam{}^3
- \tfrac34\,  g_\sigma{\!}^2 \lambda_\sigma - \tfrac12\,  g_\sigma \tlam  \lambda_\sigma
+ \tfrac{7}{72} \,   \tlam{}^2  \lambda_\sigma \big )  g_\rho{\!}^2
\nn \\
&{}+ n \, \tfrac{1}{m} \big ( \tfrac{1}{12} (2m^2 - m + 2) \,  g_\sigma -
 \tfrac{1}{8} (m^2  + 3 m -6 ) \,  \tlam \big ) g_\rho{\!}^4
\nn \\
&{} - \tfrac{1}{144}(11m^2 +11m +174) \, \tlam{}^5 + \tfrac{1}{72}(7m^2 +7m -90) \, \tlam{}^4 \lambda_\sigma
- \tfrac{101}{72} \, \tlam{}^3 \lambda_\sigma{\!}^2 - \tfrac{1}{18}  \tlam{}^2 \lambda_\sigma{\!}^3 \nn \\
&{}  - n\, \tfrac{1}{8m} (m-2)(m+4)  \big ( 3\,  g_\rho{\!}^3 g_\sigma \lambda_\rho
+   g_\rho{\!}^3 \tlam \lambda_\rho + \tfrac12  \,  g_\rho{\!}^2 g_\sigma \lambda_\rho{\!}^2 - \tfrac{7}{12} \,
 g_\rho{\!}^2 \tlam \lambda_\rho{\!}^2 \big ) \nn \\
 &{}-  \tfrac{1}{72m} (m-2)(m+4)  \big ( 67\, \tlam{}^3 \lambda_\rho{\!}^2 + \tfrac{83}{4} \, \tlam{}^2 \lambda_\rho{\!}^2
 \lambda_\sigma \big )
 \nn \\
 &{}- \tfrac{1}{384m^2} (m-2)(m+4)(11 m^2 + 46 m - 280 ) \, \tlam \lambda_\rho{\!}^4  \, .
\end{align}
\endgroup
For the anomalous dimensions
\begin{align}
\gamma_\Phi{\!}^{(2)} ={}& - \tfrac{1}{432} (11 N - 26) \,  g_\sigma{\!}^4 +  \tfrac19 \, g_\sigma{\!}^3 \lambda_\sigma
-  \tfrac{11}{432} \, g_\sigma{\!}^2 \lambda_\sigma{\!}^2 \nn \\
&{}+ \tfrac{1}{864m}(m-1)(m+2) \big (- 11m\,  g_\sigma{\!}^2 \tlam{}^2  + 52 \, g_\rho{\!}^2 g_\sigma{\!}^2
- 22 \, g_\rho{\!}^2 \tlam{}^2 + 144 \, g_\rho{\!}^2  g_\sigma \tlam \big ) \nn \\
&{} + \tfrac{1}{3456\, m^2}(m-1)(m-2)(m+2) (m+4) \big ( 48 \, g_\rho{\!}^3 \lambda_\rho
- 11\, g_\rho{\!}^2 \lambda_\rho{\!}^2 \big )   \nn \\
&{} - \tfrac{1}{864m^2}(m-1)(m+2) \big ( 11 (N + m^2) -13( m-2) \big )  \, g_\rho{\!}^4\, , \nn \\
\noalign{\vskip 4pt}
\gamma_\sigma{\!}^{(2)} ={}& \tfrac{1}{9} \big ( N ( \tfrac{1}{24} \, g_\sigma{\!}^4
 + \, g_\sigma{\!}^3\lambda_\sigma  -  \tfrac{11}{48} \, g_\sigma{\!}^2 \lambda_\sigma{\!}^2 )  +
\tfrac{13}{48} \, \lambda_\sigma{\!}^4 \big )\nn \\
&{} +  \tfrac{1}{18}(m-1)(m+2) \big ( n ( \tfrac{1}{24} \, g_\sigma{\!}^2 g_\rho{\!}^2
+  g_\sigma  g_\rho{\!}^2 \tlam - \tfrac{11}{48} \,  g_\rho{\!}^2 \tlam{}^2  ) +
    \tfrac{1}{24} \, \tlam{}^4 + \tlam{}^3 \lambda_\sigma
- \tfrac{11}{48} \, \tlam^2 \lambda_\sigma{\!}^2  \big ) \nn \\
&{}+ \tfrac{13}{3456\, m}(m-1)(m-2)(m+2) (m+4) \, \tlam{}^2 \lambda_\rho{\!}^2 \, , \nn \\
\noalign{\vskip 4pt}
\gamma_\rho{\!}^{(2)}  ={}& \tfrac{1}{432}\,  n  \big ( 2\, g_\sigma {\!}^2 g_\rho{\!}^2  + 96 \,   g_\rho {\!}^2 g_\sigma \tlam
- 11 \,  g_\rho {\!}^2 \tlam{}^2 - 11 m \,  g_\sigma {\!}^2 \tlam{}^2  \big ) + \tfrac19 \, \tlam{}^3 \lambda_\sigma
- \tfrac{11}{432} \, \tlam{}^2 \lambda_\sigma{\!}^2   \nn \\
&{}-  \tfrac{1}{432m}  ( 11 m^2 - m +2 ) \,  n \, g_\rho {\!}^4 - \tfrac{1}{864}  ( 11 m^2 + 11 m  -74  ) \, \tlam {}^4 \nn \\
&{}+ \tfrac{1}{1728m}(m-2)(m+4) \big (  n ( 48\, g_\rho{\!}^3 \lambda_\rho - 11 \,   g_\rho{\!}^2 \lambda_\rho{\!}^2 )
+ 87 \,  \tlam{}^2 \lambda_\rho{\!}^2 \big ) \nn \\
&{}
 +\tfrac{1}{6912m^2}(m-2)(m+4) (  m^2 + 26 m - 200)  \, \lambda_\rho{\!}^4  \, .
\end{align}
The corresponding three-loop results, obtained using
\eqref{betaV33},  for the anomalous dimensions and $\beta$-functions of the theory defined by \eqref{Vsix} are included in an ancillary file.

When $m<4$ there are fixed points with just $\lambda_\rho$ non zero
\be
\lambda_{\rho\hskip 0.5 pt *}{\!}^2 = \frac{8m}{(m+10)(4-m)} \, \vep \, , \quad
\begin{pmatrix} \pr_{\lambda_\rho} \beta_{\lambda_\rho}  & \pr_{\lambda_\rho} \beta_{g_\rho} \\
\pr_{g_\rho} \beta_{\lambda_\rho} & \pr_{g_\rho} \beta_{g_\rho}  \end{pmatrix}
= \begin{pmatrix} \vep & 0 \\ 0 & \frac{m^2+5m-32}{6(m+10)(4-m)}\, \vep \end{pmatrix} \, .
\ee
This example was considered in \cite{Priest}, the fixed point is stable for $m>3.68$.

Fixed points determined by vanishing of the $\beta$-functions for the couplings
$g_\sigma,g_\rho, \lambda_\sigma,\lambda_\rho,\tlam$ belong to equivalence
classes as a consequence of \eqref{bequiv}. For large $n$, to leading order in $1/n$,
there are three inequivalent fixed points similar to those in the $O(N)$ model
where the non zero couplings are ${\rm O}(n^{-\frac12})$. To ${\rm O}(\vep^2)$
these are given by
\begin{align}
H)& \quad g_{\rho*} = \lambda_{\rho*} = \tlam_* = 0 \, , \quad g_{\sigma*}{\!\!}^2 = \frac{6}{N} \, \vep
\Big ( 1 + \tfrac{44}{N}
- \tfrac{155}{3N} \, \vep \Big ) \, ,
\ \  \lambda_{\sigma*}{\!\!}^2 = \frac{216}{N} \, \vep  \Big (  1  + \tfrac{324}{N}
 - \tfrac{215}{N} \, \vep \Big )  \, , \nn \\
-)& \quad g_{\sigma*} = \lambda_{\sigma*} = \tlam_* =  0 \, , \nn \\
& \hskip 2cm g_{\rho*}{\!\!}^2 = \frac{6}{n} \, \vep \Big ( 1 + \tfrac{1}{N}\, (7m^2+22m-80)  -
\tfrac{1}{6 N} \, (50m^2 + 155 m -598) \, \vep \Big ) \, , \nn \\
& \hskip 2cm \lambda_{\rho*}{\!\!}^2  = \frac{216}{n} \, \vep  \Big (
1  + \tfrac{27}{N}(m-4)(m+10 )
- \tfrac{1}{2N }  \, (50  m^2 +215 m -1726)\, \vep \Big ) \, , \nn \\
+)& \quad g_{\sigma*}{\!\!}^2 = \frac{6}{N} \, \vep
\Big ( 1   + \frac{1}{n} \, (m+1) \big ( 22 - \tfrac{155}{6}\,  \vep \big ) \Big ) \, , \quad
\lambda_{\sigma*}{\!\!}^2  = \frac{216}{N} \, \vep  \Big ( 1  + \tfrac{1}{n} \, (m+1)
\big ( 162 - \tfrac{215}{2} \, \vep \big ) \Big )  \, , \nn \\
& \hskip 0.38cm \tlam_*{\!}^2 =  \frac{216}{N} \, \vep \Big ( 1 + \tfrac{1}{n} \big (
18(4m+9) - \tfrac{5}{2} \, (21m+43 ) \, \vep \big ) \Big ) \, , \nn \\
& \hskip 0.38cm  g_{\rho*}{\!\!}^2 = \frac{6}{n} \, \vep \Big ( 1  +
\frac{1}{n} \big ( 7 m+22 -  \tfrac{5}{6} \, (10m+31) \, \vep \big )  \Big )\, , \nn \\
& \hskip 0.38cm  \lambda_{\rho*}{\!\!}^2 = \frac{216}{n} \, \vep
\Big (  1  + \tfrac{1}{n} \big ( 27 (m+6) - \tfrac{5}{2} \,(10m+43 ) \, \vep \big ) \Big )
\, .
\end{align}
The eigenvalues of the stability matrix to first order in $\vep$ at large $n$ are then
\begin{align}
& H) \ \ \big ( 1, \, 1 -\tfrac{420}{N}, \, \tfrac{40}{N}, \, -\tfrac12, -\tfrac12 -\tfrac{4}{N} \big ) \vep \, , \nn \\
&-) \ \ \big (1, \, 1 - \tfrac{30}{N}(m^2 + 7m-15),  \, \tfrac12 - \tfrac{2}{N}( 10m^2 + 25m-122), \, -\tfrac12, \,
- \tfrac12 -\tfrac{20}{N} (m-1)(m+2) \big ) \vep \, , \nn \\
&+) \ \ \big ( 1 , \, 1+\tfrac{40}{n}(m+1), \, 1+ \tfrac{10}{N}\,x_1 ,\,
1+ \tfrac{10}{N}\, x_2 , \,1 +\tfrac{10}{N}\,x_3 \big ) \vep \, , \nn \\
&f(x_i)=0 \, , \ f(x)= x^3 - 3m(m-6)x^2 - 9m^2(2m^2 +17m+99) x - 54m^3 (m^2 -55m -2) \, .
\end{align}
Clearly the last fixed point is stable and there are in general three real roots, for large $m$, $x_1\approx - 3m^2, \ x_2 \approx - 3m , \
x_3 \approx 6m^2$.
The corresponding leading non zero contributions to the anomalous dimensions for each case
are  given by, to  order  $1/n$ and ${\rm O}(\vep^3)$ from three loops,
using \eqref{betaV33},
\be
\gamma_{\Phi\, H,+,-} =  k_{H,+,-}  \, \tfrac{1}{n}   \big ( \vep - \tfrac{11}{12} \, \vep^2 - \tfrac{13}{144} \, \vep^3 \big )  \, ,
\ee
where
\be
\quad k_H= \tfrac{1}{m}  \, , \quad  k_- = \tfrac{1}{2m} (m-1)(m+2) \, , \quad k_+ = \tfrac12 (m+1) \, .
\label{kHpm}
\ee
The anomalous dimensions for $\sigma,\, \rho$ in the same limit are
\begin{align}
& H) \  \gamma_{\sigma\hskip 0.5pt H} =  \tfrac12 \, \vep  +  k_H\, \tfrac{8}{n}
\big ( 5\,   \vep - \tfrac{13}{3} \, \vep^2  - \tfrac{11}{36} \, \vep^3 \big ) \, , \nn \\
& -) \  \gamma_{\rho\, -} =  \tfrac12 \, \vep  - \gamma_{\Phi\, -}+ \tfrac{1}{N} (m-2)
\big ( 3( 3m+13) \vep - \tfrac{1}{4} ( 33 m + 139) \vep^2
- \tfrac{1}{16} ( 5m + 31)\vep^3 \big ) \, , \nn \\
& +) \  \gamma_{\sigma\, +} =   \tfrac12 \, \vep + k_+ \, \tfrac{8}{n} \big ( 5 \,   \vep
- \tfrac{13}{3}   \, \vep^2- \tfrac{11}{36} \, \vep^3  \big ) \, , \nn \\
& \hskip 0.6cm \gamma_{\rho\, +} =  \tfrac12 \, \vep    - \gamma_{\Phi\, +}+ \tfrac{1}{n} \big ( 3(3m+7) \, \vep
- \tfrac{1}{4} (33m+73) \, \vep^2 - \tfrac{1}{16} ( 5m+21) \, \vep^3 \big ) \, .
\label{gamma6d}
\end{align}
These results are in accord with large $n$ expansions
\cite{Pelissetto,Gracey3,Gracey4}. The operators $\sigma, \rho$ have canonical dimension
$\tfrac12(d-2)$ and the $\tfrac12 \, \vep$ terms in \eqref{gamma6d} ensure
that they have leading dimension 2 at large $n$.

\section{Scalar Theories with Reduced Symmetry in
\texorpdfstring{$\mathbf{4\boldsymbol{-}\boldsymbol{\vep}}$}{4-epsilon} Dimensions}
\label{redSymFour}

For $N$-flavour scalar theories in $4-\vep$ dimensions the fixed point
structure  even with just two essential couplings can be rather non trivial.
Many such theories with different symmetry groups
have been discussed in the literature \cite{Brezin}.
A large class can be encompassed by considering a quartic potential with two
couplings of the form
\be
V(\phi) = \tfrac{1}{24} \, \lambda_{ijkl} \, \phi_i \phi_j \phi_k \phi_l
= \tfrac18 \, \lambda \, (\phi^2)^2 + \tfrac{1}{24} \, g \,
d_{ijkl} \phi_i \phi_j \phi_k \phi_l \, ,
\label{VfourN}
\ee
where $d_{ijkl} $ is symmetric and traceless on any pair of indices
and is assumed to be the unique tensor invariant under some subgroup
$H  \subset O(N)$
(in general there are, for arbitrary $N$,
$\tfrac{1}{24} (N-1)N(N+1)(N+6)$ such symmetric traceless tensors).
For $g=0$ this reduces to the usual $O(N)$ symmetric theory and the
extra term breaks $O(N)$ to the subgroup $H$. To first order in $g$
\eqref{VfourN} describes a perturbation of the $O(N)$ symmetric theory
by an approximately marginal harmonic operator \cite{Wallace}. In general
there are bounds such that $- \delta_- \,  (\phi^2)^2 \le d_{ijkl} \phi_i \phi_j \phi_k \phi_l
\le \delta_+  (\phi^2)^2$, $\delta_\pm >0$ for $N\ge 2$, which lead to constraints on $\lambda,\, g$ to ensure
the potential is bounded below. For $\lambda$  positive then for $g>0$ stability requires
$3\lambda/g > \delta_-$ while for $g<0$, $- 3\lambda/g > \delta_+$.

A general quartic potential would contain a term $\tfrac14 \, \phi^2 d_{ij} \phi^i \phi^j$,
with $d_{ij}$ symmetric and traceless, in addition to \eqref{VfourN}. For cases
considered here the only quadratic in $\phi$ invariant under $H$ is $\phi^2$ and
so such terms can be dropped. The restriction to \eqref{VfourN} does not exclude
any fixed points relevant for condensed matter systems considered in the literature  but an example
in which the perturbation is extended by a further coupling and where the RG
flow realises new IR fixed points is described in appendix  \ref{altfp}.

To two loop order it is sufficient to ensure the RG flow is restricted
to the two dimensional space parameterised by $\lambda,g$ to impose
\be
d_{ijmn} \, d_{klmn} = \tfrac{1}{N-1} \, a
\big ( \tfrac12 N (\delta_{ik} \delta_{jl}   +  \delta_{il} \delta_{jk})   -
\delta_{ij} \delta_{kl}  \big ) + b \, d_{ijkl}  \, .
\label{ddrel}
\ee
Tensors $d_{ijkl}$ related by $O(N)$ transformations define equivalent theories,
manifestly $H$ always contains ${\mathbb Z}_2$ induced by $\phi_i \to -\phi_i$
for all $i$.
For any non zero $d_{ijkl}$, $a>0$ and we may choose $b>0$ by changing
the sign of $d_{ijkl}$ if necessary, the value of $b^2/a$ is independent of
the choice of normalisation. As a consequence of \eqref{ddrel} $d_{iklm}d_{jklm}
= \tfrac12(N+2)a \, \delta_{ij}$, as expected if the only $H$ invariant quadratic in $\phi$ is
$\phi^2$. There is then only one strongly relevant ${\mathbb Z}_2$ invariant operator
which needs to be tuned to zero to attain an IR fixed point under RG flow.
For \eqref{ddrel} the subgroup $H$ is discrete.
More general possibilities than \eqref{ddrel} allowing continuous $H$ and also
more than one such tensor are considered later.

For $N=2$ \eqref{ddrel} requires $b=0$, which can be seen by
taking $d_{1111} = d_{2222} = - d_{1122}=1, \, d_{1112} = - d_{1222} = d$, so that
$d_{ijkl} \phi_i \phi_j \phi_k \phi_l  = \frac12 ( 1 - i d) (\phi_1 + i \phi_2)^4 +
\frac12 ( 1 + i d) (\phi_1 -i \phi_2)^4$,
and then \eqref{ddrel} requires $a = 2 (d^2 + 1) $ as well as $ b=0$.\footnote{The
results  reduce to the two flavour theory in section \ref{redSymFour} with $\lambda_1 = \lambda_2 = 3 \lambda+g, \,
\lambda_0 = \lambda - g, \, g_1 = - g_2 = d \, g$.  In terms of \eqref{fix4} case a) corresponds to taking
$g^2 = (d^2+1) g^2$  and the $O(2)$ theory of course requires $g=0$.} For
$N>2$ there is an inequality obtained in appendix \ref{boundsTP}
\be
a \ge 2\, \frac{N-1}{(N-2)^2} \, b^2 \, .
\label{inequal}
\ee
For $N=3$ the inequality is saturated so that $a=4\, b^2$.
Possible solutions for $a,b$ are further constrained by considering the eigenvalue
problem
\be
d_{ijkl} \, v_{kl} = \mu \, v_{ij} \, ,  \  v_{ij}=v_{ji} \, , \ v_{ii}=0 \quad \Rightarrow  \quad  \mu^2 - b \, \mu
-  \tfrac{N}{N-1} \, a =0 \, ,
\label{rhoe}
\ee
from \eqref{ddrel}. For the two solutions $\mu_\pm$ with corresponding degeneracies $d_\pm$
then we must have
\be
d_+ \, \mu_+ + d_- \, \mu_- = 0 \, , \qquad  d_+ + d_-=  \tfrac12 (N-1)(N+2) \, .
\label{dmu}
\ee
The traceless
condition then leads, as in the discussion in section \ref{redSymSix}, to
\be
1 + \frac{4N}{N-1} \, \frac{a}{b^2} = \bigg ( \frac{\tfrac12 (N-1)(N+2) } {\tfrac12 (N-1)(N+2) - 2\, d_+}
\bigg )^{\! 2} \ge \frac{(N+2)^2}{(N-2)^2} \, ,
\ee
using the inequality \eqref{inequal}. This requires $d_\pm$ are restricted to the finite range
\be
N-1 \le  d_\pm \le \tfrac12 N(N-1) \, .
\ee
The spaces of dimension $d_\pm$ must form representations of $H$. Tensors $d_{ijkl}$
are equivalent under $d_+ \leftrightarrow d_-$. Thus for $N=3$ there is just one case
when $d_\pm=2, \, d_\mp =3$, for $N=4$ there are two possibilities $d_\pm=3, \, d_\mp =6$
and $d_\pm=4 , \, d_\mp =5$.

For the general scalar theory the lowest order $\beta$-function is
determined, after rescaling $\lambda_{ijkl} \to (4\pi)^2 \lambda_{ijkl}$, by
the gradient flow equation
\cite{WallaceG}
\be
\beta_{ijkl} = \frac{\pr}{\pr \lambda_{ijkl}} A \, , \qquad
A = - \tfrac12 \, \vep \, \lambda_{ijkl} \lambda_{ijkl} +
\lambda_{ijkl} \lambda_{klmn} \lambda_{mnij} \, ,
\label{Afour}
\ee
and at any fixed point at this order \cite{VicariZ}
\be
A_* = - \tfrac16 \, \vep \,
\lambda_{*\hskip 0.5pt ijkl}\lambda_{*\hskip 0.5pt ijkl}
= - 2 \, \vep \, \tr(\gamma_*) \, ,
\label{agamma}
\ee
with $\gamma_*$ the lowest order $\phi$ anomalous dimension matrix at the fixed point
as given in \eqref{BVfour}.
For positive $\lambda_{ijkl}$ $A$ is bounded below so that $A$ has at
least one non zero perturbative minimum in the $\vep$ expansion so long
as the RG flow does not destabilise the vacuum.

With the constraint \eqref{ddrel} the lowest order $\beta$-functions become
\begin{align}
\beta_\lambda = {}& - \vep \, \lambda + (N+8) \lambda^2 +    a\, g^2 \, , \nn \\
\beta_g = {}& - \vep \, g + 12 \, \lambda \, g + 3 \, b \, g^2 \, .
\label{bgh1}
\end{align}
The associated $a$-function is given by specialising \eqref{Afour}
\be
A = N(N+2)\big ( - \tfrac32 \, \vep \, \lambda^2 - \tfrac14\, \vep \, a \,g^2 + (N+8) \lambda^3
+ 3\, a \, \lambda g^2 + \tfrac12 \, a\, b \, g^3 \big ) \, ,
\label{Afour2}
\ee
so that, with the $\beta$-functions in \eqref{bgh1},
\be
\frac{\pr}{\pr \lambda} A = 3N(N+2) \, \beta_\lambda \, , \qquad
\frac{\pr}{\pr g} A = \tfrac12 N(N+2) \, a \, \beta_g \, .
\ee

To lowest order in the  $O(N)$ invariant theory, obtained when $g=0$,
the Heisenberg fixed point is given by
\be
\lambda_{*\hskip 0.5ptH} = \frac{1}{N+8} \, \vep \, , \qquad
A_{*\hskip 0.5pt H} = - \frac{N(N+2)}{2(N+8)^2} \, \vep^3
= \frac{(N-4)^2}{48(N+8)^2} \, \vep^3  - \tfrac{1}{48}\, N \, \vep^3 \, .
\label{Hfp}
\ee

For theories described by the lowest order $\beta$-functions in \eqref{bgh1} there are
potentially, besides the trivial Gaussian fixed point and the $O(N)$ symmetric fixed point
with $g=0$,  two additional fixed points.
To analyse these fixed points it is convenient to write in \eqref{bgh1}
$\beta_\lambda = - \vep \, \lambda + \lambda^2 f_\lambda(x) , \,
\beta_g = - \vep\, g + \lambda^2 f_g(x)$ for $x=g/\lambda$ and $f_\lambda(x) =  N+8 + a\, x^2, \
f_g(x) = 12 \, x + 3 \, b \, x^2$. Then the fixed point
condition requires
\be
f(x)=0 \quad \mbox{for} \quad f(x) = x f_\lambda(x) - f_g(x)
= x ( a\, x^2 - 3b \, x + N-4) \, .
\label{f4x}
\ee
The cubic polynomial $f(x)$ has simultaneous roots for $N=4$ or
$b^2/a = \tfrac49 (N-4)$.  The solution $x=0$ of course corresponds to the
$O(N)$ symmetric theory with $g=0$.
For $g=0$ $\lambda_*= \vep /(N+8)$ but if $N>4$ $g$ is the coupling to a
relevant operator. So long as
\be
F(N) = 16 - 4 N +9\, b^2/a  > 0 \,,
\label{defFN}
\ee
(the bound \eqref{inequal} gives $F(N) < \tfrac{(N+2)^2}{2(N-1)}$)
there are  two extra real fixed  points determined by the roots
\be
\sqrt{a} \, x_\pm = X_\pm \, , \qquad X_\pm =
\tfrac12 \big ( 3 \, b/\sqrt a \pm F(N)^{\frac12} \big )
= \tfrac12 \Big ( \big (4(N-4) + F(N)\big )^\frac12 \pm F(N)^{\frac12} \Big ) \, .
\label{xpm}
\ee
To leading order in $\vep$ the additional fixed points are given by
\be
\lambda_{*\, \pm} = \frac{1}{N+8 + X_\pm{\!}^2} \, \vep \, , \qquad
\sqrt a \, g_{*\, \pm} = \frac{X_\pm }{N+8 + X_\pm{\!}^2} \, \vep \, ,
\label{lgone}
\ee
For the stability matrix
\be
M = \begin{pmatrix} \pr_\lambda \beta_\lambda & \pr_g
\beta_\lambda \\ \pr_\lambda \beta_g & \pr_g \beta_g\end{pmatrix}
\bigg |_{\lambda=\lambda_*,g=g_*} \ ,
\ee
then for either $x_*= x_\pm$ or $x_*=0$
\be
\det M = - \vep \, \lambda_* f'(x_*) \, , \quad \tr M =
\vep -  \lambda_* f '(x_*) \, , \quad  x_* = g_* /\lambda_* \, , \quad
f(x_*)= 0 \, .
\label{detM}
\ee
At a fixed point $\lambda_* = \vep /f_\lambda(x_*)$.
From \eqref{detM} the eigenvalues of $M$ to first order in $\vep$ are just
\be
\kappa_0 = \vep\, , \qquad \kappa_1 = - \lambda_* f'(x_*) \, .
\label{eigen4}
\ee
just as in \eqref{eigen4}.
In consequence the fixed point is stable if $f'(x_*)<0$. Since $a>0$ there is one possible root
$x_*$ of the cubic $f(x)$
with negative $f'(x_*)$ and two which are positive. For $N<4$ the $O(N)$ symmetric root
at $x=0$, giving $\lambda_* = \vep /(N+8)$, $g_*=0$, has $f'(0)=N-4 < 0$ and so is stable.
The other roots corresponding to non zero $g_*$ are then unstable.
Otherwise $f'(0)>0$ and  $f'(x_\pm) \gtrless 0$, assuming $b>0$.
The perturbation is then always relevant when $N>4$ and leads to an RG flow along which
the $O(N)$ symmetry is broken.

Applying \eqref{lgone} in \eqref{Afour2} gives
\be
A_{*\, \pm} = - \tfrac12 N(N+2) \, \frac{1 + \tfrac16 \, X_\pm{\!}^2}{(N+8 + X_\pm{\!}^2)^2 }\, \vep^3
\ge - \tfrac{1}{48} \, N \, \vep^3 \, .
\label{Abound}
\ee
The bound arises when
\be
X_{\pm}{\!}^2  =  N - 4 \quad \Rightarrow \quad F(N) = 0\, .
\ee
The bound is attained at the Heisenberg fixed point \eqref{Hfp} when $N=4$.
In general $F(N)$ vanishes when the fixed points merge, for small $F$
\be
A_{*\, \pm} = - \tfrac{1}{48} \, N \,\Big ( 1 - \tfrac{N-4}{4(N+2)^2} F(N)
\mp \tfrac{3\sqrt{N-4}}{2(N+2)^3} \, F(N)^\frac32+ {\rm O}\big ( F(N)^2 \big ) \Big ) \vep^3 \, ,
\ee
so that for $N>4$ the least $A_*$, and a stable fixed point, is expected to be achieved for the smallest $F(N)$
and for the fixed point correspond to $X_-$, as observed above. In general for $N>4$, $A_{*,H} > A_{*\,\pm}$.

The condition $F(N)=0$ or
\be
\frac{a}{b^2} = \frac{9}{4(N-4)} \, ,
\ee
determines a bifurcation point at lowest order. The eigenvalue equation \eqref{rhoe} becomes
$(\mu/b - \frac12)^2 = \frac{(N+2)^2}{4(N - 1 )(N - 4)}$. In order to satisfy \eqref{dmu} for integer
$d_\pm$ it is necessary that the solutions $\mu_\pm/b$ be rational. This requires
\be
(N-1)(N-4) = k^2  \  \Rightarrow  \ ( 2N - 5 -2k)(2N-5 + 2k) = 9 \  \Rightarrow  \ k=2, \, N=5 \, , \ \mbox{or} \ k=0  , \,  N=1,4 \, .
\ee
The solution $N=1$ is spurious since the couplings are not then independent
 but $N=4,5$ are realised in particular examples as shown later.

At two loops
\begin{align}
\beta_\lambda{\!}^{(2)} = {}& - 3 (3N+14) \, \lambda^3 -
\tfrac{1}{6} (5N+82) \, a\, \lambda g^2\
- 2 \,a b \, g^3 \, , \nn \\
\beta_g{\!}^{(2)} = {}& - (5N+82) \,  \lambda^2  g +
\tfrac{1}{6(N-1)} ( N^2 -17N +34) \, a\, g^3
- 6 \, b ( 6 \, \lambda g^2 + b \, g^3 ) \, ,
\label{bgh2}
\end{align}
and the anomalous dimension matrix is $\gamma \, {\mathds 1}$ with
\be
\gamma^{(2)} =\tfrac14
( N+2)\big ( \lambda^2 + \tfrac{1}{6} \, a \, g^2 \big ) \, ,
\label{anom2}
\ee
and from \eqref{Afour2} to lowest order in $\vep$
\be
A_*=A \big |_{\beta_\lambda = \beta_g = 0} = - 2 \, N \, \gamma_*{\!}^{(2)} \vep \, .
\label{Astar}
\ee
At the Heisenberg fixed point given by \eqref{Hfp}
\be
\gamma_{H} = \frac{N+2}{4(N+8)^2}\, \vep^2 \,  .
\ee

At three loops results further conditions are necessary to restrict the
RG flow to two couplings. In addition to \eqref{ddrel} we require
\begin{align}
d_{ijkp}\,d_{ilmq}\, d_{ljnr}\, d_{nmks} = {}& \tfrac14 \,  \A \big (
\delta_{pq}\delta_{rs} + \delta_{pr}\delta_{qs}
+ \delta_{ps}\delta_{qr} \big ) + c \, d_{pqrs} \, , \nn \\
\A= {}&
\frac{1}{N-1} \, a \big ( (N-2)a + 2(N-1)\, b^2 \big ) \, ,
\label{crel}
\end{align}
where $c^2/a^3$, or $c/b^3$, does not depend on the arbitrary normalisation of $d_{ijkl}$.
This gives for the three loop $\beta$-functions
\begin{align}
\beta_\lambda{\!}^{(3)} = {}&  \tfrac{1}{8} (33N^2+922N+2960) \, \lambda^4
+ 12 (5N+22) \zeta_3\, \lambda^4 \nn \\
&{} +\tfrac{1}{16} (N^2+500N+3492)\, a\, \lambda^2 g^2
+ 12(N+14)\zeta_3 \, a\, \lambda^2 g^2 \nn  \\
&{} + \tfrac{1}{8} (27N  + 470)  \, a\, b \, \lambda g^3 \nn
+ 48 \zeta_3 \,a\, b \, \lambda g^3  \\
&{} - \tfrac{1}{16(N-1)} (  7  \,N^2-33 \,N  +114 )\, a^2  \, g^4
+ \tfrac{13}{2}\,
a\, b^2 \, g^4 + 3 \, \A \,\zeta_3 \, g^4 \, , \nn \\
\beta_g{\!}^{(3)} = {}& - \tfrac14 (13N^2 -368N -3284) \,  \lambda^3  g
+ 48 (N+14) \zeta_3 \,  \lambda^3  g \nn \\
&{} + \tfrac{3}{8} (43 N +1334) \, b\, \lambda^2 g^2
+ 432 \zeta_3 \, b\, \lambda^2 g^2 \nn \\
&{}+ \tfrac{3}{N-1} \, \big ( 3\, N^2 +33\, N  - 50  \big )
a  \, \lambda g^3 + 156 \, b^2 \, \lambda g^3
+ 72 \, \A /a \,  \zeta_3  \, \lambda g^3  \nn \\
&{} - \tfrac{1}{16(N-1)}( 11\, N^2 -289  \, N  + 626) \, a \, b \, g^4 + \tfrac{39}{2} \, b^3 g^4
+ 12\zeta_3 \, c \, g^4  \, ,
\label{bgh3}
\end{align}
and
\be
\gamma^{(3)} = -\tfrac{1}{16}( N+2)(N+8) \, \lambda^3
- \tfrac{1}{32}(N+2) ( 6  a \, \lambda  g^2  + a\, b \, g^3 ) \, .
\ee

The two and three loop results for $\beta$-functions determine corrections as an expansion
in $\vep$ to the structure of fixed points found at one loop. The perturbation given by the
coupling $g$ in \eqref{VfourN} becomes relevant in the $O(N)$ symmetric theory, and so induces
an RG flow in which the $O(N)$ symmetry is broken, when $\pr_g \beta_g \big |_{\lambda=\lambda_*,g=0}<0$.
With results to three loops this requires
\be
N >  N_{{\rm crit},H} =4 - 2\, \vep + \tfrac52( \zeta_3 - \tfrac16) \vep^2 \, .
\label{NcritH}
\ee
The results in \eqref{Hfp} may be straightforwardly extended to higher order, for large $N$
they simplify to
\be
\lambda_{*\hskip 0.5pt H} \sim \tfrac{1}{N} \, \vep - \tfrac{1}{N^2} \big ( 8 \, \vep - 9\, \vep^2 + \tfrac{33}{8} \, \vep^3 \big )\, ,
\ee
leading to
\begin{align}
& \gamma_{H}  \sim  \tfrac{1}{4N} \big ( \vep^2 - \tfrac14 \, \vep^3 \big ) \, , \qquad
\gamma_{\sigma\, H} \sim \vep - \tfrac{2}{N}\big ( 3 \, \vep - \tfrac{13}{4} \, \vep^2 + \tfrac{3}{16} \, \vep^3 \big )\, , \nn \\
& \pr_g \beta_g \big |_{\lambda=\lambda_{*\hskip 0.2pt H} ,g=0 } \sim   - \vep + \tfrac{1}{N} \big ( 12 \, \vep - 5 \, \vep^2 - \tfrac{13}{4} \, \vep^3 \big ) \, .
\end{align}

In the two coupling theory considered here the broken symmetry fixed points merge at critical point
$N=N_{\rm crit}$. To lowest order in an $\vep$-expansion  $N_{\rm crit}=N_0$ with $N_0$ determined by  $F(N_0)=0$,
with $F$ is defined in \eqref{defFN}. In the neighbourhood of the critical point, $N\approx N_{\rm crit}$,
the straightforward  $\vep$-expansion breaks down but if $F(N)$ has a simple zero at $N=N_0$ we can write for
$N>4$, $N \approx N_0$,
\begin{align}
\lambda_{*\,+} - \lambda_{*\,-} ={}& a_\lambda (N,\vep) \, {\tilde F}(N,\vep )^{\frac12} \big ( 1 + {\rm O}( {\tilde F}) \big )\, ,
\nn \\
g_{*\,+} - g_{*\,-} = {}&  a_g (N,\vep) \, {\tilde F}(N,\vep )^{\frac12} \big ( 1 + {\rm O}( {\tilde F}) \big ) \, , \nn \\
 {\tilde F}(N,\vep ) = {}& F(N)  - F_1(N) \, \vep - F_2(N) \, \vep^2 + {\rm O}(\vep^3) \, ,
\end{align}
where $F_1, \, F_2$ are determined by ensuring that any singular terms in the $\vep$-expansion
as $\tilde F\to 0$ are cancelled.
Including higher loop corrections the critical $N$ is then determined by iteratively solving
\be
{\tilde F}(N_{\rm crit},\vep) = 0 \, , \qquad N_{\rm crit} = N_0 +{\rm O}(\vep) \, ,
\label{Ncrit}
\ee
as an expansion in $\vep$. For fixed points to be present ${\tilde F}(N,\vep) >0$,
for $N >  N_{{\rm crit},H} $ and ${\tilde F}(N,\vep) <0$ the RG flow does attain fixed points
accessible in the $\vep$-expansion.
Using the two and three loop results for the $\beta$
functions\footnote{Here $ a_\lambda =  - \sqrt{N-4} \, \vep / 2 (N+2)^2 + {\rm O}(\vep^2), \,
\sqrt{a}\, a_g =  3 \, \vep /(N+2)^2 + {\rm O}(\vep^2).$}
\begin{align}
F_1(N) = {}& \frac{2(N+2)^2}{3(N-1)} \, , \nn \\
F_2(N) =  {}& \frac{(N+2)^2 (2N+7) }{36(N-1)^2} \nn \\
&{} + \bigg ( \frac{8N^4 - 155 N^3 + 534 N^2 -932 N -184}{3 (N-1)(N+2)^2}
- \frac{32(N-4)^3}{9(N+2)^2}\, \frac{c}{b^3}\bigg ) \zeta_3\, .
\label{Ncrit12}
\end{align}

The general theory described by \eqref{VfourN} may be extended to include
relevant operators quadratic in $\phi$ by letting
\be
V(\phi) \to V(\phi) + \tfrac12 \, \sigma \, \phi^2 + \tfrac12 \, \rho_{ij} \, \phi_i \phi_j \, ,
\qquad \rho_{ii}= 0 \, .
\ee
At a fixed point the couplings of such operators must be tuned to zero but the general
formula for $\beta_V(\phi)$, obtained from  \eqref{betaV3} and \eqref{BVfour},
has contributions linear in $\sigma, \, \rho$ which take the form
\begin{align}
\beta_\sigma{\!}^{(1)} = {}& (N+2) \, \lambda \, \sigma \, , \qquad
\beta_\sigma{\!}^{(2)} = -\tfrac52(N+2)(  \lambda^2 + \tfrac16 a \, g^2 )  \, \sigma \, , \nn \\
\beta_\sigma{\!}^{(3)} = {}& \tfrac{1}{16}(N+2) \big (12(5N+37) \lambda^3 + (N+164) \, a \, g^2 \lambda
+ 27 \, ab \, g^3\big) \, \sigma \, , \nn \\
\beta_{\rho,ij}{\!}^{(1)} = {}& 2 \lambda \, \rho_{ij} + g \, d_{ijkl} \rho_{kl} \, , \nn \\
\beta_{\rho,ij}{\!}^{(2)} = {}&
- \tfrac{1}{2} \big ( (N+10)\lambda^2 - \tfrac{N^2-5N +10}{6(N-1)} \,a \,  g^2 \big  ) \rho_{ij}
- ( 4 \lambda g + b \, g^2 )  \, d_{ijkl} \rho_{kl} \, , \nn \\
\beta_{\rho,ij}{\!}^{(3)} = {}&
-\tfrac14 \big (\tfrac12 (5N^2-84N-444)\lambda^3 - \tfrac{5N^2 +65N-82}{N-1} \,a \,  g^2  \lambda
+\tfrac{ N^2-35N+54}{4(N-1)}\, ab \, g^3 \big  ) \rho_{ij} \nn \\
\noalign{\vskip -1pt}
& +\tfrac18 \big ( 3(9N + 146 )g \,\lambda^2 + 192 \, b \, g^2 \lambda - \tfrac{3N^2 -25 N+66}{2(N-1)} \,a \,  g^3
+ 32 \, b^2 \, g^3 \big  )  \, d_{ijkl} \rho_{kl} \, .
\label{bsigrho}
\end{align}
In general $\beta_\sigma = \gamma_\sigma \, \sigma$.
The anomalous dimensions for $\rho$ are then determined by the eigenvalues in \eqref{rhoe}.

For application here it is necessary to consider possible tensors $d_{ijkl}$
satisfying \eqref{ddrel} as well as \eqref{crel}.
If $b=0$ solving the fixed point equations at lowest order using  \eqref{bgh1} requires $N<4$ or $g=0$.

\subsection{Fixed Points with discrete
  \texorpdfstring{$\boldsymbol{H\subset O(N)}$}{H subset of O(N)}}
\subsubsection{Hypercubic fixed points}
An extension of the simple $N=2$ case is obtained by taking, as was considered long ago \cite{Aharony,Aharony2,Wallace2}
and more recently in \cite{FeiSphere},
\be
d_{ijkl} \phi_i \phi_j \phi_k \phi_l = \sum_i \phi_i{\!}^4
- \frac{3}{N+2} \, (\phi^2)^2 \, ,
\label{cubic}
\ee
where $- \tfrac{2(N-1)}{N(N+2)} (\phi^2)^2 \le d_{ijkl} \phi_i \phi_j \phi_k \phi_l \le \tfrac{N-1}{N+2}(\phi^2)^2 $.
This breaks $O(N)$ to the discrete  hypercubic group
$H_{\rm cubic} \simeq BC_N\simeq {\cal S}_N  \ltimes {\mathbb Z}_2{\!}^N$,
which ensures no other couplings than $\lambda,g$ are necessary for
renormalisability. If $\lambda=\frac{1}{N+2}\, g$ in \eqref{VfourN} this
reduces to $N$ decoupled $\phi^4$ theories. In \eqref{ddrel} and \eqref{crel}
\be
a = 2\, \frac{N-1}{(N+2)^2} \, , \qquad b = \frac{N-2}{N+2} \, , \qquad
c = \frac{N(N-2)(N-4)}{(N+2)^3} \, ,
\label{abc4}
\ee
and the solutions of \eqref{rhoe} and \eqref{dmu} give
\be
\mu_+ = \frac{N}{N+2} \, , \quad d_+ = N-1 \, , \qquad
\mu_- = - \frac{2}{N+2} \, , \quad d_- = \tfrac12 N(N-1) \, .
\ee
Clearly \eqref{abc4} saturates the inequality \eqref{inequal}.
In \eqref{defFN}
\be
F(N)_{\rm cubic} = \frac{(N+2)^2}{2(N-1)} \, .
\ee
which is positive for any $N> 1$.

For this case the roots of $f(x)=0$ are
$x_-= \tfrac12(N-4)(N+2)/(N-1)$, $x_+=N+2$ as well as $x=0$.
At these roots $f'(x_\pm) = \pm x_\pm$ and $f'(0)=N-4$. Thus
for $N>4$ $x_*=x_- $ is the stable fixed point, for $N<4$ $x_*=0$.
At lowest order there are then fixed points for $N>1$ for
$x_*=x_-$, $x_* = x_+$ respectively
\begin{align}
\lambda_{*-} = {}& \frac{2(N-1)}{3N(N+2)}\, \vep  \, ,
\quad g_{*-} = \frac{1}{3N}(N-4) \, \vep\, , \quad
\gamma_{*-} = \frac{(N-1)(N+2)}{108\, N^2} \, \vep^2\, , \nn \\
\lambda_{*+} = {}& \frac{1}{3(N+2)}\, \vep  \, , \qquad
g _{*+} = \frac13\, \vep \, , \qquad \gamma_{*+} = \frac{1}{108}\, \vep^2 \, ,
\label{Nfix}
\end{align}
with the anomalous dimension at the fixed point obtained from \eqref{anom2}.
From \eqref{eigen4} the eigenvalues of the stability matrix to lowest order
are given by
\be
\kappa_{0,\pm} = \vep \, , \qquad \kappa_{1,-} = \frac{N-4}{3N} \, \vep \, , \qquad
\kappa_{1,+} = - \frac13 \, \vep \, .
\label{Kcube}
\ee
For the second case the $(\phi^2)^2$ interaction is cancelled
so this describes $N$ decoupled $\phi^4$ theories.
For $N=2$ the two fixed points are equivalent and correspond to case a) in
\eqref{fix4} with $\lambda_0 = \lambda_1$ or $\lambda_0=0$. For $N=4$
the results in \eqref{Nfix} corresponding to $x_*= x_-$ are identical with \eqref{Hfp}.
The higher order results for $\beta$-functions allow an extension of \eqref{Kcube}
as an expansion  in $\vep$. The results for  $\kappa_{0,+} , \, \kappa_{1,+} $ match those for the Ising model as expected
and as given in \eqref{Ising}. To the next order
\begin{align}
\kappa_{0,- } ={}&  \vep - \tfrac{(N-1)(17N^2 - 4N +212)}{27N^2 (N+2)} \, \vep^2 \, , \quad
\kappa_{1,-} = \tfrac{N-4}{3N} \, \vep - \tfrac{(N-1)(19N^3-72N^2 - 660N + 848)}{81N^3 (N+2)} \, \vep^2 \, .
\end{align}
The critical point is stable for $\kappa_{1,-} >0$. Including three loop contributions this arises for exactly the
same $N$ as in \eqref{NcritH} when the $O(N)$ fixed point is unstable. In consequence the $O(N)$ theory
with a hypercubic perturbation has a unique  stable fixed point for all $N$. For the $a$-function from \eqref{Afour2}
$A_{*\,-{\rm cubic}} = - \tfrac{1}{48}  N \vep^3 + \tfrac{(N-4)^2 }{432 N}\, \vep^3$ which attains the lower bound when $N=4$.

Although the fixed point at $x_* = x_- $ is stable for $N \gtrapprox 4$ there may be
slightly relevant $\phi^4$ perturbations. At this fixed
point the anomalous dimensions to lowest order are determined by the eigenvalues of
operator
\be
\D = \vep \, \tfrac{1}{3N} \big ( \tfrac12 \, \phi^2 \, \pr^2 + ( \phi \cdot \pr)^2 - \phi\cdot \pr
+ \tfrac12 (N-4) \, {\ts \sum_i}\,  \phi_i{\!}^2 \, {\pr_i}{\!}^2 \big ) \, .
\ee
The $\tfrac{1}{24}N(N+1)(N+2)(N+3)$ $\phi^4$ operators may be decomposed into irreducible  representations of the hypercubic
symmetry group to obtain
\begin{align}
&\D \, \phi_i \phi_j \phi_k \phi_l =  \vep \, \tfrac{4}{N} \, \phi_i \phi_j \phi_k \phi_l  \, , \qquad
i\ne j\ne k \ne l \, ,  \quad \dim \ \tbinom{N}{4}\nn \\
& \D \, \phi_i \phi_j ( \phi_k \phi_k - \phi_l \phi_l ) =  \vep \, \tfrac{N+8}{3N} \, \phi_i \phi_j (\phi_k \phi_k - \phi_l\phi_l)   \, , \ \
i\ne j\ne k \ne l \, , \ \ \dim \ \tfrac12 N(N-1)(N-3) \, , \nn \\
& \D \, S_{ij} =   \vep \, \tfrac{2(N+2)}{3N} \, S_{ij} \, , \quad  {\ts \sum_i}\, S_{ij}=0 \, , \quad \dim \ \tfrac12 N(N-3) \, , \nn \\
& \ \ \  S_{ij} = \phi_i{\!}^2 \phi_j{\!}^2 - \tfrac{1}{N-2} \big ( ( \phi_i{\!}^2  + \phi_j{\!}^2 )\phi^2 - \phi_i{\!}^4  - \phi_j{\!}^4 \big  )
+\tfrac{1}{(N-1)(N-2)}  \big ( (\phi^2)^2 - {\ts \sum_i} \,  \phi_i{\!}^4 \big ) \, , \ i\ne j \, ,  \nn \\
& \D \big ( \phi_i \phi_j{\!}^3 - \phi_i{\!}^3 \phi_j \big ) = \vep \big ( \phi_i \phi_j{\!}^3 - \phi_i{\!}^3 \phi_j \big ) \, , \ \ i\ne j\, ,
 \quad \dim \ \tfrac12 N(N-1)\, , \nn \\
& \D \begin{pmatrix} \phi_i \phi_j \, \phi^2\\ \phi_i \phi_j{\!}^3 + \phi_i{\!}^3 \phi_j \end{pmatrix} =
\vep\,\frac{1}{3N}  \begin{pmatrix} 2(N+6) & 2(N-4) \\  6 & 3N \end{pmatrix}
\begin{pmatrix} \phi_i \phi_j \, \phi^2\\ \phi_i \phi_j{\!}^3 + \phi_i{\!}^3 \phi_j \end{pmatrix} \, , \ \ i\ne j \, ,  \  \dim \ N(N-1)\, , \nn \\
& \D \begin{pmatrix} (\phi_i{\!}^2 - \phi_j{\!}^2)\phi^2 \\ \phi_i{\!}^4 - \phi_j{\!}^4 \end{pmatrix} =
\vep\,\frac{1}{3N}  \begin{pmatrix} 3N+8  & 4(N-4) \\  6 & 6(N-2) \end{pmatrix}
\begin{pmatrix} (\phi_i{\!}^2 - \phi_j{\!}^2)\phi^2 \\ \phi_i{\!}^4 - \phi_j{\!}^4 \end{pmatrix}  \, , \ \ i\ne j \, , \ \ \dim \ 2(N-1) \nn \\
& \D \begin{pmatrix}  (\phi^2)^2\\ \sum_i \phi_i{\!}^4 \end{pmatrix} =
\vep\,\frac{2}{3N}  \begin{pmatrix} 2(N+2) & 2(N-4) \\  3 & 3(N-2) \end{pmatrix}
\begin{pmatrix}  (\phi^2)^2\\ \sum_i \phi_i{\!}^4 \end{pmatrix} \, , \quad \dim \ 2 \, .
\label{pertcube}
\end{align}
For any eigenvalue $\mu$ of $\D$ the associated eigenvector determines a scaling $\phi^4$ operator with
scaling dimension $2(d- 2) +\mu$ so that if $\kappa=\mu- \vep<0 $ it represents a relevant perturbation.
For each operator in \eqref{pertcube} the values of $\kappa/\vep$ in turn
are $\tfrac{4-N}{N}, \, \tfrac{2(4-N)}{3N}, \, \tfrac{4-N}{3N}, \, 0$,
$\tfrac{1}{6N} \big ( 12-N \pm \sqrt{N^2+24N -48} \big ), \, 0, \,
\tfrac{3N-4}{3N}, \,
1, \, \tfrac{N-4}{3N}$.
The last two values correspond to the two perturbations which preserve
hypercubic symmetry and just give \eqref{Kcube}.
However other perturbations, which break the hypercubic symmetry,
become relevant for $N>4$.

\subsubsection{Hypertetrahedral fixed points}
An additional case \cite{ZiaW} is obtained by considering $N+1$ vectors
$e_i{\!}^\alpha$, $\alpha =1,2 , \dots , N+1$,
forming the vertices of a $N$-dimensional hypertetrahedron and
satisfying, with a convenient normalisation,
\be
{\sum_\alpha} \, e_i{\!}^\alpha  = 0 , \, \qquad
{\sum_\alpha} \, e_i{\!}^\alpha e_j{\!}^\alpha = \delta_{ij} \, , \qquad
e_i{\!}^\alpha e_i{\!}^\beta = \delta^{\alpha\beta} - \frac{1}{N+1} \, .
\label{tetra}
\ee
We may then define
\be
d_{ijkl} \phi_i \phi_j \phi_k \phi_l = {\sum_\alpha} \, \big (e_i{\!}^\alpha \phi_i \big )^4
- \frac{3N}{(N+1)(N+2)} \, (\phi^2)^2 \, .
\label{tetra2}
\ee
In this case  $- \tfrac{2(N-1)}{(N+1)(N+2)} (\phi^2)^2 \le d_{ijkl} \phi_i \phi_j \phi_k \phi_l
\le \tfrac{(N-1)(N-2)}{N(N+2)}(\phi^2)^2 $.\footnote{The bounds are
equivalent to those for
$\sum_{\alpha = 1}^{N+1} x_\alpha{\!}^4 /
( \sum_{\alpha = 1}^{N+1} x_\alpha{\!}^2)^2$
subject to $\sum_{\alpha = 1}^{N+1} x_\alpha =0$. For $N$ even the lower bound
is modified.}
The symmetry group is $H_{\rm tetrahedral} \simeq A_N \otimes {\mathbb Z}_2 \simeq {\cal S}_{N+1} \otimes {\mathbb Z}_2$,
with $ {\cal S}_{N+1}$ corresponding to the
permutations of ${\underline e}^\alpha$ or the vertices of the hypertetrahedron and $ {\mathbb Z}_2$ to reflections
$\phi_i \to - \phi_i$. This
expression for $d_{ijkl}$ satisfies \eqref{ddrel} with
\be
a = 2 \, \frac{(N-1)(N-2)}{(N+1)(N+2)^2} \, , \qquad
b = \frac{N^2-3N - 2}{(N+1)(N+2)} \, ,
\ee
and in \eqref{crel}
\be
c =  \frac{N^6 - 9 N^5 + 20 N^4 + 30 N^3 - 60 N^2 - 120 N - 48}
{(N+1)^3 (N+2)^3} \, .
\ee
For this case $a- 2\, \tfrac{N-1}{(N-2)^2} \, b^2 = 2 \, \tfrac{(N-1)(N-3)}
{(N+1)^2 (N-2)^2}$.
The solutions of \eqref{rhoe} and \eqref{dmu}  are then
\be
\mu_+ = \frac{N-2}{N+2} \, , \quad d_+ = N\, , \qquad
\mu_- = - \frac{2N}{(N+1)(N+2)} \, , \quad d_- = \tfrac12 (N-2)(N+1) \, .
\ee
For $N=2$, $d_{ijkl}=0$ and for $N=3$,  \eqref{tetra2} reduces, up to
an overall scale, to \eqref{cubic} which has cubic symmetry
(${\cal S}_4 \subset {\cal S}_3  \ltimes {\mathbb Z}_2{\!}^3 \simeq
 {\cal S}_4  \ltimes {\mathbb Z}_2 $, with ${\cal S}_4$ corresponding to
permutations of the diagonals of the cube, in this case $b^2/a = \tfrac14, \, c/b^3=-3$,
for other $N$ the hypertetrahedral
symmetry group is not a subgroup of the hypercubic group). For large
$N$ $a,b,c$ have the same limits as in the cubic case so the fixed
points approach each other. However, unlike the cubic case,  the
 hypertetrahedral theory contains an $H$ invariant cubic operator
${\sum_\alpha} \, \big (e_i{\!}^\alpha \phi_i \big )^3$.

The fixed points are determined by the roots of $f(x)=0$ in \eqref{f4x}.
Apart from $x=0$  these are $x_- = (N+2)/2$,
$x_+= (N-4)(N+1)(N+2)/(N-1)(N-2)$ and
$f'(x_-) =- \frac12(N-5)(N+2)/(N+1)$, $f'(x_+) = (N-5)(N-4)(N+2)/(N-1)(N-2)$.
For $N>5$ $x_* = x_- $ is then the stable fixed point.
These roots are coincident for $N=5$. In \eqref{defFN}
\be
F(N)_{\rm tetrahedral} = \frac{(N+2)^2(N-5)^2}{2(N-1)(N-2)(N+1)}
< F(N)_{\rm cubic} \ \mbox{for} \  N>3 \, .
\label{Ftetra}
\ee

The  couplings at the fixed points and the lowest order anomalous dimensions are then
for $x_*=x_-$
\begin{align}
\lambda_{*\, -} =  \frac{2(N+1)}{3(N+2)(N+3)}\, \vep  \, ,
\quad g_{*\, -} = \frac{N+1}{3(N+3)} \, \vep\, , \quad
\gamma_{*\, -} = \frac{(N+1)(N+7)}{108\, (N+3)^2} \, \vep^2\, ,
\label{lgtet}
\end{align}
and for $x_* = x_+$
\begin{align}
\lambda_{*\, +} = {}& \frac{(N-1)(N-2)}{3(N+2)(N^2 - 5N + 8 )}\, \vep  \, , \qquad \quad
g _{*\, +}= \frac{(N-4)(N+1)}{3(N^2 - 5N + 8 )}\, \vep \, , \nn \\
\gamma_{*\, +}= {}& \frac{(N-1)(N-2)(N^2-6N+11)}{108\,(N^2 - 5N + 8 )^2}\, \vep^2 \,.
\label{Nfix2}
\end{align}
In this case from \eqref{eigen4}
\be
\kappa_{0,\pm} = \vep \, , \qquad \kappa_{1,-} = \frac{N-5}{3(N+3)} \, \vep \, , \qquad
\kappa_{1,+} = - \frac{(N-4)(N-5)}{3(N^2-5N+8)} \, \vep \, .
\ee
For the $a$-function $A_{*\,-{\rm tetrahedral}} = - \tfrac{1}{48}  N \vep^3 + \tfrac{(N-5)^2N}{432(N+3)^2}\, \vep^3$ and
\be
A_{*\,-{\rm tetrahedral}} - A_{*\, -{\rm cubic}} = - \frac{(N-3)(N^2-3N-6)}{54N(N+3)^2} \, ,
\ee
so that for $N\ge 5$ the tetrahedral fixed point has a lower $A$ and also
a lower symmetry than the cubic one. The difference vanishes when $N=3$ since
in this case the cubic and tetrahedral theories coincide. The marginal perturbations
at this fixed point are analysed in appendix \ref{pertsHT}.

The result \eqref{Ftetra} has a double zero at $N=5$ when \eqref{lgtet} and \eqref{Nfix2}
coincide. Taking into account  higher order contributions
for $N\approx 5$ the two fixed points have an expansion where
\begin{align}
\lambda_{*\,+} - \lambda_{*\,-} ={}& a_\lambda (N,\vep) \,  {\tilde N}(N,\vep)^\frac12 \, , \qquad
g_{*\,+} - g_{*\,-} =   a_g (N,\vep) \,  {\tilde N}(N,\vep)^\frac12 \, , \nn \\
 {\tilde N}(N,\vep) = {}&  \big ( N- 5 - 6 \,\vep  -  \tfrac{281}{32}\, \vep^2
 + \tfrac{61}{8} \, \zeta_3 \,\vep^2 \big )^2 - 24\, \vep - \tfrac{289}{4} \, \vep^2
 +  30 \, \zeta_3 \, \vep^2 \, , \nn \\
 a_\lambda (N,\vep)\approx {}& - \tfrac{1}{168} \big ( 1 - \tfrac{1}{56} ( N-5) \big ) \vep\, , \qquad
 a_g (N,\vep) \approx \tfrac{1}{8} \big ( 1 - \tfrac{7}{12} ( N-5) \big ) \vep\, ,
 \label{Ncrit5}
\end{align}
where $a_\lambda (N,\vep), \, a_g (N,\vep)$ have no singularities as $N\to 5$. $ {\tilde N}(N,\vep) =0$
determines  $N_{\rm crit}$ where new fixed points emerge. From \eqref{Ncrit5}, to ${\rm O}(\vep^2)$,
\be
N_{{\rm crit} \pm} =  5 + 6 \,\vep  +  \tfrac{281}{32}\, \vep^2 - \tfrac{61}{8} \, \zeta_3 \,\vep^2
 \pm \sqrt{24\, \vep  + \tfrac{289}{4} \, \vep^2 - 30 \, \zeta_3 \, \vep^2} \, .
 \ee
For $ N_{{\rm crit}-} < N <  N_{{\rm crit}+}$ there are no real fixed points but there are nearby complex fixed points.
For $N=5$ then there are conjugate fixed points which may be found as a series in $\vep^{\frac12} $ using the
three loop $\beta$-functions as\footnote{The degeneracy of the one loop fixed points for $N=5$ is removed by higher loop contributions but  it is necessary to consider a perturbative
 expansion in $\vep^{\frac12}$ and the fixed point becomes complex. The $\beta$-functions reduce to
 $$
 \beta_\lambda = - \vep \, \lambda+ 13 \, \lambda^2 + \tfrac{4}{49} \, g^2 + f_\lambda \, , \qquad
\beta_g = - \vep   \, g + 12 \, \lambda g + \tfrac47\, g^2 + f_g \, ,  \eqno{(a)}
$$
where $f_\lambda , \, f_g$ correspond to contributions at two or more loops. Writing for the fixed point couplings a Puiseaux seies
$$
\begin{pmatrix} \lambda_* \\ g_* \end{pmatrix} = \begin{pmatrix} \frac{1}{14} \\  \frac14 \end{pmatrix} \vep
+ \sum_{n\ge 3} \begin{pmatrix} a_n  \\  b_n \end{pmatrix} \vep^{\frac12 n}  \, ,   \eqno{(b)}
\label{explg}
$$
then the equations for a vanishing $\beta$-functions at order $\vep^{\frac12 n+1}$  take the form
$$
M   \begin{pmatrix} a_n  \\  b_n \end{pmatrix}  =   \begin{pmatrix} c_n  \\  d_n \end{pmatrix} \, , \qquad
M = \begin{pmatrix} \frac67 & \frac{2}{49} \\ \noalign{\vskip 1pt} 3&  \frac17 \end{pmatrix} \, , \qquad c_3=d_3=0\, ,  \eqno{(c)}
$$
where $c_{2n},\, d_{2n}, \, c_{2n+1}, \, d_{2n+1}$ are determined by contributions to $\beta_\lambda, \, \beta_g$ for up to $n$-loops
and depend non linearly on $a_{p}, \, b_q$ for $p,q<2n$.
 Since $M$ is singular it is necessary that $7 \,c_n - 2 \, d_n =0 $ with a corresponding freedom in the solutions  of the form
 $a_n \sim a_n + x_n, \, b_n \sim b_n - 21\, x_n$. At lowest order $b_3 =- 21\, a_3$.
This freedom may be used to ensure that $c_{n+1}, \, d_{n+1}$  satisfy the solvability constraints for $a_{n+1}, \, b_{n+1}$
at the next order. To fully determine $ a_{2n},\, b_{2n}$ it is then necessary to
know the $\beta$-functions to $n$ loops whereas for $a_{2n+1} , \, b_{2n+1} $  $n+1$ loops are required.  The presence
of the $\vep^{\frac32}$ terms in $(b)$ are required to ensure  $7\, c_4 - 2\, d_4 =0$ so that the ${\rm O}(\vep^3)$ terms
 in the expansion of $\beta_\lambda, \, \beta_g$ can be set to zero. Thus if in $(a)$
 $f_\lambda \to {\hat c}_4 \, \vep^3, \, f_g \to {\hat d}_4 \, \vep^3$
 then $7^3 a_3{\!}^2 = 2\, {\hat d}_4 -  7\, {\hat c}_4$.  This leads to conjugate
 imaginary solutions for $a_3, \, b_3$ and hence for all $ a_{2n+1},\, b_{2n+1}$.}
 \begin{align}
\lambda_* = {}&  \tfrac{1}{14} \, \vep + \tfrac{1}{24} \, \vep^2 \pm  \tfrac{1}{28\sqrt{6}}
 \, i \,  \vep^{\frac32}  \big ( 1 + \tfrac{1}{64}( 79- 40 \zeta_3 ) \vep \big ) + \dots \, , \nn \\
 g_* = {}&  \tfrac{1}{4} \, \vep + \tfrac{21}{64} \, \vep^2 \mp  \tfrac{1}{4\sqrt{6}}
 \, i \,  \vep^{\frac32}  \big ( 3  + \tfrac{1}{64}( 181 -  120 \zeta_3 ) \vep \big ) + \dots \, .
\end{align}
Such almost real fixed points can have dramatic effects on the RG flow \cite{Gorbenko}.

It is natural to consider extensions of \eqref{tetra} based on regular
polytopes for general $N$, generalisations of the Platonic solids,
which define discrete subgroups $H$ of $SO(N)$.
However for $N\ge 5$ we have, apart from hypertetrahedrons or $N$-simplices, just
hypercubes, with
$2^N$ vertices $(\pm 1 , \pm 1, \dots , \pm 1)$, and hyperoctahedrons with $2N$
vertices $(0,0,\dots, \pm 1 , \dots, 0)$. These are dual by interchanging
faces and vertices and have the same symmetry. In either case extending
\eqref{tetra2} is equivalent to \eqref{cubic}. More generally vertex transitive or
isogonal polytopes also define discrete subgroups of $SO(N)$. An example are
demihypercubes with $2^{N-1}$ vertices $(\pm 1 , \pm 1, \dots , \pm 1)$, with an
even number of $-1$'s. For $N=3$ this is just a tetrahedron.
For general $N$ the symmetry group
is $D_N  \simeq{\cal S}_N  \ltimes {\mathbb Z}_2{\!}^N/{\mathbb Z}_2$. However for quartic
polynomials ${\sum_\alpha} \, \big (e_i{\!}^\alpha \phi_i \big )^4 =- 2^N \sum_i \phi_i {\!}^4 +
 3 \times 2^{N-1} (\phi^2)^2$ except when $N=4$ and there is an additional term
 $192\, \phi_1 \phi_2 \phi_3 \phi_4$ which is invariant under $D_4$.
 Nevertheless this is equivalent to the hypercubic
case for all $N$, for $N=4$, ${\sum_\alpha} \, \big (e_i{\!}^\alpha \phi_i \big )^4 =
32 \sum_i \vphi_i {\!}^4$ taking
 $\vphi_1 = \tfrac12(\phi_1+\phi_2+\phi_3+\phi_4), \,
\vphi_2 = \tfrac12(\phi_1-\phi_2-\phi_3+\phi_4), \,
\vphi_3 = \tfrac12(\phi_1-\phi_2+\phi_3-\phi_4), \,
\vphi_4 = \tfrac12(\phi_1+\phi_2-\phi_3-\phi_4)$.

\subsection{Fixed Points with continuous
  \texorpdfstring{$\boldsymbol{H\subset O(N})$}{H subset of O(N)}}
Theories in which the RG flow involves just two couplings can
also be obtained by extending \eqref{ddrel} to\footnote{For the RG flow
for the theory defined by \eqref{VfourN} to be closed at one loop it is
necessary to impose $d_{ijmn} \, d_{klmn} + d_{ikmn} \, d_{jlmn} + d_{ilmn} \, d_{jkmn}
= a ( \delta_{ij} \delta_{kl} + \delta_{ik} \delta_{jl} + \delta_{il} \delta_{jk}) + 3b\,  d_{ijkl}$. In
that case we may define $e^u\, w_{u,ijkl} = \tfrac13 (2 \, d_{ijmn} \, d_{klmn} - d_{ikmn} \, d_{jlmn} - d_{ilmn} \, d_{jkmn} )
+ \tfrac{N+2}{6(N-1)}\, a ( 2 \, \delta_{ij} \delta_{kl} - \delta_{ik} \delta_{jl} - \delta_{il} \delta_{jk})$.}
\be
d_{ijmn} \, d_{klmn} = \tfrac{1}{N-1} \, a
\big ( \tfrac12 N (\delta_{ik} \delta_{jl}   +  \delta_{il} \delta_{jk})   -
\delta_{ij} \delta_{kl}  \big ) + e^u\, w_{u,ijkl} + b \, d_{ijkl}  \, ,
\label{ddrel2}
\ee
assuming $w_{u,ijkl}$ are a basis of $H$-invariant tensors satisfying\footnote{If $c_{ikjl}
= w_{ijkl} - w_{kjil}$ then $c_{ikjl} = - c_{kijl} = c_{jlik}$ and $c_{i[kjl]}=0, \, c_{ijjl}=0$
and so has the symmetries of the Weyl tensor. $w_{ijkl} = (c_{ikjl}+c_{jkil})/3$.}
\be
w_{u,ijkl} = w_{u,ji\hskip 0.5ptkl}  = w_{u,kl\hskip 0.5 pt ij} \, , \qquad w_{u,i(jkl)} = 0 \, ,
\qquad  w_{u,ii\hskip 0.5 pt kl} =0 \, .
\label{wrel}
\ee
Such tensors may be present for $N\ge 4$.
For calculations beyond one loop we require
\begin{align}
d_{ijmn} \, w_{u,klmn} = {}& w_{u,ijmn} \, d_{klmn}
=  f_u{\!}^v\, w_{v,ijkl} + h_u\, d_{ijkl}  \, , \nn \\
w_{u,ijmn} \, w_{v,klmn} = {}& w_{v,ijmn} \, w_{u,klmn} \nn \\
 = {}& \tfrac{1}{N-1} \, a'{\!}_{uv}
\big ( \tfrac12 N (\delta_{ik} \delta_{jl}   +  \delta_{il} \delta_{jk})   -
\delta_{ij} \delta_{kl}  \big ) + e'{\!}_{uv}{\!}^w\, w_{w,ijkl} + b'{\!}_{uv} \, d_{ijkl}  \, .
\label{dwrel}
\end{align}
There are various consistency conditions such as
\be
e^u a'{\!}_{uv} = h_v \, a  \, , \quad \ f_u{\!}^v a'{\!}_{vw} = b'{\!}_{uw} \, a \, , \quad
 f_u{\!}^v h_v = b'{\!}_{uv} \hskip 0.5pt e^v \, , \quad
e'{\!}_{uv}{\!}^x \, a'{\!}_{xw} = e'{\!}_{uw}{\!}^x \, a'{\!}_{xv} \, .
\label{consis}
\ee
The eigenvalue equation \eqref{rhoe} is extended to
\be
d_{ijkl} \, v_{kl} = \mu \, v_{ij} \, ,   \quad w_{u,ijkl} \, v_{kl} = \mu_u \, v_{ij} \,\, ,
\label{rhoe2}
\ee
requiring
\be
\mu^2 = e^u \mu_u +  b \, \mu +  \tfrac{N}{N-1} \, a  \, , \quad
\mu \, \mu_u = f_u{}^v \mu_v + h_u \, \mu \, , \quad
\mu_u \,\mu_v = e'{\!}_{uv}{}^w \mu_w +  b'{\!}_{uv}  \, \mu +  \tfrac{N}{N-1} \, a'{\!}_{uv} \, .
\label{eiggen}
\ee

 Assuming  \eqref{ddrel2} with \eqref{wrel}
the one loop $\beta$-functions remain as in \eqref{bgh1}. At two loops
in \eqref{bgh2}, using $w_{u,imjn}\,d_{klmn}= - \tfrac12\, w_{u,ijmn}\, d_{klmn}$,
and  \eqref{bsigrho}
\be
\Delta \, \beta_g{\!}^{(2)} = 3 \, e^u h_u \, g^3 \, , \qquad
\Delta \beta_{\rho,ij}{\!}^{(2)} =  \tfrac12\, g^2  e^u w_{u,ijkl} \, \rho_{kl} \,  .
\label{dbgh2}
\ee
and, since in \eqref{crel} $\Delta \A = - a\, e^u h_u$, at three loops \eqref{bgh3} is modified by
\begin{align}
\Delta \, \beta_\lambda{\!}^{(3)} = {}& - a\,  e^u h_u  \big ( \tfrac{11}{2}  \, + 3 \,\zeta_3 \big ) g^4 \, , \nn \\
\Delta \, \beta_g{\!}^{(3)} ={}&
- \tfrac34  \big ( 29 \, b \,  e^u h_u  - 8 \,  e^u f_u{\!}^v h_v \big )   g^4
- 6 \, e^u h_u \big (  7  + 12\, \zeta_3  \big )  \lambda g^3 \, , \nn \\
\Delta \beta_{\rho,ij}{\!}^{(3)} = {}& -\tfrac{11}{4} \, e^u h_u \,  g^3 \, d_{ijkl} \, \rho_{kl}
  - \big  ( 3 \, g^2 \lambda  + \tfrac 54 \, b \, g^3 \big )  e^u w_{u,ijkl} \, \rho_{kl}
  + g^3 \, e^u f_u{}^v w_{v,ijkl} \, \rho_{kl}  \, .
\label{dbgh3}
\end{align}
As a consequence of the changes in the $\beta$-functions in \eqref{dbgh2} and \eqref{dbgh3}
then in \eqref{Ncrit12} there are additional terms
\begin{align}
\Delta F_1(N) = {}& - \frac{6(N-4)}{N+2} \, \frac{e^uh_u}{a}  \, , \nn \\
\Delta F_2(N) = {}&
- \frac{3(N^3 + 35 N^2 -82N -8)}{4(N-1)(N+2)^2} \, \frac{e^uh_u}{a}
+  \frac{9(N-4)^2(N+8)}{4(N+2)^4} \, \Big ( \frac{e^uh_u}{a} \Big )^2\nn \\
&{} -\frac{16(N-4)^3}{9 (N+2)^2} \,  \frac{e^u f_u{\!}^v h_v}{b^3} - \frac{3(N-4)(N-28)}{(N+2)^2} \, \frac{e^uh_u}{a} \, \zeta_3 \, .
\label{Ncrit2}
\end{align}

A particular example arises when $d_{ijkl}$ is the symmetric traceless rank 4 tensor
for a Lie group when there is an independent quartic Casimir.
For a simple Lie algebra with a basis $\{ t_i \}$ where we assume, with indices
$i,j,k,l=1,\dots , N $,
 \be
 [ t_i, \, t_j ] = f_{ijk}\, t_k \, , \quad f_{ikl} f_{jkl} = C \, \delta_{ij} \, , \quad
 \tr ( t_i t_j ) = - R \, \delta_{ij} \, , \quad t_i t_i = - C_R \, {\mathds 1} \, ,
 \label{LieA}
 \ee
 then we may take
 \be
 d_{ijkl} \phi_i \phi_j \phi_l \phi_l =  \tr \big ( t_\phi{\!}^4 \big  ) + y \, (\phi^2)^2 \, ,\qquad
 t_\phi = t_i \phi_i \, ,
 \label{dC}
 \ee
 with $y$ determined to ensure $d_{ijkl}$ is traceless.
 Using $t_i t_\phi t_i = (\tfrac12 C - C_R) t_\phi$ this gives
 \be
 y = - \frac{3\,  C_R - \tfrac12\,  C}{N+2} \, R .
 \ee
 Similarly we may define, if $\tr( t_i \{ t_j , t_k \} ) = 0$,
 \be
 w_{ijkl}\phi_k\phi_l = -  \tr  \big ( [ t_i ,t_\phi ] \, [t_j , t_\phi ] \big )
+ z ( \delta _{ij} \phi^2 -  \phi_i   \phi_j ) \, , \quad z = - \frac{C\, R}{N-1} \, .
\ee

 For the case of $O(n)$ then $t_i$ are a basis of $n\times n$ antisymmetric
 matrices  so that the theory reduces to antisymmetric  tensor fields
 $\vphi_{ab} = (t_\phi)_{ab}, \, a,b=1,\dots,n$, as discussed in \cite{Antonov,Antonov2}.
 In this case in \eqref{LieA}
 \be
 C= 2(n-2) \, , \quad R = 2 \,  , \quad C_R = n-1 \, , \quad N= \tfrac12 n(n-1) \, ,
 \ee
 so that $y=-2(2n-1)/(N+2), \, z= - 8/(n+1)$. In this case for $d_{ijkl} \phi_i \phi_j \phi_l \phi_l$
given in  \eqref{dC}
  $- \tfrac{2(n^2-4)}{n(N+2)} (\phi^2)^2 \le d_{ijkl} \phi_i \phi_j \phi_k \phi_l
\le \tfrac{(n-2)(n-3)}{N+2}(\phi^2)^2 $ (for $n$ odd the lower bound is modified).
 With the completeness  relation $ t_i  \tr (t_i Y ) = - R \, Y$  for any $Y^T = -Y$
 and also $t_i M t_i = (C_R - \tfrac12 C ) M^T - {\mathds 1} \, \tr(M) $.  For this case the
 definition \eqref{dC} gives $d_{ijkl}=0$ for $n=2,3$ and with the above results we may verify
 \eqref{ddrel2} with
 \be
 a = \frac{2(n-2)(n-3)(n+1)(n+2)}{3(N+2)^2} \, , \quad
 b = \frac{(n-5)(n+4)(2n-1)}{9(N+2)} \, , \quad e = \tfrac{2}{27}(n+1)\, ,
 \label{abn}
 \ee
and, with more complexity,
\be
c= \frac{n^8- 84\,n^7 +302\,n^6 +2212\,n^5-6215\,n^4 -13944\,n^3
+ 27456\,n^2 -40448\,n+ 15616}{216(N+2)^3} \, .
\label{cn}
\ee
With $a,b$ given by \eqref{abn}  then in \eqref{defFN} $F(N)<0 $ for $n> \frac14(2 + 3 \sqrt{22})$
so that there are only  fixed points for real non zero couplings $\lambda,g$ when $n=4$.
In \eqref{dwrel}
\begin{align}
h ={}&  \frac{8 (N+2)}{3(n+1)}\, , \qquad \qquad f=\frac{ (n-8)(n+1)^2}{9(N+2)} \, , \nn \\
a'= {}& \frac {24(n-2)(n-3)(n+2)}{(n+1)(N+2)}\, ,  \qquad b'= 4 (n-8) \, , \qquad
e'=\frac{4(N+11)}{3(n+1)} \, ,
\end{align}
which in conjunction with \eqref{cn} determine higher loop contributions to
$\beta$-functions.\footnote{The results to three loops are in numerical agreement
with \cite{Antonov2} in terms of couplings $g_1 = 6 \lambda + 2 y \, g, \, g_2 =8g$.}

\subsubsection{\texorpdfstring{$\boldsymbol{O(m)\times O(n)}$}{O(m) x O(n)} fixed points}
A theory with non trivial fixed points is obtained by considering a symmetry breaking
where $O(N) \to O(m) \times O(n) /{\mathbb Z}_2$, $N=m\, n$ as was considered in
 \cite{Kawamura,Pelissetto}.
This case can be discussed in the present context by taking
$\phi_i = \Phi_{ar}$ with $a=1,\dots , m$, $r= 1,\dots ,n$ and, with the usual
summation convention for repeated paired indices,
\be
d_{ijkl} \phi_i \phi_j \phi_k \phi_l = \Phi_{ar} \Phi_{br} \Phi_{as} \Phi_{bs}
- \frac{m+n+1}{N+2} \, (\Phi^2)^2 \, .
\ee
This vanishes for $m=1$ or $n=1$ and for $m\le n$, satisfies the bounds
$- \tfrac{(m-1)(m+2)}{m(N+2)} \, (\Phi^2)^2
\le d_{ijkl} \phi_i \phi_j \phi_k \phi_l  \le \tfrac{(m-1)(n-1)}{N+2} \, (\Phi^2)^2$.
In this case there are two tensors $w_{u,ijkl}$, $u=1,2$,  which are determined by
 \begin{align}
 w_{1,ijkl} \phi_k \phi_l = {}&  \delta_{rs}  \Phi_{at} \Phi_{bt} -  \delta_{ab} \Phi_{cr} \Phi_{cs}
 - \frac{n-m}{N-1} \big ( \delta_{ab} \delta_{rs} \, \Phi^2 - \Phi_{ar} \Phi_{ bs} \big )\, , \nn \\
 w_{2,ijkl} \phi_k \phi_l = {}&  \delta_{rs} \Phi_{at} \Phi_{bt} +  \delta_{ab}  \Phi_{cr} \Phi_{cs}
 - 2 \, \Phi_{as}\Phi_{br}
 - \frac{m+n-2}{N-1} \big ( \delta_{ab} \delta_{rs} \, \Phi^2 - \Phi_{ar} \Phi_{ bs} \big )\, .
 \end{align}
 In \eqref{ddrel}
 \begin{align}
 a= {}& \frac{(m-1)(m+2)(n-1)(n+2)}{3(N+2)^2} \, , \nn \\
 b ={}& \frac{D_{mn}}{9(N+2)} \, , \qquad D_{mn} = mn (m+n) +4\,mn -10(m+n)-4 \, ,\nn \\
 e^1 ={}& \tfrac{1}{18}{(n-m)} \, , \qquad e^2 = \tfrac{1}{54} ( m+n + 4) \, ,
 \label{abmn}
 \end{align}
 and in \eqref{crel}
 \begin{align}
 c = {}& \frac{1}{216(mn+2)^3}\, \big ( m n - 4 (m +n) -2 \big )\nn \\
 &{}\times \big ( 5\, m^3 n^3 + 10\, m^2 n^2 (m+n) - 40\, mn (m^2+n^2)
 + 20\, m^2 n^2 - 88 \,mn (m+n) \nn \\
 \noalign{\vskip- 4pt}
 &\hskip 1cm+ 16 (m^2 +n^2 ) + 284\, mn - 24 (m+n) + 240 \big ) \, ,
 \label{cmn}
 \end{align}
and in \eqref{wrel}
\begin{align}
f_1{\!}^1 = {}& \tfrac16 (n+m+2) - \frac{2(m+n+1)}{3(N+2)}\, , \qquad f_1{\!}^2 = \tfrac{1}{18}(n-m) \, ,\nn \\
f_2{\!}^1 = {}& \tfrac16(n-m) \, , \qquad
f_2{\!}^2 = \tfrac{1}{18} (n+m-2) - \frac{2(m+n+1)}{3(N+2)}\, ,  \nn \\
 h_1 = {}& (n-m) \bigg ( \frac13 + \frac{1}{N-1} \bigg ) \, , \quad
 h_2 =  (n+m-2) \bigg ( \frac13 + \frac{1}{N-1} \bigg ) \, , \nn \\
a'{\!}_{11}={}& \frac{(m-1)(n-1)(2 N + 3m +3n +4)}{(N-1)(N+2)} \, , \quad
a'{\!}_{12}=  a'{\!}_{21}= 3\, \frac{(m-1)(n-1)(n-m)}{(N-1)(N+2)} \, , \nn \\
a'{\!}_{22}={}& 3\, \frac{(m-1)(n-1)(2 N + m +n - 4)}{(N-1)(N+2)} \, , \nn \\
b'{\!}_{11} ={}& n+m \, , \qquad  b'{\!}_{12}=  b'{\!}_{21}= n-m \, , \qquad
b'{\!}_{22}=  m+n-8 \, .
\label{fhmn}
\end{align}
The coefficients $e'{\!}_{uv}{\!}^w$ are also determined but not
quoted here. For $m=n$ only the $w_2$ tensor is relevant.

These results suffice to determine the $\beta$-functions to three
loops.\footnote{The $\beta$-functions to three loops obtained from
\eqref{bgh1}, \eqref{bgh2}, \eqref{bgh3} in conjunction with \eqref{dbgh2},
\eqref{dbgh3} and \eqref{abmn}, \eqref{cmn}, \eqref{fhmn} are identical
with the results of Pelissetto {\it et al} \cite{Pelissetto} where their results are expressed in
terms of the couplings $u=6\lambda + 2 (1-(m+n+1)/(mn +2)) g$,
$v=2g$.}
Solving the eigenvalue equations \eqref{eiggen} for the $O(m)\times O(n)$
case gives four possibilities for  $(\mu, \, \mu_1, \, \mu_2)$
\begin{align}
({\rm i}) \quad & \big ( \tfrac{(m-1)(m+2) n}{3(N+2)}, \, - \tfrac{(m^2-1)n}{N-1}, \, \tfrac{(m-1)^2 n}{N-1} \big ) \, ,
& & \dim = \tfrac12 (n-1)(n+2) \, ,
\nn \\
({\rm ii}) \quad & \big ( \tfrac{ m(n-1)(n+2) }{3(N+2)}, \,  \tfrac{m(n^2-1)}{N-1}, \, \tfrac{m(n-1)^2 }{N-1} \big ) \, ,
& &  \dim = \tfrac12 (m-1)(m+2) \, ,
\nn \\
 ({\rm iii}) \quad & \big ( \tfrac{N-2m-2n}{3(N+2)}, \,  \tfrac{n-m}{N-1}, \, -\tfrac{2N-m-n}{N-1} \big ) \, ,
& & \dim = \tfrac14 (m-1)(m+2)(n-1)(n+2) \, ,
\nn \\
({\rm iv}) \quad & \big (- \tfrac{(m+2)(n+2) }{3(N+2)}, \,  \tfrac{n-m}{N-1}, \, \tfrac{2N + m+n -4}{N-1} \big ) \, ,
& & \dim = \tfrac14 m(m-1)n(n-1)\, ,
\label{eigenOmn}
\end{align}
corresponding to the decomposition of the eigenvectors $v_{ar,bs}$ into irreducible
representations of $O(m)\times O(n)$, $v_{ar,as}, \ v_{a r , b r}, \
v_{a r , bs} + v_{br ,as} - \frac{2}{m} \delta_{ab} v_{cr,cs} - \frac{2}{n} \delta_{rs}v_{a t , b t}$
and $v_{ar,bs}- v_{br,as}$. In \eqref{defFN}
\begin{align}
F(m,n)_{O(m)\times O(n)} = {}&\frac{1}{9a}\,  R_{mn}\, , \quad
R_{mn} =  m^2 + n^2 - 10 \, m n - 4(m+n) + 52 \, ,
\end{align}
with notation borrowed from \cite{Kawamura}.
Positivity of $F$  then requires  $R_{mn}>0$ so that $n$ is restricted by $n< 5m+2 - 2\sqrt{6(m-1)(m+2)}$ or $n>
5m +2 + 2\sqrt{6(m-1)(m+2)}$. This excludes $m=n$ except when  $m=n=2$.
With the results in \eqref{Ncrit2} these boundaries are modified in the $\vep$ expansion to
\begin{align}
n_{{\rm crit}\pm} = {}& 5m +2 \pm 2  \sqrt{6(m-1)(m+2)} - (5m+2)\vep
\mp\tfrac{25m^2 +22m-32}{2\sqrt{6(m-1)(m+2)}}\, \vep \nn \\
& {}+  \Big ( \tfrac{8m^5 + 31m^4 -426m^3-1376m^2+1184 m +1632}{8(m-7)(m+8)(m-1)(m+2)}
\pm  \tfrac{20m^4 + 73 m^3 -1230 m^2 -2960m+6176}{8(m-7)(m+8)\sqrt{6(m-1)(m+2)}}\Big ) \zeta_3 \, \vep^2\nn \\
& {}+ \tfrac{4m^7 + 21 m^6 - 417 m^5 -1883m^4 +9288 m^3 +45444m^2 +18880 m -33024}{
24(m-7)^2(m+8)^2(m-1)(m+2)} \, \vep^2\nn \\
&{} \pm \tfrac{5 m^8 + 58m^7 -389  m^6 - 5854 m^5 + 7756m^4 + 139240 m^3 + 176192m^2 - 160256 m -253292}
{32(m-7)^2(m+8)^2(m-1)(m+2)\sqrt{6(m-1)(m+2)} }\, \vep^2 \, .
\end{align}
To ${\rm O}(\vep)$ this agrees with \cite{Kawamura} and to ${\rm O}(\vep^2)$ when $m=2$ with \cite{Antonenko}
and for any $m$ with \cite{Pelissetto}. For $n> n_{{\rm crit}+} $ and $n< n_{{\rm crit}-}$ there are  two
fixed points with non zero $g$. The $O(N)$ fixed point becomes unstable when $n>N_{{\rm crit},H}/m$ as given in
\eqref{NcritH}.

At the fixed point to lowest order
\begin{align}
\lambda_{*\, \pm} =  \frac{1}{2(N+2)}\, \frac{D_{mn} \pm (N+2)\sqrt{R_{mn}}}{ D_{mn}\pm 6\sqrt{R_{mn}} }\, \vep \, ,
\qquad g_{*\, \pm} = \frac{3(N-4)}{ D_{mn}\pm 6 \sqrt{R_{mn}} }\, \vep \, ,
\end{align}
and from \eqref{eigen4} the stability matrix eigenvalues are
\be
\kappa_{0,\pm} = \vep \, , \qquad \kappa_{1,\pm} =  \pm \frac{(N-4)\sqrt{R_{mn}}}{ D_{mn}\pm 6\sqrt{R_{mn}} }\, \vep \, .
\ee
From \eqref{Astar}
\be
A_{*\pm} =  -  \frac{1}{48} \, N \, \vep^3  + \frac{(N-4)^2 \, R_{mn}}{ 48\big ( D_{mn}\pm 6\sqrt{R_{mn}} \big )^2 } \, N \, \vep^3 \, .
\ee
For a fixed point $R_{mn}>0$ and so this respects the bound \eqref{Abound}.
In application to condensed matter systems the two fixed points,
denoted by $+,-$, are referred to as chiral and anti-chiral \cite{Kawamura}.
Since $A_{*+} < A_{*-}$ the chiral fixed point  is therefore the stable one.

At higher order the results are complicated but for $n \gg m$ and to order $\vep^3$,
taking into  account two and three loop contributions,
\begin{align}
\lambda_{*\, \pm}
\sim {}& \begin{cases}\frac{1}{N} \big ( 1-\frac{2}{N} \big )  \, \vep
+ \frac{3}{2N n} (m+1) \big ( -2\, \vep + 3 \,\vep^2 - \frac{11}{8} \, \vep^3 \big )
\\ \noalign{\vskip 2pt}
\frac{1}{N^2}( m-1)(m+2) \,\big ( 3 \,  \vep - 2 \, \vep^2  - \tfrac32 \, \vep^3
\big )  \end{cases} \hskip -0.4 cm  , \nn \\
\noalign{\vskip 4pt}
 g_{*\, \pm}
\sim {}& \begin{cases} \tfrac{3}{n} \Big (  \vep  + \frac{1}{n}\big ({ -(m+4)} \vep +  \frac32(m+3)\vep^2 - \frac{1}{16}(13m+33)\vep^3 \big )\Big )
\\
\noalign{\vskip 4pt}
 \tfrac{3}{n}\Big ( \vep\\
\noalign{\vskip -4pt}
 \hskip 0.5cm + \frac{1}{N} \big ({ - (m-2)(m+6)}\vep + \frac12(3m^2 +9m -34) \vep^2
-\frac{1}{16}(13m^2 +33m -162) \vep^3\big ) \Big )
\end{cases} \hskip -0.4cm .
\end{align}
Correspondingly in this limit, with notation from \eqref{kHpm} for $k_\pm$,
\be
\gamma_{ \pm} \sim k_{\pm} \,  \tfrac{1}{4n} \,  \big ( \vep^2 - \tfrac14\, \vep^3 \big ) \,  , \quad
\gamma_{\sigma\, \pm} \sim \begin{cases} \vep - k_+\, \frac{2}{n} \big ( 3 \, \vep - \tfrac{13}{4}\,  \vep^2 +  \tfrac{3}{16}\, \vep^3 \big ) \\
\noalign{\vskip 2pt}
k_- \, \frac{2}{n} \big  (3 \, \vep - \tfrac{13}{4}\,  \vep^2 +  \tfrac{3}{16}\, \vep^3   \big )  \end{cases} \hskip -0.4cm ,
\ee
There are also similar results for $\gamma_{\rho}$ which decomposes into four cases depending on the eigenvalues
in \eqref{eigenOmn}
\begin{align}
({\rm i}) \quad & \gamma_{\rho_1\, \pm} \sim 2\, \gamma_\pm - \tfrac{2}{n} \, k_\pm \,
\big  ({ \vep} - \tfrac{1}{2}\,  \vep^2 -  \tfrac{1}{4}\, \vep^3   \big )  \, , \nn  \\
\noalign{\vskip 2pt}
({\rm ii}) \quad & \gamma_{\rho_2\, \pm} \sim \vep -  \begin{cases}  \frac{1}{n} \big ( {(m+3)\,\vep} - \tfrac{1}{4}(5m+13)\,  \vep^2 +  \tfrac{3}{16}(m+1)\, \vep^3   \big ) \\
\noalign{\vskip 2pt}
\frac{1}{N} \big  ({(m-2)(m+5)} \, \vep - \tfrac{1}{4}(5m^2 +13m -50) \,  \vep^2 + \tfrac{1}{16}(3m^2+3m-22)\, \vep^3   \big )  \end{cases} \hskip -0.5cm ,
\nn \\
({\rm iii}) \quad & \gamma_{\rho_3\, \pm} \sim 2\, \gamma_\pm +
 \begin{cases}  \frac{1}{n} \big ( {\vep} - \tfrac{1}{2}\,  \vep^2 -  \tfrac{1}{4}\, \vep^3   \big ) \\
\noalign{\vskip 2pt}
\frac{1}{N} (m-2) \big ( \vep - \tfrac{1}{2} \,  \vep^2 -  \tfrac{1}{4}\, \vep^3   \big )  \end{cases} \hskip -0.4cm ,
\nn \\
({\rm iv}) \quad & \gamma_{\rho_4\, \pm} \sim 2 \, \gamma_\pm -  \begin{cases}  \frac{1}{n} \big ({ \vep} - \tfrac12\,  \vep^2 -  \tfrac{1}{4}\, \vep^3   \big ) \\
\noalign{\vskip 2pt}
\frac{1}{N} (m+2) \big  ({  \vep} - \tfrac{1}{2}\,  \vep^2 -  \tfrac{1}{4}\, \vep^3   \big )  \end{cases} \hskip -0.4cm .
\label{largemn}
\end{align}
At the fixed point $\Delta_{\sigma,\rho} = d-2 + \gamma_{\sigma,\rho*}$.
For case (ii) $\Delta_{\rho*\pm}, \, \Delta_{\sigma*+} = 2 + {\rm O}(1/n)$,
the large $n$ analysis is based on couplings $\rho_{ab}\, \Phi_{ar}\Phi_{br}$,
with $\rho_{ab}$ symmetric and traceless, and $\sigma\, \Phi^2$ which
have dimension $d$, for the anti-chiral fixed point only the $\rho_{ab}$
coupling is relevant.  The results in \eqref{largemn} show the universal
behaviour at large $n$ expected by large $n$ results \cite{Pelissetto,Gracey3,Gracey4}.

\subsubsection{\texorpdfstring{$\boldsymbol{MN}$}{MN} fixed points}
An alternative breaking of $O(N)$ symmetry (referred to in some literature as the $MN$ model \cite{RGrev}) is realised when
$O(N) \to {\cal S}_n  \ltimes O(m)^n$ for $N=m \, n$,
$\phi_i$  is decomposed as $n$ $m$-vectors ${\vec \vphi}_{r}$  and
\be
d_{ijkl} \phi_i \phi_j \phi_k \phi_l =  \textstyle{\sum_r}( {\vec \vphi}_{r}{\!}^2)^2
- \frac{m+2}{N+2} \, (\vec \vphi {\,}^2)^2 \, , \qquad  \vec \vphi{\,}^2 = \textstyle{\sum_r}{\vec \vphi}_{r}{\!} ^2 \, ,
\label{dOmn}
\ee
which satisfies $-\tfrac{2(n-1)}{n(N+2)} (\vec \vphi {\,}^2)^2 \le  d_{ijkl} \phi_i \phi_j \phi_k \phi_l
\le  \tfrac{(m-1)n}{N+2}(\vec \vphi {\,}^2)^2  $.
This expression reduces to the hypercubic model in \eqref{cubic} for $m=1$ and
if in \eqref{VfourN} $\lambda = \frac{m+2}{3(N+2)}\, g$ to $n$ decoupled $O(m)$
theories. Past discussions include \cite{Mukamel2,Shpot,Shpot2}. The $\beta$-functions for such theories have been determined
to four loops \cite{Mudrov,Mudrov2} for $m=2$ and are physically relevant for
$m=2,\, n=2,3$. For $m=n=2$ this case with $O(2)^2$ symmetry is identical with the $O(2)\times O(2)$ symmetric
theory, taking ${\vec \vphi}_1 = \frac{1}{\sqrt 2}( \Phi_{11}-\Phi_{22}, \, \Phi_{12} + \Phi_{21})$,
 ${\vec \vphi}_2 = \frac{1}{\sqrt 2}( \Phi_{11} + \Phi_{22}, \, \Phi_{12} - \Phi_{21})$, $ \vec \vphi{\,}^2 = \Phi^2$,
$ \textstyle{\sum_r}( {\vec \vphi}_{r}{\!}^2)^2  = -   \Phi_{ar} \Phi_{br} \Phi_{as} \Phi_{bs} + \tfrac32 \, (\Phi^2)^2$.
 The $\beta$-functions in the two cases are related by
\be
\beta_\lambda(\lambda,g)\big |_{O(2)\times O(2)} =  \beta_\lambda(\lambda,-g )\big |_{O(2)^2} \, , \qquad
\beta_g(\lambda,g)\big |_{O(2)\times O(2)} =  - \beta_g(\lambda,-g )\big |_{O(2)^2} \, .
\ee

In the framework described here there is a single mixed symmetry tensor, non zero for $m>1$,
\be
 w_{ijkl}\phi_i \phi_j  \phi'{\!}_k \phi'{\!}_l \to
  \textstyle{\sum_r} \big ( {\vec \vphi}_{r}{\!}^2 \, {\vec \vphi}{\,}'{\!}_{r}{\!}^2 - ( {\vec \vphi}_r \!  \cdot  \! {\vec \vphi}{\,}'{}{\!}_r )^2 \big )
  - \tfrac{m-1}{N-1} \, \big ( \vec \vphi {\,}^2 \, \vec \vphi  {\,}^{\prime \hskip 0.5 pt 2}  - ( {\vec \vphi} \cdot  {\vec \vphi}{\,}' )^2 \big ) \, .
  \ee
  This gives
  \begin{align}
&  a = \frac{2\, m (m+2)(n-1)}{3 (N+2)^2} \, ,  \hskip 0.6cm
  b= \frac{m(m+8)(n-1)  + (m+2)(m-4)}{9(N+2)} \, , \quad  e = \frac{2}{27} (m+2 ) \, , \nn \\
  &  c = \frac{(m n - 2m -2)}{27 ( m n +2)^3} \,\big ( 5\,  m^3 n^2 + 22\, m^2 n^2 - 10\,  m^3 n - 44\, m^2 n + 4\, m^2 -12 \, m +8 \big ) \, ,
  \nn \\
  & f = \frac{2 (N-1)(m+2)}{9(N+2)} \, , \hskip 0.9cm h = \frac{(N+2)(m-1)}{3(N-1)} \, , \nn \\
  & a' = \frac{3(m-1)m(n-1)}{(N-1)(N+2)} \, , \hskip 0.6cm   b' = m-1 \, , \quad e' = \tfrac13(2m-5)
  +  2 \, \frac{m-1}{N-1} \, .
  \label{cubicmn}
  \end{align}
  For eigenvectors of the $d,w$ tensors with eigenvalues $(\mu,\mu')$
there are now  three cases
  \begin{align}
  &\big ( \tfrac{N(m+2)}{3 (N+2)}, \tfrac{N(m-1)}{N-1} \big ) \, ,
&  & \dim = n-1 \, ,   &&  \nn \\
  & \big (- \tfrac{2(m+2)}{3 (N+2)}, \tfrac{m-1}{N-1} \big )  \, ,&
& \dim = \tfrac12 \,  m^2 n(n-1) \, ,  && \nn \\
&     \big ( \tfrac{2m (n-1)}{3 (N+2)}, -\tfrac{m(n-1)}{N-1} \big )\, ,  &
 & \dim = \tfrac12(m-1)(m+2)n\, . &&
 \label{MNeig}
 \end{align}
 The results in \eqref{cubicmn} are sufficient to determine the $\beta$-functions in
 this case to three loops for general $m,n$.\footnote{The results are in accord with
 \cite{Mudrov} for $m=2$ expressed in terms of couplings, up to a scale factor,
 $u= 3\lambda - (m+2)/(mn+2)\, g, \ v=g$. In general $\beta_u |_{u=0} =0$ reflecting
 the decoupled fixed point.}

The polynomial  $f(x)$, as defined  in  \eqref{f4x}, now has roots, apart from $x=0$,
given by $x_-= (N-4)(N+2) /2m (n-1), \,  x_+ = 3(N+2)/(m+2)$, $x_+>x_-$ for $m<4$. In this
case $f'(x_\pm) = \pm \tfrac13(4-m) x_\pm$. According to \eqref{eigen4}
a stable fixed point requires $f'(x_*)<0$. For $N>4$, when the $O(N)$ symmetric
fixed point has relevant perturbations, the stable  fixed point is given by
$x_* = x_-$ or $x_*=x_+$ depending  on whether $m<4$ or  $m>4$.
In \eqref{defFN}
\be
F(m,n)_{O(m)^n} = \frac{(m-4)^2(N+2)}{6\, m (m+2)(n-1)} \, ,
\ee
so that for this breaking there are always non trivial additional fixed points which
coincide when $m=4$.
The two fixed points extending \eqref{Nfix} are then given by
\begin{align}
\lambda_{*\, -} = {}&
\frac{6(N-m)}{(N+2)\big ((m+8)N - 16(m-1)\big )}\, \vep  \, ,
\quad g_{*\, -} = \frac{3(N-4)}{(m+8)N - 16(m-1)} \, \vep\, , \nn \\
\lambda_{*\, +} = {}& \frac{m+2}{(m+8)(N+2)}\, \vep  \, , \hskip 3.25cm
g_{*\,+} = \frac{3}{m+8} \, \vep \, ,
\label{Nmfix}
\end{align}
with corresponding anomalous dimensions
\be
\gamma_{-} =  \frac{(N-m)\big ( (m+2)N - 10m +16 \big )}
{4 \big ( (m+8)N - 16(m-1) \big )^2 } \, \vep^2\, , \qquad \gamma_{+} = \frac{m+2}{4(m+8)^2}\, \vep^2
\ee
The stability matrix eigenvalues at lowest order are then
\be
\kappa_{0,\pm} = \vep \, , \qquad
\kappa_{1,+} = \frac{m-4}{m+8} \, \vep \, , \qquad
\kappa_{1,-} = - \frac{(m-4)(N-4)}{ (m+8)N - 16(m-1)} \, \vep \, .
\label{kmn}
\ee
For $m = 2, \, n=2,3$ results to four loops are given in \cite{Mudrov2} (where $\kappa = - 2 \omega$).
The fixed point realised for $x_*= x_+$ corresponds to decoupled
$O(m)$ theories, this is the stable fixed point for $m>4$.
Using \eqref{agamma}
\begin{align}
A_{*+} = {}& - \frac{1}{48} \, N \, \vep^3+ \frac{(m-4)^2}{48(m+8)^2}\, N \vep^3
\, , \nn \\
A_{*-} = {}& - \frac{1}{48} \, N \, \vep^3 + \frac{(m-4)^2 ( N-4)^2}
{48 \big ( (m+8)N - 16(m-1) \big )^2 } \, N  \vep^3 \, .
\label{Amn}
\end{align}

For a positive potential at the $x_*=x_-$ fixed point we should require
$(m+8)N > 16(m-1)$ so that $\lambda_{*\, -}>0$ but this is satisfied for any $m,n>1$.
For $g_{*\, -}>0$ positivity requires $3n (N+2) \lambda_* > 2(n-1) g_*$ but with \eqref{Nmfix}
this provides no constraint.
Stability of the $\lambda_{*\, -}, \, g_{*\, -} $ fixed point requires $m<4$ or $m=1,2,3$, $m=1$
of course reduces to the hypercubic fixed point in \eqref{Nfix}.
The minimum value for $A_*$ is attained at the bifurcation point when $m=4, \, N=4n$.

For this theory the anomalous dimensions of quadratic operators may easily be determined to three
loops using \eqref{bsigrho} together with \eqref{dbgh2}, \eqref{dbgh3} and the eigenvalues given in
\eqref{MNeig}. For simplicity at large $m$ at the non decoupled fixed point,
\be
\gamma_- \sim \tfrac{1}{4N} (n-1)\big (  \vep^2 - \tfrac{1}{4} \, \vep^3 \big ) \, ,
\ee
and
\begin{align}
\gamma_{\sigma\,-} \sim  {}&  \tfrac{2}{N} (n-1)\big ( 3\,  \vep - \tfrac{13}{4} \, \vep^2 + \tfrac{3}{16} \, \vep^3 \big ) \, , \nn \\
\gamma_{\rho_1\,-} \sim {}& \vep - \tfrac{2}{N} \big ((n-1) (3\,  \vep - \tfrac{13}{4} \, \vep^2 + \tfrac{3}{16} \, \vep^3 ) -
 2\,  \vep  + 3 \, \vep^2 +  \tfrac{1}{2} \, \vep^3 \big ) \, , \nn \\
\gamma_{\rho_2\,-} \sim {}& 2\, \gamma_- - \tfrac{2}{N} \big (  \vep - \tfrac{1}{2}\, \vep^2 - \tfrac{1}{4} \, \vep^3 \big ) \, , \nn \\
\gamma_{\rho_3\,-} \sim {}& 2\, \gamma_- +  \tfrac{2}{N} (n-1) \big (\vep - \tfrac{1}{2} \, \vep^2 - \tfrac{1}{4} \, \vep^3 \big ) \, .
\end{align}
The results reflect the existence of a large $m$ limit for the $MN$ model which is analogous to that for $O(m)\times O(n)$.

\section{Scalar theories in \texorpdfstring{$\mathbf{3\boldsymbol{-}\boldsymbol{\vep}}$}{3-epsilon} dimensions}
\label{threeD}

An $\vep$ expansion starting from three dimensional renormalisable theories
determines the properties of tricritical (since there are at least two
relevant operators if ${\mathbb Z}_2$ symmetry is imposed for a single
component field) fixed points \cite{Tricrit,Lewis3d}. Such $\phi^6$ theories have
been considered from a CFT perspective in \cite{Basu,Nii}.

For our discussion we assume $N$-component scalar fields $\phi_i$ and
initially a general sextic potential $V(\phi) = \frac{1}{6!} \, \lambda_{iklmnp}
\phi_i \phi_j \phi_k \phi_l \phi_m \phi_n + \dots$. After rescaling $V\to
(8\pi)^2 V$ the
lowest order, two loop, result for the $\beta$-functions for the
couplings contained in $V$  is determined by
\be
\beta_V(\phi) = \vep \big (  V(\phi) - \tfrac12 \, \phi_i V_i(\phi) \big )
+  \, \tfrac13 \, V_{ijk}(\phi) V_{ijk}(\phi) \, ,
\label{Bone}
\ee
with $V_i(\phi) = \pr_i V(\phi)$. For the next order, at four loops,\footnote{An equivalent
expression is given in appendix A of \cite{Dwyer}.}
\begin{align}
\beta_V(\phi)^{(4)} ={}&  \tfrac16 \,
V_{ij}(\phi) V_{iklmn}(\phi) V_{jklmn}(\phi) - \tfrac43 \,
V_{ijk}(\phi)  V_{ilmn}(\phi) V_{jklmn}(\phi )  \nn \\
&{} - \tfrac{\pi^2}{12} \,
V_{ijkl}(\phi)  V_{klmn}(\phi) V_{ijmn}(\phi )
+ \phi_i \gamma_{ij}{\!}^{(4)} V_j (\phi) \, ,
\label{Bfour}
\end{align}
with the four loop anomalous dimension matrix
\be
\gamma_{ij}{\!}^{(4)} = \tfrac{1}{90} \,
\lambda_{iklmnp} \lambda_{jklmnp} \, .
\label{gfour}
\ee

For the $O(N)$ symmetric case the potential is restricted to the form
\be
V(\phi) = \frac{\lambda}{48} \, (\phi^2)^3 \, .
\label{VO6}
\ee
From \eqref{Bone},
\be
\beta_\lambda = - 2\vep \, \lambda + 4(3N+22) \lambda^2 \, ,
\label{bet3N1}
\ee
and from \eqref{gfour}
\be
\gamma_{ij}= \gamma_\phi \, \delta_{ij} \, , \quad
 \gamma_\phi{\!}^{(4)} = \tfrac16 (N+2)(N+4) \lambda^2 \, .
\label{gamma4}
\ee
The four loop result gives
\be
\beta_\lambda{\!}^{(4)} = - 4 (53N^2 + 858 N + 3304) \lambda^3
- \tfrac12 \pi^2 (N^3 + 34 N^2 +620 N +2720) \lambda^3 \, .
\label{bet3N}
\ee
Equivalent results were obtained in \cite{Pisarski} and extended to six loops in \cite{Hager0,Hager}.
At the fixed point to lowest order
\be
\lambda_* = \frac{\vep}{2(3N+22)} \, , \qquad
\gamma_{\phi *} = \frac{(N+2)(N+4)}{24(3N+22)^2} \, \vep^2 \, .
\label{lambdaon}
\ee

For large $N$  the perturbative  contributions to the $\beta$-function and the anomalous dimension $\gamma_\phi$
at higher orders have a leading $N$ dependence of the form
$(N\lambda )^{p}N^{\lfloor \frac12(p-3) \rfloor} $  and $(N\lambda )^{p}N^{\lfloor \frac12(p-2) \rfloor} $,  $p=2,3,\dots$,
respectively.\footnote{This counting is the minimal form agreeing with explicit results up to six loops and satisfying the consistency
conditions $N (N\lambda )^{p-1}N^{\lfloor \frac12(p-3) \rfloor} \, (N\lambda )^{p'}N^{\lfloor \frac12(p'-3) \rfloor}
\le  (N\lambda )^{p+p'-1}N^{\lfloor \frac12(p+p'-4) \rfloor}$ corresponding to inserting  a $\lambda^{p'}$ vertex graph
for one of the vertices in the $\lambda^p$ graph.}
Hence  with a rescaled coupling
 \be
\beta_\tlam = - 2\vep \, \tlam + \frac{1}{N} \Big (  12 \, \tlam^2 - \tfrac12 \pi^2 \tlam^3 \Big ) + {\rm O} \Big ( \frac{1}{N^2} \Big ) \, ,
\qquad \tlam = N^2 \lambda \, .
\label{betaN}
\ee
A large $N$ IR fixed point is given, at leading order, by
\be
N \vep = 6 \, \tlam_* - \tfrac14 \pi^2 \, \tlam_*{\!}^2 \le \tfrac{36}{\pi^2} \quad \Rightarrow
\quad
\tlam_* = \tfrac{4}{\pi^2} \big ( 3 - \sqrt{ 9 - \pi^2 N \vep/4}\,\big ) \approx \tfrac{1}{6} N \vep
+ \tfrac{\pi^2}{864} \, (N\vep)^2 \, .
\label{3dfp}
\ee
The bound \cite{Pisarski} determines the radius of convergence of the $\vep$-expansion.
At the fixed point
\be
\beta_\tlam{}'(\tlam_*) = 4 \, \vep - \tfrac{12}{N} \, \tlam_* \, ,
\ee
which is positive for $N \vep < \tfrac{36}{\pi^2}$ and vanishes when  $N \vep = \tfrac{36}{\pi^2}$.
In this large $N$ limit the $\beta$-function results in \eqref{betaN} also imply a UV fixed point
\be
\tlam_{*{\rm UV}} \approx  \tfrac{4}{\pi^2} \big ( 3 + \sqrt{ 9 - \pi^2 N \vep/4}\,\big )
\to \tfrac{24}{\pi^2} \quad \mbox {as}  \quad \vep\to 0 \, ,
\label{fpUV}
\ee
which for $\vep=0$ was considered in \cite{Pisarski}.\footnote{A UV fixed point for large $N$ was also proposed in \cite{Townsend,Appelquist}. However
\cite{Bardeen,David} showed  for  $N\to \infty$ and $d=3$, when the $\beta$-function vanishes,  that a one
loop effective potential  was bounded below  only when $\eta = 8 \pi^2  \tlam  \le 16\pi^2 $ or $\tlam  \le 2$.
The UV fixed point in \eqref{fpUV} for $\vep=0$ is then in an unstable regime and is no longer relevant
in this limit and there is a new scale invariant fixed point at $\tlam=2$.
The situation for large but finite $N$ is less clear \cite{Semenoff}.} The result for $\tlam_{*{\rm UV}}$ extends to $\vep<0$
although then at the IR fixed point $\tlam_*<0$  which is presumably unstable.

The $N$ dependence here is in contrast to the $O(N)$ model in $4-\vep$ dimensions where the perturbative
expansion gives contributions to $\beta_\lambda$ and $\gamma_\phi, \, \gamma_\sigma$ beyond one loop of the form
$\lambda^2 (N\lambda )^p$ and $\lambda (N\lambda )^p$, $p=1,2,\dots$. For large $N$ in this case defining $\tlam = N \lambda$
then, to first order in $1/N$,
$\beta\raisebox{-1.5 pt} {$\scriptstyle{\tlam}$} = - \vep \, \tlam + \tlam^2 + \tfrac{1}{N} \tlam f(\tlam)$,
$\gamma_\phi \sim \tfrac{1}{N} f_\phi(\tlam)$, $\gamma_\sigma \sim \tlam + \tfrac{1}{N} f_\sigma(\tlam)$ and at the fixed point
$\tlam_* \sim \vep - \tfrac{1}{N} f(\vep)$ so that $\gamma_{\phi *} \sim \tfrac{1}{N} f_\phi(\vep), \,
\gamma_{\sigma*} \sim \vep+ \tfrac{1}{N}\big (- f(\vep) +  f_\sigma(\vep) \big )$ where $ f_\phi(\vep) =  {\rm O}(\vep^2)$. The fixed
point here is also unstable when $\vep<0$.

Extending the discussion to lower dimension operators we may consider
\begin{align}
V(\phi) \to V(\phi) &{} + \tfrac12 \, \sigma \, \phi^2 + \tfrac12 \, \rho_{ij} \, \phi_i \phi_j
+ \tfrac12 \, \chi_i \, \phi_i \, \phi^2 + \tfrac16 \, \tau_{ijk} \,\phi_i \phi_j \phi_k  \nn \\
&{}+ \tfrac18 \, \xi \, (\phi^2)^2 + \tfrac14\, \mu_{ij} \, \phi_i\phi_j \phi^2 +
\tfrac{1}{24} \, \nu_{ijkl} \, \phi_i \phi_j \phi_k \phi_l  \, ,
\end{align}
with $\rho_{ij}, \, \tau_{ijk}, \,  \mu_{ij}, \,  \nu_{ijkl} $ symmetric and traceless. The results in \eqref{Bone} and
\eqref{Bfour} determine the anomalous dimensions. For the $O(N)$ symmetric theory determined
by \eqref{VO6}
\begin{align}
\gamma_\sigma{\!}^{(4)} = {}& \tfrac{16}{3} ( N+2) (N+4) \lambda^2 \, ,  \hskip 2cm
\gamma_\rho{\!}^{(4)} =  \tfrac{4}{3} (N+4)(N+8)  \lambda^2 \, ,  \nn \\
\gamma_\chi{\!}^{(2)} = {}& 2 ( N+4 )\lambda \, ,  \hskip 3.7cm
\gamma_\tau{\!}^{(2)} = 4  \, \lambda \, , \nn \\
\gamma_\chi{\!}^{(4)} = {}& - \tfrac12(N+4)( 25N + 242) \lambda^2 \, ,  \hskip 1cm
\gamma_\tau{\!}^{(4)} =  \tfrac12(7 N^2  -6 N - 424) \lambda^2 \, , \nn \\
\gamma_\xi{\!}^{(2)} = {}& 8 ( N+4 )\lambda \, , \qquad \gamma_\mu{\!}^{(2)} =  4 ( N+8 )\lambda \, , \qquad
\gamma_\nu{\!}^{(2)} =  16 \, \lambda \, , \nn \\
\gamma_\xi{\!}^{(4)} = {}& - (N+4)\big ( \tfrac43(85N+566) + \tfrac12(N^2 +18N+116) \pi^2 \big ) \lambda^2 \, ,\nn \\
\gamma_\mu{\!}^{(4)} = {}& - \big ( \tfrac83(14N^2+ 267N+1132) + (N^2 +44N+232) \pi^2 \big ) \lambda^2 \, ,\nn \\
\gamma_\nu{\!}^{(4)} = {}&  \big ( \tfrac43(5N^2 -78 N -968) - 6 (N+16) \pi^2 \big ) \lambda^2 \, .
\end{align}
For large $N$ the leading terms in $\gamma_\sigma, \, \gamma_\xi$ are expected to be of the form
$(N\lambda )^{p}N^{\lfloor \frac12(p-1) \rfloor} $, $(N\lambda )^{p}N^{\lfloor \frac12 p \rfloor} $.
In the large $N$ limit for fixed $\tlam$ these results, together with \cite{Hager}, imply
\begin{align}
\gamma_\phi = {}&  \tfrac{1}{N^2} \, \tfrac16 \, \tlam^2 + {\rm O} \big ( \tfrac{1}{N^3} \big )  \, , \qquad
\gamma_\sigma = \tfrac{1}{N^2} \big (  \tfrac{16}{3} \, \tlam^2  - \tfrac12 \pi^2 \tlam^3 \big ) + {\rm O} \big ( \tfrac{1}{N^3} \big )  \, , \nn \\
\gamma_\xi ={}&  \tfrac{1}{N} \big (  8 \, \tlam - \tfrac12 \pi^2 \tlam^2 \big ) + {\rm O} \big ( \tfrac{1}{N^2} \big ) \, .
\end{align}
At the fixed point the leading results for these anomalous dimensions are given by taking $\tlam\to \tlam_*$ as in \eqref{3dfp}.
The  large $N$ limit here is very different from that which interpolates between four and six dimensions.\footnote{However
with the  large $N$ dependence assumed above there might
be an alternative large $N$ limit where $\lambda'= N^{\frac32 } \lambda$ \cite{Appelquist2} and $\beta_{\lambda'} = - 2 \vep \, \lambda'
+ \lambda' f(\lambda'{}^2) + {\rm O}(N^{-\frac12}) $ where
$ f(\lambda'{}^2) = - \frac12 \pi^2 \lambda'{}^2 + \dots$. For $\vep < 0$ there is a UV fixed point
in which $\lambda'{\, \!}_* \sim \sqrt{ - \tfrac{4}{\pi^2} \, \vep} + {\rm O}(\vep)$. The positive square root is chosen to ensure
a stable potential   and this fixed point is identical with \eqref{fpUV}  continued to  $\vep<0$
and taking $-N\vep \gg 1$.  In  \cite{Pisarski2} it was suggested that
$f(\lambda'{}^2)$ should have a zero leading to an IR fixed point in  the vicinity of three dimensions. Such a fixed point is not accessible with
perturbative arguments but is motivated by the requirement that the UV fixed point should be annihilated by merging with
an IR point for some $0<-\vep<1$ so as to ensure a trivial theory in four dimensions.}
Only a finite number of terms contribute to any given order in $1/N$ for fixed  $N \vep$.
As is clear from \eqref{3dfp} the $O(N)$ tricritical IR fixed point is  not
present  for $N\vep > \tfrac{36}{\pi^2}$.  As argued in \cite{Delamotte} there is a
 critical line $N=N_c(d)$ which   involves another fixed point and  \cite{Delamotte}  showed $N_c(d) \propto 1/\vep$ as $\vep\to 0$.
With tensorial scalar fields transforming under $O(N)^5$
there may be large $N$ limiting theories which may be analysed directly \cite{Giombi}.

For two flavours the discussion in sections \ref{twoFlavSix},
\ref{twoFlavFour} can be extended by taking a general
potential of the form
\begin{align}
V(\phi) =  {}&
\tfrac{1}{6!} \big (  \lambda_1 \, \phi_1{\!}^6 + \, \lambda_2 \, \phi_2{\!}^6 \big )
+ \tfrac{1}{2\times 4!} \, \big ( g_1\,    \phi_1{\!}^4\, \phi_2{\!}^2 + g_2 \,
 \phi_1{\!}^2\, \phi_2{\!}^4 \big ) \nn  \\
& + \tfrac{1}{5!}  \, \big (  h_1 \, \phi_1{\!}^5 \, \phi_2 +  h_2 \, \phi_1 \, \phi_2{\!}^5 \big )
+ \tfrac{1}{3!^2}  \,   h \, \phi_1{\!}^3 \, \phi_2{\!}^3  \, ,
\label{Vphi6}
\end{align}
with seven couplings $G_I = ( \lambda_1, \, \lambda_2 , g_1 , g_2 , h_1 , h_2 ,h)$.
Decomposing into representations of $O(2)$ give
\begin{align}
I_1 ={}& \lambda_1 + \lambda_2 + 3\, g_1 + 3 \, g_2 \, , \hskip 1.1cm
v_6 = \begin{pmatrix}  \tfrac12 \big (\lambda_1 - \lambda_2  -15( g_1-g_2) \big ) \\
3(h_1+h_2) - 10\, h \end{pmatrix} \, , \quad\nn \\
v_2 = {}& \begin{pmatrix}  \tfrac12 \big ( \lambda_1 - \lambda_2 + g_1-g_2 \big )  \\
h_1+h_2 + 2\, h \end{pmatrix} \, , \quad
v_4 =  \begin{pmatrix} \frac14\big ( \lambda_1 + \lambda_2 - 5(  g_1+ g_2)\big ) \\
h_1-h_2  \end{pmatrix} \, , \quad
\end{align}
where $I_1$ is an invariant and $v_2,v_4,v_6$ are two-vectors transforming as in \eqref{fix4}.
The $O(2)$ invariant theory as in \eqref{VO6} corresponds to taking $v_2=v_4=v_6=0$
implying  $h_1 = h_2 = h=0$, $\lambda_1 = \lambda_2 = 5 g_1 = 5g_2 = 15\lambda$.
In a similar fashion to \eqref{Ifour} there are five $O(2)$ invariants, in addition to $I_1$,
which can be constructed from $v_2,v_4,v_6$ using the transformation \eqref{fix3}.

Imposing restrictions on the $O(2)$ representation space formed by the couplings
may reduce the RG flow to a sub manifold. Thus setting $v_2=v_4=0$ gives, since
tensor products of $v_6$ cannot generate vectors transforming as $v_2,v_4$,
\begin{align}
& \beta_I(G) 
= G_I \,  a(g_1 + g_2, h^2 - g_1g_2 )
+ G^0{\!}_I \, b(g_1+ g_2 ,h^2 - g_1g_2) \, , \nn \\
G_I = {}& ( 2g_1+3g_2,\, 3g_1+2g_2,\,g_1,\,g_2,\, -h \, ,-h ,\, h) \, , \quad
G^0{\!}_I  = ( 5, \, 5, \, 1,\, 1,\, 0 , \, 0, \, 0) \, ,
\label{G24}
\end{align}
where $g_1+g_2, h^2 - g_1g_2$ are two $O(2)$  invariants formed from
$I_1, \, v_6{\!}^2$. Due to the restriction on the representations of the couplings
the anomalous dimension matrix is diagonal and we must have
\be
\gamma_{ij} = \gamma\big (g_1+g_2, h^2 - g_1 g_2 \big ) \, \delta_{ij} \, .
\ee
 Perturbatively
\begin{align}
a(x,y) = {}& - 2 \, \vep + \tfrac{40}{3} \, x - \big ( \tfrac{5536}{15} + 45\, \pi^2)\,  x^2
+\tfrac{16}{15} \, y \, , \nn \\
b(x,y) = {}&  \tfrac43 ( 3\, x^2 + 4\, y ) - x \big ( 15(12+\pi^2) \, x^2 + 4 (74+9\,\pi^2)\,  y \big ) \, , \nn \\
\gamma(x,y) ={}&  \tfrac{1}{45} \big ( 7 \, x^2 + 8 \, y \big ) \, .
\label{ab24}
\end{align}
For $a(x,y)=b(x,y)=0$ we must take $y=-\tfrac34\, x^2 + {\rm O}(x^3)$ and
at the fixed point then to ${\rm O}(\vep^2)$
\be
x_* = \tfrac{3}{20}\, \vep + \tfrac{9\times 1387}{20000} \, \vep^2
+ \tfrac{3^5}{3200} \,\pi^2 \, \vep^2 \, ,
\qquad y_*  =- \tfrac{27}{1600} \, \vep^2 \, .
\ee
However for real couplings $h,g_1,g_2$ we must have $-y \le \tfrac14 \, x^2$ so this solution
is not possible for this case

A similar reduction holds if $v_2=v_6=0$. In this case
\begin{align}
 \beta_I(G)
= {}&  G_I \,  a( \lambda_1 + 3 g_1, y )
+ G^0{\!}_I \, b(\lambda_1+ 3 g_1 , y) \, ,  \quad y =  h_1{\!}^2 - (\lambda_1- g_1)g_1 \, , \nn \\
G_I = {}& ( \lambda_1 ,\, \lambda_1 ,\,g_1,\,g_1,\,h_1 \, ,-h_1 ,\, 0) \, , \qquad
G^0{\!}_I  = ( 5, \, 5, \, 1,\, 1, \,  0 , \, 0, \, 0) \, ,
\label{G26}
\end{align}
with $\lambda_1 + 3g_1, \, y$ invariants formed from $I_1, \, v_4{\!}^2$.
Perturbative results give
\begin{align}
a(x,y) = {}& - 2 \, \vep + \tfrac{20}{3} \, x - \tfrac{1}{30} \big ( 2248  + 225\, \pi^2) \, x^2
- \tfrac{1}{5} \big ( 848 + 75\, \pi^2 \big ) \, y \, , \nn \\
b(x,y) = {}&  4 \, y  -  2 ( 28 + 3\, \pi^2)  \, x\, y \, .
\label{ab26}
\end{align}
The fixed point is realised to this order at
\be
y_* = 0 \, , \qquad x_* = \tfrac{3}{10} \, \vep + \tfrac{9}{20000}(2248 +225 \, \pi^2 ) \vep^2 \, .
\ee
The constraint $y=0$ can be realised by taking $h_1=g_1=0$ which corresponds to
two decoupled theories or by taking $h_1=0, \, \lambda_1 = g_1$, so that
$ \lambda_1+ 3 g_1 = 4 g_1$,
which gives an equivalent theory.
The anomalous dimension is again necessarily diagonal
\be
\gamma_{ij} = \gamma\big (\lambda_1+ 3 g_1, h_1{\!}^2  - (\lambda_1 -g_1) g_1 \big )
\, \delta_{ij} \, ,
\ee
and from \eqref{gfour} to lowest order
\be
\gamma(x,y) = \tfrac{1}{90} \big ( x^2 + 6 \, y \big ) \, .
\label{ga2}
\ee

For both \eqref{G24} and \eqref{G26} there is a fixed point when $G_I \propto G^0{\!}_I$
giving the $O(2)$ invariant theory. In the first case it is necessary to take $g_1 = g_2, \, h=0$,
so that in \eqref{ab24} $x=2g_1, \, y= - g_1{\!}^2$, and in the second case $h_1=0 , \, \lambda_1
= 5 g_1$, and in \eqref{ab26} $ x= 8g_1, \, y = -  4 g_1{\!}^2$. The results can be reduced to
$\beta_\lambda$ for $\lambda_1 = 5 g_1 = 15 \lambda$ and agrees with \eqref{bet3N1}
and \eqref{bet3N} for $N=2$ and \eqref{ab24}, \eqref{ga2} both give
$\gamma = 4 \, \lambda^2$ in agreement with \eqref{gamma4} when $N=2$.

In a similar fashion to  four dimensions  we may consider perturbations by operators
$d_{ijklmn} \phi^i \phi^j \phi^k \phi^l \phi^m \phi^n$ and  $\phi^2 d_{ijkl} \phi^i \phi^j \phi^k \phi^l$
with $d_{ijklmn} , \,  d_{ijkl} $ symmetric and traceless. For the former to lowest order at
the $O(N)$ symmetric fixed point $\kappa = \Delta -d =  - \tfrac{2(3N+2)}{3N+22} \, \vep$,
so this is always relevant. For the latter $\kappa =  - \tfrac{2(N-14)}{3N+22} \, \vep$ so this
is also relevant when $N>14$. As an example we consider a breaking
$O(N)\to {\cal S}_N  \ltimes {\mathbb Z}_2{\!}^N$,
where, for a closed RG flow,  it is necessary to introduce two additional couplings
\be
V(\phi) = \tfrac{1}{48} \, \lambda \, (\phi^2)^3 +   \tfrac{1}{48} \, g \, \phi^2 \,  {\ts \sum_i}\, \phi_i{\!}^4
+   \tfrac{1}{6!} \, h \, {\ts \sum_i}\,  \phi_i{\!}^6   \, .
\label{VS6}
\ee
The $\beta$-functions may be read off from \eqref{Bone},
\begin{align}
\beta_\lambda ={}&  - 2\vep \, \lambda + 4\big ( (3N+22)\,  \lambda^2
+ 12\,  \lambda \, g + g^2 \big ) \, , \nn \\
\beta_g ={}&  - 2\vep \, g  + 4\big ( 2(N+18) \, \lambda \, g + 2 \, \lambda \, h
+ 13 \, g^2 + \tfrac23 \, g \, h \big ) \, , \nn \\
\beta_h = {}& -2 \vep \, h + 20 \big ( 4 \,  \lambda \, h
+ (N+32) \, g^2 + 8 \, g \, h + \tfrac13  \, h^2 \big ) \, .
\label{beta3d}
\end{align}
As a consequence of the ${\cal S}_N  \ltimes {\mathbb Z}_2{\!}^N$ symmetry the anomalous
dimension matrix is proportional to $\delta_{ij}$ and \eqref{gamma4} is extended to
\be
 \gamma^{(4)} = \tfrac16 (N+2)(N+4) \lambda^2 + (N+4)\, \lambda \, g + \tfrac16(N+14) \, g^2
 + \tfrac13 ( \lambda + g ) h + \tfrac{1}{90} \, h^2 \, .
\label{gfourgh}
\ee

For this theory there is of course a fixed point with $O(N)$ symmetry
for $g_*=h_*=0$ and where $\lambda_*$ is given by \eqref{lambdaon}.
For this the stability matrix from \eqref{beta3d} becomes
\be
M = - 2\vep \, {\mathds 1} + 8 \lambda_*  \begin{pmatrix} 3N + 22 & 0 & 0 \\ 6 & N+18 & 0 \\
0 & 1 &10 \end{pmatrix} \, .
\label{stabO}
\ee
The eigenvalues are $2\vep , \, - \tfrac{N-14}{3N+22} \, 2 \vep , \, - \tfrac{3N+2}{3N+22}\, 2\vep$
with left eigenvectors $\big ( 1 \ 0 \ 0 \big )$,
$\big  ( {- \tfrac{3}{N+2}} \ 1 \ 0 \big )$,  $\big ( \tfrac{2}{(N+8)(N+2)} \  {- \tfrac{1}{N+8}} \ 1 \big ) $
respectively.
In consequence there is always a relevant operator and the $O(N)$  fixed point is
unstable. When $N=2$ the three terms in \eqref{VS6} are not independent, in comparison
with \eqref{Vphi6}
$ \lambda_1 = \lambda_2 = 15(\lambda + g ) + h , \, g_1=g_2 = 3 \lambda + g$
with the other couplings, which break the reflection symmetry under
$\phi_i \to - \phi_i$, zero.
For this case \eqref{gfourgh} gives
$\gamma^{(4)} =\tfrac{1}{90}\, \lambda_1{\!}^2 + \tfrac16 \, g_1{\!}^2$.

There is a trivial decoupled fixed point for $\lambda_* = g_*=0, \, h_* =
\tfrac{3}{10} \, \vep$. In this case
\be
M = - 2\vep \, {\mathds 1} + 8 h_*
\begin{pmatrix} 0 & 1 & 10 \\ 0 &  \tfrac13 & 20 \\ 0 & 0 & \tfrac53
\end{pmatrix} \, ,
\label{stabD}
\ee
with eigenvalues $  2\, \vep, \, - \tfrac65 \, \vep, \, -2\, \vep$ with
eigenvectors $(0,\,0,\,1), \, (0, \, 1, \, -15), \,  ( 1, - 3, \, 30)$.

Other fixed points for non zero $\lambda,g,h$ can be found if we define
$x= g/\lambda, \, y= h/\lambda$ and then from \eqref{beta3d} we must
require
\begin{align}
F(x,y) \equiv {}& (x^2 - x + N -14) x - \tfrac23(x+3) y = 0 \, , \nn \\
G(x,y) \equiv {}& ( x^2 - 28 x + 3 N +  2 ) y - 5 (N+32) x^2 - \tfrac 53 y^2 = 0 \, .
\end{align}
For $x=-3$ then the first equation is satisfied for $N=2$ and then the second equation
determines $y=30, \, \tfrac{153}{3}$. These give $\lambda_* = \tfrac12 \vep, \, g_* = -
\tfrac32 \vep, \, h_* = 15\vep, \, \tfrac{153}{5} \vep$ implying in \eqref{Vphi6} with
$ h=h_1=h_2=0$ fixed points for $\lambda_1, g_1 = 0 ,0 $ and $ 0, \tfrac{3}{10} \vep$
respectively. For $x\ne - 3$ then $F(x,y)=0$ may be solved for $y$ and $G(x,y)$ determines
a quintic polynomial whose zeros determine the non zero $x$ giving rise to fixed points.
For $N=2$ this gives $x=-\tfrac83$ and hence $y= \tfrac{80}{3}$ and then
$\lambda_*= \tfrac{9}{56} \vep, \, g_* = - \tfrac37 \vep, \, h_* = \tfrac{30}{7} \vep$.
This corresponds to taking $\lambda_1 = 5 g_1 = \tfrac{15}{56} \vep$ which is the
$O(2)$ fixed point. For large $N$ the quintic has three real roots with $x=\pm \tfrac13 N^{\frac12}$
and $x =-18$, correspondingly $ y = \tfrac53 N$ and $y= \tfrac95 N$, giving fixed points
\begin{align}
\lambda_* ={}&  \tfrac{9}{56N} \vep, \, &  g_* = {}& \pm \tfrac{3}{56\sqrt N} \vep, \, & h_* = {}&\tfrac{15}{56} \vep \, , \nn \\
\lambda'{\!}_* = {}& \tfrac{1}{6N} \vep, \, &  g'{\!}_* ={}&  -\tfrac{3}{N} \vep, \, &  h'{\!}_* = {}& \tfrac{3}{10} \vep \, .
\label{fixlgh}
\end{align}
 The stability matrices at these fixed points, to leading order in $N$, are
given by
\be
\frac{\vep}{7} \begin{pmatrix} 13 & \pm 3 \sqrt N &  0 \\
\pm 3\big  /\sqrt N & 0 & \pm  15 \sqrt N \\ 0 & \pm 1 \big /\sqrt N & 11
\end{pmatrix} \, , \qquad
2 \vep \begin{pmatrix} 1& - \tfrac{54}{5} & 12 \\
0 & \tfrac{1}{15} & -36 \\ 0 & 0 & 1 \end{pmatrix} \, ,
\label{stabN}
\ee
with eigenvalues $2\vep , \, \tfrac17( 5 \pm \sqrt{46})\vep $ and
$2\vep , 2\vep, \tfrac{2}{15}\vep$,
which is therefore stable, respectively.
 However for $2<N < 14035$ there is just one real root, corresponding to $x \approx \tfrac13 N^{\frac12},
\, y \approx \tfrac53 N$ as $N$ becomes large, and in this case there is no stable fixed point.

For the RG flow there is a corresponding $a$-function (results for theories with fermions are
contained in \cite{Jack3d,Jack3d2}) given by
\begin{align}
A = {} N \big ( & 5 (N+2)(N+4) ( \lambda \, \beta_\lambda - \vep \, \lambda^2)
+ 5 (N+14) ( g \, \beta_g - \vep \, g^2 )
+ \tfrac13 ( h\, \beta_h - \vep \, h^2 ) \nn \\
&{} + 15 (N+4) ( \lambda \, \beta_g + g \, \beta_\lambda - 2 \, \vep
\lambda \, g ) \nn \\
&{} + 5 (  \lambda \, \beta_h + h \, \beta_\lambda
+ g \, \beta_h + h \, \beta_g - 2 \, \vep \, \lambda h
- 2  \, \vep \, gh ) \big ) \, ,
\end{align}
with $\beta_\lambda, \, \beta_g, \, \beta_h$ given by \eqref{beta3d}.
This satisfies
\be
\begin{pmatrix} \pr_\lambda \\ \pr_g \\ \pr_h \end{pmatrix} \, A
= N \begin{pmatrix} 15 (N+2)(N+4) & 45 (N+4) & 15 \\
45(N+4) & 15 (N+ 14) & 15 \\ 15 & 15 & 1 \end{pmatrix}
\begin{pmatrix} \beta_\lambda \\ \beta_g \\ \beta_h \end{pmatrix}  \, ,
\label{Athree}
\ee
and so $A$ decreases under RG flow towards IR fixed points as
the effective metric on the couplings $\lambda,g,h$ is
positive definite for $N>2$, for $N=2$ the metric has a zero eigenvalue
reflecting the redundancy of the couplings $\lambda,g,h$. From \eqref{Athree}
and \eqref{gfourgh} the lowest order results satisfy
\be
A \big |_{\beta_\lambda = \beta_g = \beta_h= 0} = - \tfrac56 \, N \,
\gamma^{(4)} \vep \, .
\ee

For the
$O(N)$ and decoupled fixed points corresponding to \eqref{stabO} and
\eqref{stabD}
\be
A_{O(N)} = - \frac{5N(N+2)(N+4)}{4 (3N+22)^2} \, \vep^3 \, , \qquad
A_{\rm decoupled} = - \frac{3N}{100} \, \vep^3 \, .
\ee
In the large $N$ limit corresponding to \eqref{fixlgh}
\be
A_* = - \frac{75 N}{448} \, \vep^3 \, , \qquad  A_*{\!}'=  - \frac{38 N}{225} \, \vep^3 <A_*,
A_{O(N)}=- \frac{5N}{36}\, \vep^3 \, .
\ee
In section \ref{redSymFour} we considered theories in $4-\vep$ dimensions based on a potential
of the form \eqref{Vphi4}.

\section{Multiple Couplings in \texorpdfstring{$\mathbf{4\boldsymbol{-}\boldsymbol{\vep}}$}{4-epsilon} Dimensions}
\label{multCoupFour}

The discussion of possible fixed points becomes tractable by limiting the number
of fields or the number of couplings. In section \ref{redSymFour} we considered theories in $4-\vep$
dimensions based on a potential of the form \eqref{Vphi4} which had two couplings.
There are many examples in the literature in which theories with three couplings
are considered in the $\vep$-expansion \cite{Mukamel2,3coupling1,3coupling2,MudrovV1, MudrovV2}.
For such examples  $V$ can be expressed as a sum of terms involving symmetric traceless
tensors $d_{u,ijkl}$ as well as an $O(N)$ symmetric term then the treatment
in section \ref{redSymFour}
 may be straightforwardly extended with couplings $\{\lambda,g^u\}$,
$u=1,\dots,p$.
 For the theory to be closed under RG flow we modify \eqref{ddrel} to
\be
d_{u,ijmn} \, d_{v,klmn} = \tfrac{1}{N-1} \, a_{uv}
\big ( \tfrac12 N (\delta_{ik} \delta_{jl}   +  \delta_{il} \delta_{jk})   -
\delta_{ij} \delta_{kl}  \big ) + b_{uv}{}^w \, d_{w,ijkl}  \, ,
\label{ddrel3}
\ee
where $[a_{uv}]$ is positive definite and
\be
b_{uvw} = b_{uv}{}^x a_{xw} = b_{(uvw)}  \, .
\label{symb}
\ee
It is convenient to use $a_{uv}$, and its matrix inverse, to lower and raise indices for
the couplings $\{g^u\}$.  Of course \eqref{ddrel3} may be extended as in \eqref{ddrel2},
but this is not pursued here.

The lowest order $\beta$-functions are just a simple modification of \eqref{bgh1}
\begin{align}
\beta_\lambda = {}& - \vep \, \lambda + (N+8) \lambda^2 +     g^2 \, , \qquad g^2 = a_{uv}\, g^u g^v \, ,\nn \\
\beta_g{}^u = {}& - \vep \, g^u + 12 \, \lambda \, g^u + 3 \,  g^v g^w\, b_{vw}{}^u \, ,
\label{bghA}
\end{align}
and associated $a$-function from \eqref{Afour2} becomes
\be
A = N(N+2)\big ( - \tfrac32 \, \vep \, \lambda^2 - \tfrac14\, \vep \, g^2 + (N+8) \lambda^3
+ 3\, \lambda\,  g^2 + \tfrac12 \, b_{uvw}\, g^u g^v g^w \big ) \, .
\label{Afour3}
\ee
At two loops we can easily extend \eqref{bgh2} to the multi-coupling case and
also at three loops \eqref{bgh3} with an appropriate generalisation of \eqref{crel}.
The couplings $(g^1, \dots , g^p) \in {\mathbb R}^p$ define a vector space spanned by
the polynomials $d_{u,ijkl} \phi^i \phi^j \phi^k \phi^l$. If there is a subspace $V\subset {\mathbb R}^p$ such that
for any $g^u,g'{}^v \in V$ then $g^u g'{}^v \,  b_{uv}{}^w \in V$ the RG flow may be restricted
to couplings belonging to $V$.

It is easy to see that there are no fixed points with $\lambda_*=0$ other than the trivial
Gaussian one. Hence it is sufficient to consider $x^u = g^u/\lambda$. Under
RG flow from \eqref{bghA} to lowest order
\be
\frac{\rmd} { \rmd t} \lambda = \lambda \big ( \vep - \lambda \, f_\lambda ( x ) \big ) \, , \qquad
\frac{\rmd} { \rmd t} x^u = \lambda \, f^u(x) \, ,
\label{rgflow}
\ee
where
\be
f_\lambda (x) = N+8 + x^2 \, , \qquad f^u(x) = (N-4 + x^2) x^u - 3 \, b_{vw}{}^u x^v x^w \, .
\label{frfun}
\ee
Higher order contributions in \eqref{rgflow}  involve contributions at $\ell$ loops of the form
$\lambda^{\ell+1} f_{\ell,\lambda}(x)$ and $\lambda^\ell f_\ell{}^u(x)$. Perturbative fixed points are
then obtained iteratively by first solving $f^u(x_*)=0$ giving $\lambda_* = \vep/f_\lambda(x_*)+{\rm O}(\vep^2)$.

 At any fixed point, where the lowest order $\beta$-functions in \eqref{bghA} vanish, there is a similar
 bound to \eqref{Abound}
\begin{align}
A_{*} = {}& - \tfrac12 N(N+2) \, \frac{1 + \tfrac16 \, x_*{\!}^2}{(N+8 + x_*{\!}^2)^2 }\, \vep^3
\ge - \tfrac{1}{48} \, N \, \vep^3  = A_{*{\rm bound}}\, , \nn \\
& \quad x_*{\!} ^u = g_*{\!}^u/\lambda_* \, , \ \
 x_*{\!}^2 = a_{uv}\, x_*{\!}^u x_*{\!}^v \, .
\label{Abound2}
\end{align}
Any  $\{\lambda,g^u\}$ such that $A< - \tfrac{1}{48} \, N \, \vep^3$ cannot then flow to
a fixed point accessible in the $\vep$-expansion.
The bound  in \eqref{Abound2} arises when
\be
x_*{\!}^2  =  N - 4 \, ,
\label{zeroe}
\ee
and so requires $N>4$.
To analyse this further we consider the stability matrix at the fixed point
\be
M = \begin{pmatrix} - \vep + 2(N+8) \lambda_* & 12\, g_*{}^v \\
\noalign{\vskip 2pt}
2 \,  g_{* \hskip 0.5pt u}&
(- \vep + 12 \, \lambda_*)\delta_u{}^v + 6 \, b_{uw}{}^v g_*{\!}^w \end{pmatrix} \, .
\ee
The eigenvalue problem may be simplified by noting that
\be
M \begin{pmatrix} \lambda_* \\ \frac16 \, g_{* \hskip 0.5pt v} \end{pmatrix} =
\vep \begin{pmatrix} \lambda_* \\ \frac16 \, g_{* \hskip 0.5pt u} \end{pmatrix} \, , \qquad
\begin{pmatrix} \lambda_* & g_{*}{\!}^u \end{pmatrix} M = \vep \,
\begin{pmatrix} \lambda_* & g_{*}{\!}^v \end{pmatrix} \, ,
\label{vepv}
\ee
so that $\vep$ is always an eigenvalue for this class of theories, and also
\be
M \begin{pmatrix} - g_*{\!}^w v_w  \\ \lambda_* \, v_v \end{pmatrix} =
\kappa  \begin{pmatrix} - g_*{\!}^w v_w  \\ \lambda_* \, v_u \end{pmatrix}   \, , \qquad
 \begin{pmatrix} - \tfrac16\, u^u g_{* \hskip 0.5pt u} & \lambda_* \, u^u  \end{pmatrix} M
 = \kappa \,  \begin{pmatrix} - \tfrac16\, u^u g_{* \hskip 0.5pt u} & \lambda_* \, u^v  \end{pmatrix} \, ,
\label{muv}
\ee
where $\kappa$ and $v_r, u^r$ are solutions of the reduced eigenvalue problems
\be
- \lambda_* \, \pr_u f^s(x_*) v_v = \kappa \, v_u \, , \qquad
- \lambda_* \, u^u \pr_u f^v(x_*)  = \kappa \, u^v \, ,
\label{fmu}
\ee
for $f^u(x)$ given in \eqref{frfun}
and where $x_*$ is determined by $f^u(x_*)=0$, corresponding to a fixed point where \eqref{bghA} vanish.
The results \eqref{vepv} and \eqref{muv} with \eqref{fmu} generalise \eqref{eigen4}
in the two coupling case. In general there are $p$ eigenvalues $\kappa$ and associated eigenvectors.
To verify \eqref{muv} it is necessary to show that \eqref{fmu} implies
\be
\lambda_* \big ( 4 - N + x_*{\!}^2 \big ) x_*{\!} ^u v_u = \kappa \,  x_*{\!} ^u v_v \, .
\ee
In consequence when \eqref{zeroe}  is satisfied there is a zero eigenvalue of the stability matrix
as expected when two fixed points annihilate.

Assuming $\lambda_* >0$ it is easy to see
that stable fixed points correspond to finding local maxima of
\be
F(x) = \tfrac12(N-4) x^2 + \tfrac14 \, (x^2)^2 - b_{rst}\, x^r x^s x^t \, .
\label{FM}
\ee
For $N<4$ manifestly there is a maximum at $x=0$. In general when $N>4$
and the origin is a minimum and $F(x_*) = \tfrac16(N-4) x_*{\!}^2 - \tfrac{1}{12}( x_*{\!}^2)^2
\le \tfrac{1}{12}(N-4)^2$. The function
$F(x)$ on ${\mathbb R}^p$ can reduced to one on $S^p \simeq {\mathbb R}^p \cup \{ \infty \}$
by modifying $F(x)$ for very large $x^2$ so that as $x\to \infty$ $F(x) \to C$,  a finite constant,
and then $F(\infty)$ becomes the maximum of $F$.
On $S^p$ $F(x)$  is  a smooth bounded function with at least one maximum and one minimum.
We may use baby Morse theory to constrain the stationary points of $F$
(in embryonic form analogous results were found in \cite{Cayley,Maxwell}). Assuming $F$ has $C_t$ stationary points
at finite $x$  with non degenerate Hessians and $t= 0, 1, \dots ,p$ negative eigenvalues ($C_0, \, C_p$ are the number
of minima, maxima) then
\be
C_t - C_{t-1} + \dots +(-1)^t C_0 \ge (-1)^t \, , \ t<p \, , \quad
{\ts \sum_{t=0}^p} (-1)^{t}C_t  +(-1)^p = 1 + (-1)^p \, ,
\label{Morse}
\ee
where we add in the second identity the contribution of the maxima at infinity separately.
A perfect Morse function on $S^p$, saturating the inequalities in \eqref{Morse},  has just
one minimum and one maximum and no other stationary points.
This is realised if there is  just a single minimum for finite $x$ in \eqref{FM} giving $C_0= 1, \, C_p=0$.
If $b_{{\rm max}} = \max ( b_{rst}\, e^r e^s e^t) |_{e^2=1}$ this holds if $N>4$ and $b_{{\rm max}}{\!}^2 < \tfrac49(N-4)$.
For a stable fixed point it is necessary that $C_p \ge 1$. If a new fixed point is generated by varying the parameters
this must correspond to $\Delta C_q = \Delta  C_{q+1}=1$ for some $q$.

As an illustration we may consider a theory \cite{Mukamel1,3coupling1}
with $p=2$ where $\phi_i = ( \vphi_r , \psi_r )$, $r=1,\dots ,n$,
$N=2n$, $\vphi^2 = {\ts \sum_r}  \vphi_r{\!}^2, \psi^2  = \sum_r \psi_r{\!}^2 $, by taking
\begin{align}
d_{1,ijkl} \phi_i \phi_j \phi_k \phi_l = {}& {\sum_r} \big ( \vphi_r{\!}^4 + \psi_r{\!}^4 \big )
- \frac{3}{2(n+1)} \big ( \vphi^2 + \psi^2 \big ) ^2 \, , \nn \\
d_{2,ijkl} \phi_i \phi_j \phi_k \phi_l ={}& {\sum_r} \, 2\, \vphi_r{\!}^2\, \psi_r{\!}^2
- \frac{1}{2(n+1)} \big ( \vphi^2 + \psi^2\big )^2 \, .
\label{dtwo}
 \end{align}
This extends the cubic fixed point theory where both expressions  in \eqref{dtwo} have the symmetry
$ {\cal S}_n  \ltimes D_4{\!}^n$, with the order 8 dihedral group $D_4$ generated by
$(\vphi_r,\psi_r) \to (-\vphi_r,\psi_r) , \, (\psi_r,\vphi_r)$.

By considering $\pi/4$ rotations of $(\vphi_r,\psi_r)$
theories related by transformations of the couplings of the form
\be
\begin{matrix}\begin{pmatrix} g^1 & g^2 \end{pmatrix} \\ \mbox{} \end{matrix}
\ \begin{matrix} \to \\ \mbox{} \end{matrix} \
\begin{matrix} \begin{pmatrix} g^1 & g^2 \end{pmatrix} \\ \mbox{} \end{matrix}
\begin{pmatrix} \frac12 &\  \frac32 \\ \noalign{\vskip 2pt} \frac12  & - \frac12  \end{pmatrix}\, ,
\label{gequiv}
\ee
are equivalent. It is straightforward to obtain $a_{rs}=a_{sr}, \, b_{rs}{}^t= b_{sr}{}^t$ in \eqref{ddrel3}
for this case which are given by\footnote{For this case there is a single $w$-tensor given by
$$w_{ijkl}\phi^i \phi^j \phi'{}^k \phi'{}^l  ={\ts  \sum_r }\big (   \vphi_r{\!}^2 \psi'{\!}_r{}^2 + \psi_r{\!}^2 \vphi'{\!}_r{}^2
- 2\, \vphi_r \vphi'{\!}_r \psi_r \psi'{\!}_r\big ) - \tfrac{1}{2n-1} \big ( (\vphi^2 + \psi^2)(\vphi'{}^2 + \psi'{}^2)- ( \vphi \cdot \vphi' + \psi \cdot \psi')^2 \big )\, ,$$
and \eqref{ddrel2} is valid with  $e_{11}= 0, \, e_{12} = e_{21}=  \frac29, \, e_{22} = - \frac{4}{27}$. Extending \eqref{dwrel} we obtain
$h_1{\!}^1 = \tfrac{1}{2n-1}, \, h_1{\!}^2= 1, \, h_2{}^1 = \tfrac13, \, h_2{}^2 = -\tfrac{4n-5}{3(2n-1)}, \, f_1 = \tfrac{2n-1}{3(n+1)}, \, f_2= \tfrac{2n-1}{9(n+1)}$
and $a'= \tfrac{3(n-1)}{(n+1)(2n-1)}, \, b'{}^1= b'{}^2= 1,\, e' = -  \tfrac{2n-7}{3(2n-1)}$. These satisfy the corresponding version of \eqref{consis}.
}
 \begin{align}
& a_{11} = \frac{2n-1}{2(n+1)^2} \, , \qquad a_{12} = - \frac{1}{2(n+1)^2} \, , \qquad
a_{22} = \frac{2n+1}{6(n+1)^2} \, , \nn \\
& b_{11}{\!}^1 = \frac{n-1}{n+1} \, , \ b_{11}{\!}^2= 0 \, , \ b_{12}{\!}^1 =-\frac{1}{3(n+1)} \, , \
b_{12}{\!}^2  = \frac{n-2}{3(n+1)} \, , \ b_{22}{\!}^1 = \frac19\, , \ b_{22}{\!}^2 = \frac{2(2n-1)}{9(n+1)} \, .
\end{align}
These results satisfy \eqref{symb}.
The functions $f^u(x^1,x^2)$ in \eqref{frfun} are given by
\begin{align}
f^1(x,y)={}& \tfrac{1}{(n+1)^2}\big (\tfrac12(2n-1)x^2 + \tfrac16(2n+1) y^2 - x\, y \big )x \nn \\
&{} + \tfrac{1}{n+1}\big ( 2\,y - 3(n-1)x\big ) x - \tfrac13 \, y^2 + 2(n-2) x \, , \nn \\
\noalign{\vskip 2pt}
f^2(x,y)={}& \tfrac{1}{(n+1)^2}\big (\tfrac12(2n-1)x^2+ \tfrac16(2n+1) y^2 - x\, y \big ) y \nn \\
&{} - \tfrac{1}{n+1}\big ( 2(n-2)x + \tfrac23(2n-1)y \big )y   + 2(n-2) y \, ,
\end{align}
Finding fixed points by solving the two cubics $f^1(x^1,x^2)=f^2(x^1,x^2) =0$ gives seven solutions which, at lowest order
in the $\vep$-expansion, are given by
\begin{align}
(x_*{\!}^1,x_*{\!}^2) = {}& (0,0) \, , \nn \\
 & (2,0)(n+1) \, , && (1,3) (n+1)  \, , \nn \\
 &  ( 2, 0) \tfrac{(n-2)(n+1)}{2n-1} \, , &&
( 1, 3) \tfrac{(n-2)(n+1)}{2n-1}\, ,\nn \\
& ( 1,1)\tfrac32(n+1) \, , && ( 1,1) \tfrac{(n-2)(n+1)}{n-1} \, , \nn \\
-\big [\lambda_*\, \pr_r f^s \big ] ={}& - \left ( \begin{smallmatrix} 1&0\\0&1 \end {smallmatrix}\right ) \tfrac{n-2}{n+4} \,\vep \, , \nn \\
& -   \left ( \begin{smallmatrix} 1& 0\\ 0 & 1\end {smallmatrix}\right) \tfrac13\, \vep \, , &&
-   \left ( \begin{smallmatrix} 1& 0\\ 0 & 1\end {smallmatrix}\right ) \tfrac13\, \vep \, ,\nn \\
&  \big ( \begin{smallmatrix} 1&0\\ -2/(2n-1) &-1 \end {smallmatrix}\big ) \tfrac{n-2}{3n} \vep\, , &&
 \big ( \begin{smallmatrix} -n-1 & 3(n-2)\\ n & n+1 \end {smallmatrix}\big ) \tfrac{n-2}{3n(2n-1)}\, \vep \, , \nn \\
&-  \left ( \begin{smallmatrix} 1& 3\\ 1& - 1\end {smallmatrix}\right ) \tfrac{1}{10}\, \vep \, , &&
\big ( \begin{smallmatrix} 1& 0\\ 0 & 1\end {smallmatrix}\big) \tfrac{n-2}{5n-4}\, \vep \, ,
\nn \\
 (\kappa_1, \, \kappa_2)={}& - ( 1, \, 1 )\tfrac{n-2}{n+4}\, \vep\, , \nn \\
  &{} - (1, \,  1) \tfrac{1}{3}\, \vep \, , && - (1, \,  1) \tfrac{1}{3}\, \vep \, , \nn \\
 &{} (1, \, - 1) \tfrac{n-2}{3n}\, \vep \, , &&  (1, \, - 1) \tfrac{n-2}{3n}\, \vep \, , \nn \\
  &{} (1, \, - 1) \tfrac15\, \vep \, , &&  (1, \,  1) \tfrac{n-2}{5n-4}\, \vep \, ,  \nn \\
  A_* = {}& - \tfrac{n(n+1)}{2(n+4)^2}\, \vep^3  = \tfrac{1}{24}n \big  ({ - 1} + \tfrac{(n-2)^2}{(n+4)^2}\big  ) \vep^3 \, , \nn \\
  &  - \tfrac{n}{27}\, \vep^3\,  && - \tfrac{n}{27}\, \vep^3  \, , \nn \\
  & -\tfrac{(n+1)(2n-1)}{54n} \, \vep^3= \tfrac{1}{24}n\big ( {- 1} + \tfrac{(n-2)^2}{9 n^2}\big ) \vep^3 \, , && -\tfrac{(n+1)(2n-1)}{54n} \, \vep^3 \, , \nn \\
  &  - \tfrac{n}{25} \, \vep^3 && - \tfrac{(n-1)n(2n-1)}{2(5n-4)^2}\,  \vep^3\nn \\
  \noalign{\vskip-4pt}
 &  && {}  = \tfrac{1}{24}n\big ( {- 1} + \tfrac{(n-2)^2}{(5n-4)^2}\big ) \vep^3 \, .
  \label{fseven}
 \end{align}
  There are therefore seven non trivial fixed points such that in \eqref{Morse} $C_0=C_1=3, \, C_2=1$.
 The first fixed point is the usual $O(2n)$ invariant Heisenberg fixed point, the second two
 are equivalent under \eqref{gequiv} and correspond to $N$ decoupled Ising models,
 the next two are also equivalent and represent  the hypercubic fixed point discussed
 in the two coupling case.
 The last two fixed points are not equivalent since each is
 invariant under \eqref{gequiv}. In this case the $d$-terms combine to
 $\sum_r( \vphi_r{\!}^2 + \psi_r{\!}^2)^2 - \frac{2}{n+1} \, (\vphi^2 + \psi^2)^2$ and the symmetry is enhanced to
 $ {\cal S}_n  \ltimes O(2)^n$. These fixed points are identical to the $M\, N$ theory starting from \eqref{dOmn}
 when $m=2$, the results for $A_*$ are identical to \eqref{Amn} with $m=2$. The final fixed  point
 is stable and has the least value of $A_*$. Results for the stability
matrix eigenvalues for each fixed point above are given to three loop
order in \cite{MudrovV1}, of course  for equivalent theories the expressions are
the same.
All the fixed points in this three coupling theory correspond
to ones obtained by RG flow starting from the trivial Gaussian fixed point
 in just single or double coupling theories.

 An extension to models with arbitrarily many couplings which have stable fixed points
 can be achieved by adapting results due to Michel \cite{Michel} to the formalism
 described here (a related discussion is contained in \cite{GrinsteinL}).
 We take $\phi_i = \vphi_{a_1, \dots , a_p,\, r}$, $a_u = 1,2, \dots, m_u$,
 where $r=1,\dots , n$ and  $u=1,\dots , p$ and define
 \begin{align}
 d_{u,ijkl} \phi_i \phi_j \phi_k \phi_l = {}& \sum_{a_{u+1}, \dots , a_p , r}  \hskip - 0.2 cm
 \Big ( \, {\ts \sum_{a_1,\dots ,a_u}} \vphi_{a_1, \dots , a_p, \, r}{\!}^2 \Big )^2
 - \frac{M_u+2}{N+2} \, \big (\vphi^2 \big )^2 \, , \nn \\
  M_u = {}&  {\ts \prod_1^u}\ m_v \, ,  \quad  N= {\ts \prod_1^p} \ m_u \, n \, , \quad
  m_u\ge 2 \ \ \mbox{if} \  \ u > 1 \, .
 \label{du}
 \end{align}
 This is invariant under ${\cal S}_{N/M_u}  \ltimes O(M_u)^{N/M_u} \in O(N)$.
 There are then $p$ couplings $g^u$ which, along with $\lambda$, define a closed
 manifold under RG flow. Each $(m_1,\dots,m_p)$ represents a different theory with
 different symmetries although as will be shown they may flow to common fixed points.
 For $m_1=1, \, m_2=2$ $d_1,d_2$  are equivalent to \eqref{dtwo}, the
 couplings are related to the previous case by $g^1 \to g^1 - g^2, \, g^2 \to g^2$.
 Scalar potentials formed by linear combinations of terms of the form \eqref{du}
 include various theories which have been considered in the literature.
 In \eqref{ddrel3} the non zero contributions are given by
 \be
 a_{uv}=a_{vu} = \frac{2(M_u+2)(N-M_v)}{3(N+2)^2} \, , \quad v\ge u \, ,
 \ee
 and
 \begin{align}
 b_{uv}{}^u = {}& b_{vu}{}^u = \frac{2(N-M_v)}{3(N+2)} \, , \quad
 b_{uv}{}^v = b_{vu}{}^v = \frac{(N-4)(M_u+2)}{9(N+2)} \, , \quad v> u \, , \nn \\
 b_{uu}{}^u = {}& \frac{1}{9(N+2)} \big ( (N-4) (M_u + 2) + 6 (N-M_u) \big )\, .
 \end{align}
 It is easy to verify \eqref{symb}.

 From \eqref{frfun}
 \begin{align}
 f^u(x) ={}& x^u \Big ( N - 4 +x^2 - \tfrac{1}{3(N+2)}  \big ( (N-4) Y^u + 6\, X^u \big ) \Big ) \, , \nn \\
 X^u = {}& ( N - M_u ) x^u + 2\, {\ts \sum_{v>u}} ( N- M_v) x^v \, , \quad
  Y^u =  (  M_u +2 ) x^u + 2\,  {\ts \sum_{v<u}} ( M_v+2) x^v \, ,  \nn \\
  x^2 = {}& \tfrac{2}{3(N+2)^2} \, {\ts \sum_u} (M_u+2) \ x^u X^u \, .
  \label{fuV}
 \end{align}
 Hence the fixed points at lowest order in $\vep$ are given by
 \be
 x^u = 0 \ \ \mbox{or} \ \ N - 4 +x^2 - \tfrac{1}{3(N+2)}  \big ( (N-4) Y^u + 6\, X^u \big ) = 0
 \ \ \mbox{for each} \ \ u \, .
 \ee
 Setting  all $x^u$ to  zero gives the $O(N)$ symmetric Heisenberg fixed point. If
 $x^{u_1}, x^{u_2}, \dots , x^{u_q}$, $ u_1 < u_2 <\dots < u_q$,  $q\le p$, are non
 zero, solutions
 of the fixed point equations can be found by taking
 \be
 x_*{}\!^{u_i} = \frac{1}{B}\, \frac{(N-4)(N+2)}{2 ( M_{u_i}-4)} \, (-1)^i \, , \quad
 i = 1, \dots , q \, ,
 \label{soli}
 \ee
 where $B$ satisfies
 \begin{align}
 B^2 + \big ( \tfrac16(N+2) - 2 S_q \big ) B
 ={}& - \tfrac16 (N-4) \sum_{i=1}^q \frac{M_{u_i} + 2}{M_{u_i}-4} \bigg ( \frac{N-M_{u_i}}{M_{u_i}-4}
 + 2 \sum_{j>i} (-1)^{i+j}  \frac{N-M_{u_j}}{M_{u_j}-4} \bigg )\nn \\
 = {}&\tfrac16 (N+2) \, S_q   - S_q{\!}^2 \, ,  \qquad
 S_q = \sum_{i=1}^q (-1)^i \, \frac{N-M_{u_i}}{M_{u_i}-4} \, .
 \end{align}
 In consequence there are two solutions for $B$
 \be
 B_1 = S_q \, , \quad B_2 = S_q - \tfrac16 (N+2)\, ,
 \label{solB}
 \ee
 giving at the fixed point
 \be
 \lambda_{*1} = \frac{6 \, S_q}{12 \, S_q - N+4} \, \frac{1}{N+2} \, \vep \, , \quad
 \lambda_{*2} =   \frac{6 \, S_q - N -2 }{12 \, S_q - N - 8 } \, \frac{1}{N+2} \, \vep \, .
 \ee
 For stability it is necessary that $\lambda_{*1}$ or $\lambda_{*2}$ are positive,
  thus for the first fixed point, assuming $N>4$,
 \be
 S_q<0 \quad \mbox{or} \quad 12\, S_q > N-4 \, .
 \ee
 With the solution given by \eqref{soli} we have
 \be
 \sum_{i=1}^q \, ( M_{u_i} + 2 ) \, x_*{}\!^{u_i} = \frac{3}{B} (N+2) \begin{cases} S_q - \tfrac16 (N+2) \, , \quad & q \ \mbox{odd} \, , \\
 S_q \, , & q  \ \mbox{even} \, , \end{cases} \ .
 \ee
 Choosing the appropriate solution in \eqref{solB} we then have $\sum_{i=1}^q \, ( M_{u_i} + 2 ) \, x_*{}\!^{u_i} = 3(N+2)$. Given the
 form \eqref{du} this ensures that the $\big (\vphi^2 \big )^2 $ term in $V$ is then cancelled at this fixed point. In consequence the
 fixed point corresponds to $N/M_q$ decoupled theories.

 For the two cases in \eqref{solB}
 \be
 A_{*1} = \bigg ( \Big ( \frac{N-4}{12\, S_q - N +4} \Big )^2 -1 \bigg ) \frac{N}{48} \, \vep^3 \, , \quad
A_{*2} =  \bigg ( \Big ( \frac{N-4}{12\, S_q - N -8 } \Big )^2 -1 \bigg ) \frac{N}{48} \, \vep^3 \, .
 \ee
 In general $M_{u_1}< M_{u_2} < \dots <M_{u_q}<N$ so that if $M_{u_1}>4$ then $S_q<0$.
 For $S_q >0$ it is necessary that $M_{u_1} <4$ requiring $m_1<4$.
 For $q$ even
 \be
 S_q = (N-4) \sum_{i=1}^q (-1)^i \, \frac{1}{M_{u_i}-4} \, .
 \ee
 Besides the Heisenberg fixed point there are for each $q=1,\dots,p$
 $\binom{p}{q}$ choices and hence
 $2(2^p-1)$ fixed points with some non zero $x_*{\!}^i$.
 These results for $p=1$ are identical to \eqref{Nmfix} and \eqref{Amn} for $m_1 \to m$.

 For each fixed point given by \eqref{soli} and \eqref{solB}, $q=1,\dots , p$ we may determine
 the $p$ eigenvalues of the stability matrix by solving \eqref{fmu} with  \eqref{fuV}. The results
  take the simple form for the two cases in \eqref{solB}
 \be
 \kappa_1 =  \frac{N-4}{12 \, S_q - N+4} \, \big ( 1_{p-t} , -1_t \big ) \vep \, , \qquad \kappa_2
 = \frac{N-4}{12 \, S_q - N-8} \, \big ( 1_{p-t-1} , -1_{t+1} \big )\vep \, .
  \ee
  The number of negative modes is given by
  \be
  t= \begin{cases} p- u_1 + u_2 - \dots - u_q \, , \quad & q \ \mbox{odd} \, , \\
  u_2 - u_1 +  \dots + u_q - u_{q-1}  -1 \, , \quad & q \ \mbox{even} \, . \end{cases}
  \ee
  Thus for $q=1$, $t=p- u_1$,  for $q=2$, $t=u_2-u_1 - 1 $
 and  for $q=p$, $t=\lfloor \frac12(p-1)\rfloor$. Assuming $12 S_q > N-4$ then $A_{*1}<A_{*2}$ and $\lambda_{*1}$  corresponds to a stable
 fixed point in this multi coupling theory when $ t=0$. This  is possible only for
 $q=1$ or $2$ when $u_1=p$ or $u_2=u_1 +1$ respectively. For $S_q<0$ a stable fixed point arises for $t+1=p$ and $A_{*2} < A_{*1}$.
  It is then necessary to take $q=1$, $u_1=1$ and $m_1>4$. For $q=1, \, u_1=p$, so that only $x_*{\!}^p$ is non zero,
  \be
  \frac{N-4}{12 \, S_q - N+4} = \frac{(N-4)(4-M_p)}{8N + (N-16)M_p +16} \, ,
  \ee
  so that this positive for $M_p=1,2,3$ which may be realised when $p=2$ for $m_1=1, \, m_2=2,3$.
  If $q=2, \, u_2 = u_1+1$ then
  \be
  \frac{N-4}{12 \, S_q - N+4} = \frac{(4-M_1)(4-M_2)}{16 M_1 - 8M_2 - M_1 M_2-16} \, ,
\ee
and for positivity we may take $M_1<4, \, M_2>4$. For $p=2$ and $m_1=1, \, m_2=2,3$ this is negative
in accord with the result that there should be just one stable fixed point. It is easy to verify that
these results are in accord with \eqref{fseven} when $m_1=1, \, m_2=2$.

\section{Conclusion}

 Finding fixed points in the $\vep$-expansion is hardly {\it terra incognita}, nevertheless there is  as yet no
 {\it mappa mundi}. At one level this amounts to determining quartic, or cubic or sextic, polynomials in $\N$ variables
 invariant  under all possible subgroups of $O(\N)$.  Nevertheless   understanding RG flows between different fixed points
 depends to a large extent on constructing an $a$-function which decreases monotonically under RG flow.
 Here we considered just the lowest order expression although this can be extended, at least perturbatively,
 to higher orders. The discussions here suggest that fixed points are stable against  RG flow to other
 fixed points near a bifurcation point. However the RG flow may also lead to $a$
 decreasing indefinitely, leading in
 unitary theories to a trivial IR limit such as when all fields are massive.

\ack{Computations in this paper have been performed with the help of
\href{http://www.nikhef.nl/~form}{\texttt{FORM}} and also
\emph{Mathematica} with the package
\href{http://www.xact.es}{\texttt{xAct}}.}

We are grateful to Aninda Sinha, Bertrand Delamotte, Matteo Bertolini,  Vladimir Bashmakov  and Mikhail Kompaniets
for very helpful emails.

\begin{appendices}

\section{Bifurcations of Fixed Points in
\texorpdfstring{$\mathbf{6\boldsymbol{-}\boldsymbol{\vep}}$}{6-epsilon} Dimensions}
\label{bifSix}

The discussion of fixed points for the $O(N)$ theory may be extended to higher orders
 using results for the two and three loop $\beta$-functions. To three loop order the fixed
 points as functions of $N$ are determined by solving
\be
 f(x) + g^2 f_2(x) + g^4 f_3(x)  = 0 \, , \qquad  g^2  f_g(x) + g^4 f_{g2}(x)   = \tfrac12 \, \vep\, ,
 \ee
 with $f_g(x),\, f(x)$ defined in \eqref{fgf}, \eqref{f6d} and from the two and three loop $\beta$-functions
  \begin{align}
 & f_2(x) = \tfrac{1}{108}\big ( (-26 x^3 +78 x^2 +220 x +18)N + (97 x^4 -6x^3 -157 x^2 -90x -134)x \big ) \, , \nn \\
 &  f_{g2}(x)   = - \tfrac{1}{432} \big ( ( 11x^2  - 132 x +86)N - 13 x^4 + 24 x^3 + 628 x^2 + 360 x + 536 \big )\, , \nn \\
 & f_3(x) = \tfrac{1}{31104}\big ( (-1185 x^3 + 16974 x^2 - 59542 x + 41544) N^2 \nn \\
 \noalign{\vskip -2pt}
 & \hskip 2cm + (-9884 x^5 - 10752 x^4 + 218830 x^3 - 24408 x^2 + 109456 x + 51336) N \nn \\
 \noalign{\vskip -2pt}
 & \hskip 2cm + (52225 x^6 - 16806 x^5 - 4980 x^4 - 48888x^3 - 179240 x^2 - 9000x -125680) x \big ) \nn \\
 & \hskip 1.2cm + \tfrac{1}{12}\big ((-12x^3 + 23 x^2 + 66x -22)N
 + 7x^6 + 6 x^5 - 12 x^4 -24x^3+ 8 x^2 -36 x -4 \big ) x \, \zeta_3 \, .
 \end{align}
 To this order finding fixed points is equivalent to solving
 \be
 f(x) + \vep \, F_2(x) + \vep^2 \, F_3(x) = 0 \, , \qquad F_2(x) = \frac{f_2(x)}{2 \, f_g(x)} \, , \quad
 F_3(x) = \frac{f_3(x)}{4\, f_g(x)^2} - \frac{f_2(x) \, f_{g2}(x)}{4\, f_g(x)^3}  \, .
 \ee
 For a fixed point the $\vep$-expansion
\be
x_* = x_1 + x_2 \, \vep   + x_3 \, \vep^2 +{\rm O} (\vep^3) \, , \qquad f(x_1)=0 \, ,
\ee
is then given by
\be
x_2 = - \frac{F_2(x_1)}{f'(x_1)} \, ,
\ee
and
\be
x_3 = - \frac{1}{f'(x_1)}  \big ( F_3(x_1) + F_2{\!}'(x_1) \, x_2 +  \tfrac12\, f''(x_1)\, x_2{\!}^2 \big )
\, .
\ee

The $\vep$-expansion breaks down when two fixed points collide since then $f'(x_1)\to 0$.
If $x_{*\pm}$ are fixed points
which coincide when $N\to N_{\rm crit}$ then near the bifurcation point
$x_{*+}+ x_{*-}$ is regular but for the difference  the $\vep$-expansion may
be modified to the form
\be
x_{*+}- x_{*-} \approx \frac{x_{1+}-x_{1-}}{\sqrt{d(N)}}\, \big ( d(N) + d_1(N) \, \vep + d_2(N) \, \vep^2
\big) ^{\frac12} \, ,
\ee
where
\be
d(N)= \tfrac{1}{324} \big ( 5\, N^3 - 5196 \, N^2 + 4848\, N + 464 \big ) \, ,
\ee
is the discriminant for the cubic $f(x)$, for $d(N)>0$ there are three real roots.
When $x_{1+}\to x_{1-}$  for $N\to N_1$ then
$d(N_1)=0$ and $f'(x_{1+}) \sim - f'(x_{1-}) \to 0$.
By comparing with the $\vep$-expansion
\be
d_1(N) = 2\,  d(N)\, \frac{x_{2+}-x_{2-}}{x_{1+}-x_{1-}}  \, , \quad
d_2(N) = 2\,  d(N)\, \frac{x_{3+}-x_{3-}}{x_{1+}-x_{1-}}  + \frac{d_1(N)^2}{4 \, d(N)} \, .
\ee
Crucially $d_1(N), \, d_{2}(N)$ are finite as $N\to N_1$, although the two terms in $d_2(N)$
separately have poles.
Hence
\be
N_{\rm crit} = N_1 + N_2 \, \vep +N_3 \, \vep^2 + {\rm O}(\vep^3) \, , \qquad d(N_1)=0 \, ,
\ee
where
\be
N_2 = - d_1(N_1) /d'(N_1) \, , \quad
N_3 = -\big  ( d_2(N_1) + d_1{\!}'(N_1) \, N_2 + \tfrac12 \, d''(N_1) \, N_2{\!}^2 \big ) /d'(N_1) \, .
\ee

For the three roots of $d(N)=0$ the above formalism gives
\begin{align}
N_{\rm crit}  = {}& 1038.266  -609.840 \, \vep - 364.173 \, \vep^2 \, ,  \nn \\
N_{\rm crit}  = {}& 1.021453 + 0.0325306 \, \vep -  0.0016301 \, \vep^2 \, , \nn \\
N_{\rm crit}  = {}& - 0.0875026 +  0.3472646 \, \vep -  0.8827372 \, \vep^2 \, .
\end{align}
The results agree precisely with \cite{Fei3loop}.  Correspondingly, extending \eqref{bifurc},
\begin{align}
x_{*{\rm crit}} = {}& 8.74532  - 0.125786 \, \vep + 0.93833\, \vep^2 \, , \nn \\
x_{*{\rm crit}} = {}& 1.103394 + 0.0676068 \, \vep - 0.022110\, \vep^2 \, , \nn \\
x_{*{\rm crit}} = {}& -0.24872 + 0.70563 \, \vep^2  -1.60777 \, \vep^2 \, .
\end{align}

\section{Alternative Fixed Points in
  \texorpdfstring{$\mathbf{4\boldsymbol{-}\boldsymbol{\vep}}$}{4-epsilon}
  Dimensions}
\label{altfp}

We here discuss perturbations of the $O(N)$ theory by operators
$\tfrac14 \, \phi^2 d_{ij} \phi^i \phi^j$ with $d_{ij}$ symmetric and traceless.
At the $O(N)$ fixed point this is not relevant but we find new fixed points by
considering a potential
\be
V(\phi)
= \tfrac18 \, \lambda \, (\phi^2)^2 + \tfrac{1}{4} \, g  \, \phi^2 d_{ij} \phi^i \phi^j
+ \tfrac18 \, h \,  ( d_{ij} \phi^i \phi^j)^2 \, ,
\label{VfourNa}
\ee
where $d_{ij}$ satisfies \eqref{drel} to ensure a closed RG flow. The one loop $\beta$-functions are then
\begin{align}
\beta_{\lambda}={}& - \vep \, \lambda + (N+8) \lambda^2 + (N+16)\,g^2 + 4\, (\lambda + h + b\, g)h \, , \nn \\
\beta_{g}={}& - \vep \, g + (N+16)\, g ( \lambda+h) + 8\, b \, g^2  + 6 \, b \,h^2 + 2 \, b \, (\lambda  + b\, g)h \, , \nn \\
\beta_{h}={}& - \vep \, h + (N+4+8\, b^2) h^2 + (N+16)g^2 + 12\,  \lambda h  + 28 \, b\, g h \, .
\end{align}
A simpler form  is obtained by taking  $b=\alpha-1/\alpha$ with $\alpha=\sqrt{\tfrac n m}, \, N=m+n$ and defining
\be
g_1 = \lambda + 2 \alpha \, g + \alpha^2 \, h \, , \quad
g_3 = \lambda - \tfrac{2} {\alpha}\, g + \tfrac{1}{\alpha^2} \, h \, , \quad
g_2 = \lambda + ( \alpha -\tfrac{1}{\alpha})\, g - h \, ,
\ee
and with $\vphi=\delta_+ \phi , \, \psi = \delta_- \phi$, $\phi^2 = \vphi^2 + \psi^2, \, d_{ij}\phi^i \phi^j =
\alpha \, \vphi^2 - 1/\alpha\, \psi^2$, using definitions in \eqref{project},
\be
V(\phi)=  \tfrac18 \, g_1 \, (\vphi^2)^2 + \tfrac18 \, g_3\, (\psi^2)^2 + \tfrac14 \, g_2\, \vphi^2 \psi^2\, ,
\label{Vphid}
\ee
and the lowest order $\beta$-functions become
\begin{align}
\beta_{g_1} ={}& - \vep \, g_1 + (m+8) \, g_1{\!}^2 + n\, g_2{\!}^2 \, , \nn \\
\beta_{g_3} ={}& - \vep \, g_3 + (n+8) \, g_3{\!}^2 + m\, g_2{\!}^2 \, , \nn \\
\beta_{g_2} ={}& - \vep \, g_2 + \big ( (m+2) \, g_1 + (n+2) \, g_3 \big ) g_2   + 4\, g_2{\!}^2 \, .
\end{align}
Positivity of $V$ requires $g_1,\, g_3>0$ and $ g_1  g_3 > g_2{\!}^2$.
The theory described by \eqref{Vphid} was considered previously
in~\cite{Nelson:1974xnq, Kosterlitz:1976zza, Calabrese:2002bm} and more
recently in \cite{Bashmakov}. Two and three loop results for anomalous dimensions and
$\beta$-functions are given in an ancillary file.

Under RG flow the $h$ coupling plays the crucial role in obtaining IR fixed points. This term
may be expressed in terms of a symmetric traceless 4-tensor
\be
d_{ijkl} \phi_i\phi_j\phi_k \phi_l = ( d_{ij} \phi^i \phi^j)^2 - \tfrac{4}{N+4}\,b\,   \phi^2 d_{ij} \phi^i \phi^j -  \tfrac{2}{N+2} \, (\phi^2)^2 \, ,
\ee
which becomes relevant  for $N>4$.
Manifestly for $g_2=0$ there are two decoupled theories with $O(m)$ and $O(n)$ symmetry
and, apart from the Gaussian fixed point, there are fixed points with
$(g_{1*},\, g_{3*})= ( \tfrac{1}{m+8}, \, 0)\vep,\, (0,\,\tfrac{1}{n+8})\vep, \, (  \tfrac{1}{m+8}, \,\tfrac{1}{n+8})\vep$.
The eigenvalues of the stability matrix are in each case $(1, \, -1, \,  -\tfrac{6}{m+8})\vep$,
$\big (1, \, -1, \,  -\tfrac{6}{n+8} \big )\vep$ and $\big (1, \, 1, \,  \tfrac{(m+2)(n+2)-36}{(m+8)(n+8)} \big )\vep$. The last case is then stable
for $m=1,n>10,\, m=2,n>7,\,  m=3, n>5,\, m=4,n>4$ and also with $m\leftrightarrow n$. When $(m+2)(n+2) < 36$ $\vphi^2 \psi^2$ is a relevant
operator at the decoupled $O(m) \times O(n)$ fixed point. For these fixed points
the anomalous dimension matrix $(\gamma_\phi)_{ij}$ is no longer proportional to the unit matrix but  has two eigenvalues,
either zero or $\gamma_\vphi, \, \gamma_\psi$ for the $O(m), \,O(n)$ symmetric theories.

For $g_2 \ne 0$  there is the $O(N)$ symmetric fixed point with $g_{1*}=g_{2*}=g_{3*} = \tfrac{1}{N+8}\,\vep$
for which the stability matrix eigenvalues are $( 1 , \, \tfrac{8}{N+8}, \, \tfrac{4-N}{N+8}) \vep$ so that this is unstable for $N>4$.
Finding other fixed points with $g_2$ non zero, realising a coupled theory with $O(m)\times O(n)$ symmetry, is more involved.
For $n=m$ there is a symmetry for $\vphi \leftrightarrow  \psi$ and  $ g_1 \leftrightarrow g_3$. It is easy to see by
requiring $\beta_{g_1} - \beta_{g_3} =0$ that
we must take $g_1=g_3$ or $g_1 + g_3 = \tfrac{1}{m+8} \, \vep$. For the latter there are no real fixed points for $m>1$.
For the former we may recover the $O(N), \, N=2m$, fixed point or $g_{1*}=g_{3*} = \tfrac{m}{2(m^2+8)} \, \vep , \
g_{2*} = \tfrac{4-m}{2(m^2+8)} \, \vep$. The stability matrix eigenvalues for this case are then $\big (1, \, \tfrac{8(m-1)}{m^2+8}, \,
- \tfrac{(m-2)(m-4)}{m^2+8} \big ) \vep$, with left eigenvectors $(1, - \tfrac{m-4}{m}, 1)$, $(1,0,-1)$, $(1,\tfrac{m+2}{m-4},1)$ and
where $m=3$ is stable. The anomalous dimension matrix has a
single eigenvalue, to lowest order $\gamma_\phi = \tfrac{m(m^2-3m +8)}{8(m^2+8)^2} \, \vep^2$.
This fixed point coincides with the decoupled fixed point with $O(m)\times O(m)$ symmetry when $m=4$.
For $m=n$  the stable fixed points are then  the  $O(2m)$ symmetric case  for $m<2$,  the
coupled $O(m)\times O(m)$ fixed point for $2<m<4$ and the decoupled one for $m>4$. When $n=m$ and
$g_1 =g_3$ when  there is thr  symmetry under $\vphi \leftrightarrow \psi$  the symmetry group becomes
${\cal S}_2 \ltimes O(m)^2 $, for $m=1$, $O(1) \simeq {\mathbb Z}_2$, this is the dihedral group $D_4$.
Under the action of this group $(\vphi,\psi)$
form a single irreducible representation and there is just one quadratic invariant instead of two more generally.
For   the fixed points for this case
\begin{align}
& A_{O(2m)} =  \tfrac{m}{24} \big (  - 1 + \tfrac{(m-2)^2}{(m+4)^2} \big ) \vep^3 \, \, , \quad
A_{O(m)\times O(m)_{\rm coupled } } =  \tfrac{m}{24} \big (  - 1 + \tfrac{(m-2)^2(m-4)^2}{(m^2+8)^2} \big ) \vep^3 \, ,  \nn \\
& A_{O(m)\times O(m)_{\rm decoupled } } =  \tfrac{m}{24} \big (  - 1 + \tfrac{(m-4)^2}{(m+8)^2} \big ) \vep^3 \, .
\end{align}

For $g_2\ne 0$ and $m\ne n$ finding fixed points is simplified by letting $x=g_1/ g_2, \, y= g_3/g_2$ and
then it is sufficient  to solve
\be
\big ( (m+2) x - 6\, y +4\big ) y = m \, ,\qquad \big ( (n+2) y - 6\, x +4\big ) x = n\, .
\label{xyeq}
\ee
The couplings at a $g_2$ non zero fixed point to ${\rm O}(\vep)$ are given by
\be
g_{2*} = \vep \big / \big ( 4 + (m+2) x + (n+2)y \big ) \, , \quad
g_{1*}= x \, g_{2*} , \, \quad  g_{3*}= y\, g_{2*} \, .
\ee
The $O(N)$ solution corresponds to $x=y=1$.
Defining
\be
p= (m+2)(n+2)-36 \, ,
\label{defh}
\ee
then for $p$ small  there is a solution
\be
x \approx - \frac{4}{p}(n+8) - \frac{m(n+2)}{4(m+8)}   - \frac{3n}{2(n+8) } \, , \qquad
y \approx - \frac{4}{p}(m+8) -  \frac{n(m+2)}{4(n+8)}   - \frac{3m}{2(m+8)} \, ,
\label{xysol}
\ee
giving
\be
g_{2*} \approx - \frac{p}{4(m+8)(n+8)}\, \vep \,  , \quad g_{1*} = \frac{1}{m+8}\, \vep  + {\rm O}(p) \, , \quad
 g_{3*} = \frac{1}{n+8}\, \vep + {\rm O}(p) \, .
 \ee
As $p\to 0$ this coincides with the decoupled $O(m)\times O(n)$ fixed point. Besides $x=y=1$
\eqref{xyeq} has three solutions, for $m,n>1$ two are complex conjugates, one is real and matches with
\eqref{xysol} as $p\to 0$. For $m \to 0$, $x \to - \tfrac{3n}{n-16} , \, y \to - \tfrac{n-32}{3(n-16)}$ and similarly
for $m\leftrightarrow n , \, x\leftrightarrow y$. For $m,n $ large $x\sim a, \, y\sim 1/a$ where
$ 3 a^3 + a^2 + \tfrac{n}{m}( a+3 )=0$ giving $-\tfrac13 > a >-3$ for $0<\tfrac{n}{m} <\infty$
and $a=-1$ for $n=m$.

At higher orders the critical point where the decoupled and coupled theories coincide is shifted from
$p=0$. Imposing the requirement  that one of the eigenvalues of the stability matrix in the
decoupled  theory vanishes gives
\be
p_{\rm crit} = - 48 \, \vep + 8( 1 + 3 \zeta_3 )\vep^2 + \tfrac{72(7 +6 \zeta_3)}{m+n+16} \, \vep^2 \, ,
\ee
or in the symmetric case $n=m$
\be
 m_{\rm crit}= 4 -4 \, \vep + \tfrac12(1+7\zeta_3) \vep^2 \, .
\ee
The same condition ensures $g_2=0$ in the coupled theory.

This example may be extended further by considering fields $\{ \vphi_{ra} , \, \psi_{ra} \}$,
$r=1, \dots, n$, $a=1,\dots, m$, $N= 2\, m  n$,
\begin{align}
d_{1,ijkl} \phi_i \phi_j \phi_k \phi_l = {}& {\ts {\sum_r}} \big (  {\vphi_r}{\!}^2 + { \psi}_r{\!}^2 \big )^2
- \tfrac{2(m+1)}{N+2} \big ( \vphi^2 +  \psi^2 \big ) ^2 \, , \nn \\
d_{2,ijkl} \phi_i \phi_j \phi_k \phi_l ={}& {\ts{\sum_r}} \, 2\, {\vphi}_r{\!}^2\, {\psi}_r{\!}^2
- \tfrac{m}{N+2} \big ( {\vphi}^2 + {\psi}^2\big )^2 \, ,
\label{dtwoa}
 \end{align}
where $\vphi_r{\!}^2 = \sum_a \vphi_{ra}{\!}^2 , \,  {\psi}_r{\!}^2 = \sum_a {\psi}_{ra}{\!}^2$,
 $\vphi^2 = \sum_r  {\vphi_r}{\!}^2 , \,  {\psi}^2 = \sum_r {\psi}_r{\!}^2$. $d_{1,ijkl}\phi_i \phi_j \phi_k \phi_l $
 is invariant under ${\cal S}_n \ltimes O(2m)^n$ and is zero when $n=1$. For $d_{2,ijkl}\phi_i \phi_j \phi_k \phi_l $
 the symmetry is reduced to  ${\cal S}_{2n} \ltimes O(m)^{2n}$. For $m=1$ the potential is equivalent  to that determined by
 \eqref{dtwo} and for $n=1$ to \eqref{Vphid} for $g_1=g_3$.  When $m=n=1$ the symmetry group becomes
 ${\cal S}_2 \ltimes {\mathbb Z}_2{\!}^2 \simeq D_4$.
 The tensors in \eqref{dtwoa} satisfy \eqref{ddrel2} with
 \begin{align}
 a_{11} = {}&\tfrac{4 (m+1)(N-2m)}{3(N+2)^2} \, , \qquad a_{12}=a_{21} = \tfrac{m(N-2m)}{3(N+2)^2}  \, , \qquad
 a_{22} = \tfrac{2m(N+2-m)}{3(N+2)^2} \, ,  \nn \\
 b_{11}{\!}^1 ={}&  \tfrac{2}{9(N+2)} \big ( (m+4)N - 10 m -4 \big ) \, , \quad  b_{12}{}^1= b_{21}{}^1= \tfrac{(N-4)m}{9(N+2)} \, , \quad
 b_{22}{}^1 = \tfrac19 \, m \, , \nn \\
 b_{11}{\!}^2 = {}& 0 \, , \qquad b_{12}{}^2 = b_{21}{\!}^2 =  \tfrac{2(N-2m)}{3(N+2)} \, , \qquad
b_{22}{}^2 =   \tfrac{(4-m)(N+2) - 12m}{9(N+2)} \, .
 \end{align}
 Fixed points to lowest order in $\vep$
 are determined by the zeros of $f^1(x,y), \, f^2(x,y)$ given by \eqref{frfun} with $x= g^1/ \lambda, \, y= g^2/\lambda$.
 Apart from the trivial Gaussian fixed point there are seven solutions where
 \begin{align}
 (x_*,y_*) = {}& (0,\, 0) \, , \ \ \tfrac{3(N+2)}{2(m+1)} (1,0) \, , \ \tfrac{(N+2)(N-4)}{4m(n-1)}(1,0) \, , \  \ \tfrac{3(N+2)}{m+2} (1,-1) \, , \ \
 \tfrac{(N+2)(N-4)}{2m(2n-1)} (1,-1) \, , \nn \\
 {}& \tfrac{N+2}{2m} \big ( m , - 2(m-2) \big ) \, , \quad \tfrac{3 (N+2)(N-4)}{2(N(m^2 -3m +8) + 2(m^2 - 12 m +8) } \big ( m , - 2(m-2) \big )\, .
 \label{xyN}
 \end{align}
 For $2(m+1)x_* + m y_* = 3(N+2)$, as in cases 2,4 and 6, the fixed points correspond to decoupled theories unless $n=1$.
 The eigenvalues of the stability matrix, apart from $\vep$ in each case, are then determined from \eqref{fmu}
 \begin{align}
 (\kappa_1,\kappa_2) = {}& - \tfrac{N-4}{N+8} ( 1,1) \vep \, , \quad \tfrac{m-2}{m+4}( 1, -1) \vep \, , \quad
 -  \tfrac{(m-2)(N-4)}{(m+4)N- 8(2m-1)}( 1, 1) \vep  \, , \nn \\
{}&   \tfrac{m-4}{m+8}( 1, 1) \vep \, , \qquad \tfrac{(m-4)(N-4)}{(m+8)N- 16(m-1)}( 1, - 1) \vep  \, , \nn \\
{}& - \tfrac{(m-2)(m-4)}{m^2+8}( 1, 1) \vep \, , \qquad  \tfrac{(m-2)(m-4)(N-4)}{(m^2 +8)N + 8(m-1)(m-8)}( 1, - 1)\vep \, .
\label{Kmn}
 \end{align}
 For each case
 \begin{align}
 48 A_* + N \vep^3 = {}& \tfrac{N(N-4)^2}{(N+8)^2}\, \vep^3 \, , \quad  \tfrac{N(m-2)^2}{(m+4)^2}\, \vep^3 \, , \quad
 \tfrac{N(N-4)^2(m-2)^2}{((m+4)N- 8 (2m-1)) ^2}\, \vep^3 \, , \nn \\
 {}&  \tfrac{N(m-4)^2}{(m+8)^2}\, \vep^3 \, ,  \qquad
 \tfrac{N(N-4)^2(m-4)^2}{((m+8)N- 16 (m-1)) ^2}\, \vep^3  \, , \nn \\
{}&  \tfrac{N(m-2)^2(m-4)^2}{(m^2+8)^2}\, \vep^3 \, , \qquad
 \tfrac{N(N-4)^2(m-2)^2(m-4)^2}{((m^2+8)N +8 (m-1)(m-8)) ^2}\,  \vep^3 \, .
 \label{A7mn}
 \end{align}
 Except for the Heisenberg fixed point when $N<4$ the stable fixed points for $m>4$ and $2<m<4$ represent
 decoupled theories. For $m=1$ the results in \eqref{Kmn} and \eqref{A7mn} reduce to \eqref{fseven}.
 In general the results are identical to the discussion corresponding to \eqref{du} for $p=2$ where $M_1 = m, \, M_2=2m$
 and $d_1{}^{\rm here} = d_2, \,  d_2{}^{\rm here} = d_2- d_1$. For $n=1$ the 4th and 6th cases in \eqref{xyN}, \eqref{Kmn},
 \eqref{A7mn}   are identical to the fixed points obtained from \eqref{Vphid} when $g_1 = g_3, \, n=m$ and where the two
 cases correspond to $g_2=0$ and $g_2 \ne 0$ respectively. The 5th and 6th cases correspond to the $MN$ model results
 in \eqref{kmn} and \eqref{Amn}.

\section{Bounds  for Tensor Products}
\label{boundsTP}

The tensorial relation \eqref{ddrel} necessary at lowest order   for a simple two coupling
renormalisable theory in four dimensions introduces two parameters $a,b$.
Here we demonstrate the inequality \eqref{inequal} valid so long as $d_{ijkl}$ are real.
To this end we consider the
decomposition of the rank six tensor $d_{ijkp}\, d_{lmnp}$ into irreducible components.

A symmetric traceless tensor may be defined by
\be
S_{ijklmn} = d_{(ijk|p} \,d_{lmn)p} - \frac{9\, b}{N+8} \, \delta_{(ij} \, d_{klmn)}
- \frac{3\, a}{N+4} \, \delta_{(ij} \, \delta_{kl}\, \delta_{mn)} \, .
\ee
Using \eqref{ddrel2} and \eqref{dwrel} this has the norm
\be
S_{ijklmn} \, S_{ijklmn}  = \frac{1}{40} \, N(N+2)a  \bigg ( \frac{(N+2)(N-2)(N+14)}{(N-1)(N+4)} \, a +
\frac{18(N+2)}{N+8} \, b^2  - 9  \, e^u h_u \bigg )  \, .
\ee
A mixed symmetry tensor corresponding to a $[4,2]$ Young tableaux is given by
\begin{align}
M_{ijklmn} = {}& d_{ij(k|p} \,d_{lmn)p} - d_{i(kl|p} \,d_{mn)jp} \nn \\
\noalign{\vskip -3pt}
&{}+ \frac{b}{N-2} \big (  \delta_{ij} \, d_{klmn} -  \delta_{i(k} \, d_{lmn)j} -  \delta_{j(k} \, d_{lmn)i}
+  \delta_{(kl} \, d_{mn)ij} \big ) \nn \\
\noalign{\vskip -3pt}
&{} + \frac{a}{N-1} \big ( \delta_{ij} \, \delta_{(kl}\, \delta_{mn)} - \delta_{i(k} \, \delta_{lm}\, \delta_{n)j} \big )
- \frac{5}{N+4}  \, e^u w_{u,ij(kl}\, \delta_{mn)}  \, ,
\end{align}
satisfying $M_{ijklmn} = M_{(ij)(klmn)}, \, M_{i(jklmn)}=0$ and traceless on contraction of any
pair of indices. The norm is then
\be
M_{ijklmn} \, M_{ijklmn}  = \frac{5}{48} \, N(N+2)a  \bigg ((N+2)  \frac{N-2}{N-1} \, a -
\frac{2(N+2)}{N-2} \, b^2+ \frac{N-16}{N+4}  \, e^u h_u  \bigg )  \, .
\ee
From \eqref{consis} $ a \, e^u h_u  = e^u e^v a'{\!}_{uv} \ge 0$.
Setting the $w$ tensor contributions to zero
positivity for $N>2$  is equivalent to \eqref{inequal}.
For $N=3$ $M_{ijklmn}=0$ and we must take $e^uh_u=0$ so that $a= 4b^2$ then.
For $N<16$ the $w$-tensor contributions ensure a more stringent bound on
$b^2/a$.

The 8 index tensor formed from two $d_{ijkl}$ may also be decomposed into representations
of $O(N)$. The mixed symmetry $[4,4]$ representation gives the bound
$a\le 15 b^2$ in the absence of $w$-tensor contributions when $N=4$.

As an illustration  of similar constraints we consider a similar discussion for a  three
index symmetric traceless tensor $d_{ijk}$ which may define a single coupling renormalisable theory
in six dimensions if at one loop order it satisfies \cite{Mckane0,Mckane}
\be
d_{ikl} \, d_{jkl} = \alpha \, \delta_{ij} \, , \ \ \alpha>0 \, ,\qquad d_{ilm} \, d_{jmn} \, d_{knl} = \beta \,  d_{ijk} \, .
\label{d1loop}
\ee
In this case analogous symmetric traceless and mixed symmetry tensors are formed by
\begin{align}
S_{ijkl} = {}& d_{(ij|m} \, d_{kl)m} - \tfrac {2\alpha}{N+2} \, \delta_{(ij}\, \delta_{kl)} \, , \nn \\
M_{ijkl} = {}& d_{ij m} \, d_{klm} - d_{(i|km}\, d_{j)lm} + \tfrac{\alpha}{N-1} \big ( \delta_{ij}\, \delta_{kl}
- \delta_{(i|k} \, \delta_{j)l} \big ) \, ,
\end{align}
where $M_{(ijkl)} =0, \, M_{ijkk} = M_{ikjk} =0$. Positivity of $S_{ijkl} \,S_{ijkl} , \ M_{ijkl}\, M_{ijkl}$ gives
\be
- \tfrac{N-2}{2(N+2)} \, \alpha \le \beta \le \tfrac{N-2}{N-1}\, \alpha \, ,
\ee
where the lower and upper bounds require $S_{ijkl}=0$ and $M_{ijkl}=0$ respectively.
At two loop order it is necessary that
\be
d_{ilm} \, d_{jnp} \, d_{krs} \, d_{lnr}\, d_{mps} = \gamma \, d_{ijk}\, ,
\label{d2loop}
\ee
where
\be
S_{ijkl} = 0 \ \ \Rightarrow \ \ \gamma = - \tfrac{N^2-10N-16}{2(N+2)^2}\, \alpha^2 \, , \qquad
M_{ijkl} = 0 \ \ \Rightarrow \ \ \gamma = \tfrac{N^2 - 4N +5}{(N-1)^2}\, \alpha^2 \, .
\ee
Identities similar to \eqref{d2loop} are required at higher order for graphs of new topology, the
coefficients may be calculated for the two limiting cases in a similar fashion \cite{Pang}.
$M_{ijkl}=0$ is realised by taking $d_{ijk} = \sum_\alpha  e_i{\!}^\alpha  e_j{\!}^\alpha  e_k{\!}^\alpha$
when $\alpha= \tfrac{N-1}{N+1}\, , \beta = \tfrac{N-2}{N+1}$.  The case $S_{ijkl}=0$, with additional
assumptions, reduces to the $F_4$ family considered in \cite{birdtracks} where $N=5,8,14,26$.
Other solutions with a single tensor $d_{ijk}$, satisfying \eqref{d1loop}, \eqref{d2loop} and the various
higher order extensions, are provided by the invariant symmetric tensors for $SU(n)$, $N=n^2-1$.

\section{Perturbations at the Hypertetrahedral Fixed Point}
\label{pertsHT}

We here analyse the scaling dimensions for $\phi^4$ operators at the stable fixed point
obtained by taking $d_{ijkl} \phi^i \phi^j \phi^k \phi^l$ to be given by \eqref{tetra2}
and assuming $N>5$. It is convenient to define $\phi^\alpha = e_i{\!}^\alpha \phi_i , \,
\pr^\alpha = e_i{\!}^\alpha \pr_i$, $\alpha=1,\dots,N+1$,  with $e_i{\!}^\alpha$ satisfying \eqref{tetra}
so that $\sum_\alpha \phi^\alpha  = \sum_\alpha \pr^\alpha =0$.
Using \eqref{lgtet} the anomalous dimensions to ${\rm O}(\vep)$  at the stable fixed point
for $N\ge 5$ are determined by the eigenvalues of the operator
\be
\D = \vep \, \tfrac{1}{3(N+3)} \big ( \tfrac12 \, \phi^2 \, \pr^2 + ( \phi \cdot \pr)^2 - \phi\cdot \pr
+ \tfrac12 (N+1) \, {\ts \sum_\alpha}\,  \phi^\alpha \phi^\alpha  \,
\pr^\alpha \pr^\alpha \big ) \, .
\label{DTeig}
\ee

Under the action of the symmetry group $\S_{N+1}$ $\phi^4$ operators can be decomposed
into tensors $V^{\alpha\beta\gamma\delta}, \, V^{\alpha\beta\gamma}, \,
V^{\alpha\beta}$, $\alpha\ne \beta\ne \gamma \ne \delta$, where
$\sum_\alpha  V^{\alpha\beta\gamma\delta} = \sum_\alpha  V^{\alpha\beta\gamma}
= \sum_\alpha  V^{\alpha\beta} =0$ for each index, forming irreducible representations under permutations
of the indices.\cite{Vasseur} The possible representations are described by Young tableaux with $Q=N+1$ boxes labelled
by $[n_1,n_2,\dots,n_r]$, $n_1\ge n_2\ge \dots \ge n_r>0$.
Besides $\phi^2 = \sum_\alpha \phi^\alpha \phi^\alpha$ two $\S_{N+1}$ singlets are relevant
\be
\phi^3 = {\ts \sum_\alpha} \, \phi^\alpha \phi^\alpha \phi^\alpha \, , \qquad
\phi^4 = {\ts \sum_\alpha} \, \phi^\alpha \phi^\alpha \phi^\alpha \phi^\alpha\, .
\ee
We first define for the $[Q-4,4]$ tableaux
\begin{align}
V^{\alpha\beta\gamma\delta} ={}&  \phi^\alpha \phi^\beta \phi^\gamma \phi^\delta
+ \tfrac{1}{N-5} \, \P_{12} \,  \phi^\alpha \phi^\alpha \phi^\beta \phi^\gamma  \nn \\
&{} + \tfrac{1}{(N-4)(N-5)} \big ( 2\, \P_6 \, \phi^\alpha \phi^\alpha \phi^\beta \phi^\beta
+ 2\, \P_{12} \, \phi^\alpha \phi^\alpha \phi^\alpha \phi^\beta
- \phi^2 \, \P_6 \, \phi^\alpha \phi^\beta \big )  \nn \\
&{}
+ \tfrac{1}{(N-3)(N-4)(N-5)} \big (6\,  \P_4 \, \phi^\alpha \phi^\alpha \phi^\alpha \phi^\alpha
- 3 \,  \phi^2 \, \P_4 \, \phi^\alpha \phi^\alpha
- 2 \, \phi^3 \, \P_4 \, \phi^\alpha \big )  \nn \\
&{}+ \tfrac{3}{(N-2)(N-3)(N-4)(N-5)} \big ( \phi^2 \phi^2 - 2\, \phi^4 \big ) \, ,
\quad \dim \ \tfrac{1}{24} (N+1)N(N-1)(N-6) \, ,
\end{align}
where $\P_n$ denotes a sum over the $n$ permutations of $\alpha,\beta, \gamma,\delta$
necessary for overall symmetry of the relevant term. This satisfies
\be
\D \, V^{\alpha\beta\gamma\delta}  =  \vep \, \tfrac{4}{N+3} \,
V^{\alpha\beta\gamma\delta} \, .
\ee
For the three index case there is a symmetric three index tensor corresponding to $[Q-3,3]$
\begin{align}
V_S^{\alpha\beta\gamma} ={}& \P_{3} \,  \phi^\alpha \phi^\alpha \phi^\beta \phi^\gamma  \nn \\
&{} + \tfrac{1}{N-3} \big ( 2\, \P_3 \, \phi^\alpha \phi^\alpha \phi^\beta \phi^\beta
+ 2\, \P_{6} \, \phi^\alpha \phi^\alpha \phi^\alpha \phi^\beta
- \phi^2 \,  \P_3 \,\phi^\alpha \phi^\beta \big )  \nn \\
&{}+ \tfrac{1}{(N-2)(N-3)} \big ( 6 \,  \P_3 \, \phi^\alpha \phi^\alpha \phi^\alpha \phi^\alpha
-  3 \,  \phi^2 \, \P_3 \, \phi^\alpha \phi^\alpha  - 2 \, \phi^3 \, \P_3 \, \phi^\alpha \big )  \nn \\
&{}+ \tfrac{3}{(N-1)(N-2)(N-3)} \big ( (\phi^2)^2 - 2\, \phi^4 \big ) \, ,
\qquad \dim \ \tfrac{1}{6} (N+1)N(N-4) \, ,
\end{align}
where $\P_n$ now symmetrises over $\alpha,\beta,\gamma$, and
\be
\D \, V_S^{\alpha\beta\gamma}  =  \vep \, \tfrac{N+7}{3(N+3)} \,
V_S^{\alpha\beta\gamma} \, .
\ee
Note $(N-5) V^{\alpha\beta\gamma\delta}|_{N\to 5} = ( V_S^{\alpha\beta\gamma} + V_S^{\beta\gamma\delta} +
V_S^{\alpha\beta\delta}  + V_S^{\alpha\gamma\delta} )|_{N=5}$.
In addition there are mixed symmetry tensors $\big (V_M^{\alpha\beta\gamma} ,V_M^{\alpha\gamma\beta} \big )$,
corresponding to $[Q-3,2,1]$
determined by
\begin{align}
V_M^{\alpha\beta\gamma} ={}& \big ( \phi^\alpha \phi^\alpha \phi^\beta -
 \phi^\alpha \phi^\beta \phi^\beta \big ) \phi^\gamma  \nn \\
&{} + \tfrac{1}{N} \big ( \big ( \phi^\alpha \phi^\alpha - \phi^\beta \phi^\beta \big )  \phi^\gamma \phi^\gamma
+ \phi^2 \big ( \phi^\alpha - \phi^\beta \big ) \phi^\gamma  \big )
+ \tfrac{1}{N-2} \big (  \phi^\alpha \phi^\alpha \phi^\alpha \phi^\beta -  \phi^\alpha \phi^\beta \phi^\beta \phi^\beta \big )
\nn \\
&{}+ \tfrac{1}{N(N-2)} \big ( \big (  \phi^\alpha \phi^\alpha \phi^\alpha - \phi^\beta \phi^\beta \phi^\beta \big ) \phi^\gamma
-  (N-1)\big ( \phi^\alpha -\phi^\beta \big )\phi^\gamma  \phi^\gamma \phi^\gamma
+ \phi^3\big ( \phi^\alpha - \phi^\beta \big )  \big ) \, ,
\end{align}
with an overall dimension  $\tfrac{1}{3} (N+1)(N-1)(N-3)$ and
\be
\D \, V_M^{\alpha\beta\gamma}  =  \vep \, \tfrac{N+13}{3(N+3)} \,
V_M^{\alpha\beta\gamma} \, .
\ee
In the symmetric  two index case for tableaux $[Q-2,2]$
\begin{align}
V_1^{\alpha\beta}={}&  \phi^\alpha \phi^\alpha \phi^\alpha \phi^\beta +
\phi^\alpha \phi^\beta \phi^\beta  \phi^\beta \nn \\
&{} +  \tfrac{1}{N-1} \big ( 2\big ( \phi^\alpha \phi^\alpha \phi^\alpha \phi^\alpha
+ \phi^\beta \phi^\beta \phi^\beta  \phi^\beta \big ) - \phi^3 \big ( \phi^\alpha +\phi^ \beta \big ) \big )
- \tfrac{2}{N(N-1)} \, \phi^4 \, , \nn \\
V_2^{\alpha\beta}={}&  \phi^\alpha \phi^\alpha \phi^\beta  \phi^\beta    \nn \\
&{} +  \tfrac{1}{N-1} \big (  \phi^\alpha \phi^\alpha \phi^\alpha \phi^\alpha
+ \phi^\beta \phi^\beta \phi^\beta  \phi^\beta - \phi^2 \big ( \phi^\alpha \phi^\alpha +\phi^ \beta \phi^\beta \big ) \big )
+ \tfrac{1}{N(N-1)} \big ( (\phi^2)^2 - \phi^4 \big )  \, , \nn \\
V_3^{\alpha\beta}={}&  \phi^2  \phi^\alpha \phi^\beta
 +  \tfrac{1}{N-1} \, \phi^2 \big ( \phi^\alpha \phi^\alpha +\phi^ \beta \phi^\beta \big )
- \tfrac{1}{N(N-1)} \, (\phi^2)^2  \, ,
\end{align}
where each have dimension $\tfrac12(N+1)(N-2)$, and for $[Q-2,1,1]$ there is an antisymmetric two index tensor
\be
V_A^{\alpha\beta}=  \phi^\alpha \phi^\alpha \phi^\alpha \phi^\beta -
\phi^\alpha \phi^\beta \phi^\beta  \phi^\beta
+ \tfrac{1}{N+1}\, \phi^3 \big ( \phi^\alpha - \phi^ \beta \big )   \, , \quad \dim \ \tfrac12N(N-1)\, .
\ee
For these
\begin{align}
\D \begin{pmatrix} V_1^{\alpha\beta} \\ V_2^{\alpha\beta} \\ V_3^{\alpha\beta} \end{pmatrix}
= {}&   \vep \, \frac{1}{3(N+3)}\begin{pmatrix} 3(N+3) & - 6 & 6 \\ - 4 & 2(N+5) & 0 \\ 2(N+1) &0 & 2(N+8)
\end{pmatrix}
 \begin{pmatrix} V_1^{\alpha\beta} \\ V_2^{\alpha\beta} \\ V_3^{\alpha\beta} \end{pmatrix} \, , \nn \\
 \D \, V_A^{\alpha\beta}= {}& \vep \, V_A^{\alpha\beta} \, .
\end{align}
In addition for $[Q-1,1]$ there are three single index vectors
\begin{align}
U_1^\alpha ={}&   \phi^\alpha \phi^\alpha \phi^\alpha \phi^\alpha - \tfrac{1}{N+1} \, \phi^4 \, , \quad
U_2^\alpha = \phi^2 \, \phi^\alpha \phi^\alpha  - \tfrac{1}{N+1} \, (\phi^2)^2 \, ,
\qquad U_3^\alpha =  \phi^3 \,  \phi^\alpha \, , \nn  \\
 \D & \begin{pmatrix}  U_1^\alpha \\
U_2^\alpha  \\ U_3^\alpha \end{pmatrix} =  \vep \, \frac{1}{3(N+3)}
 \begin{pmatrix} 6(N+1) & 6 & 0\\ 4(N+1) & 3(N+5) &-4 \\ 3(N+1) & 0 & 3(N+3) \end{pmatrix}
 \begin{pmatrix}  U_1^\alpha \\ U_2^\alpha  \\ U_3^\alpha \end{pmatrix} \, ,
 \end{align}
where $U_1^\alpha, \, U_2^\alpha\, , U_3^\alpha$ each have dimension $N$, and there are also
two singlets corresponding to $[Q]$
 \be
 \D \begin{pmatrix} \phi^4 \\ (\phi^2)^2 \end{pmatrix} =  \vep \, \frac{1}{3(N+3)}\begin{pmatrix} 6(N+1) & 6 \\ 4(N+1) & 4(N+4) \end{pmatrix}
\begin{pmatrix} \phi^4 \\ (\phi^2)^2 \end{pmatrix} \, .
\ee
The dimensions of the representations listed above add up to $\tfrac{1}{24}N(N+1)(N+2)(N+3)$ as expected.

For each perturbation then for $\kappa = \Delta -d$ to lowest order $\kappa/\vep$ is given for each case in turn by
$ - \tfrac{N-1}{N+3}$, $ - \tfrac{2(N+1)}{3(N+3)}$,
$- \tfrac{2(N-2)}{3(N+3)}, \ -\tfrac{N+5}{3(N+3)}, \ \tfrac{1}{6(N+3)}( 13-N \pm \sqrt{N^2 + 22N+25}), \ 0$,
$-\tfrac{2}{N+3}, \, \tfrac{1}{6(N+3)}( 3N+9 \pm \sqrt{9N^2 + 6N+33})$ and $ 1, \, \tfrac{N-5}{3(N+3)}$.

Similar discussions can be extended to other scalar operators which involve no derivatives and whose
anomalous dimensions are determined by the action of the differential operator $\D$ in \eqref{DTeig}.
For operators quadratic in $\phi$ the decomposition into irreducible representations gives
\begin{align}
V^{\alpha\beta} ={}& \phi^\alpha \phi^{\beta} + \tfrac{1}{N-1}\, \P_2 \, \phi^\alpha\phi^\alpha - \tfrac{1}{N(N-1)} \, \phi^2 \, , \ \alpha\ne \beta\, ,
\quad \dim \ \tfrac12(N+1)(N-2)\, , \nn \\
V^\alpha ={}& \phi^\alpha \phi^\alpha - \tfrac{1}{N+1} \, \phi^2 \, , \quad \dim \ N \, , \qquad
  \phi^2 \, , \quad \dim \ 1 \, ,
\end{align}
where
\be
\D \, V^{\alpha\beta} =  \vep \, \tfrac{2}{3(N+3)} \, V^{\alpha\beta} \, , \quad
\D \, V^{\alpha} =  \vep \, \tfrac{N+1}{3(N+3)} \, V^{\alpha} \, , \quad \D \, \phi^2 = \vep\, \tfrac{2(N+1)}{3(N+3)} \, \phi^2\, .
\ee
For cubic operators the corresponding decomposition is
\begin{align}
V^{\alpha\beta\gamma} ={}& \phi^\alpha \phi^{\beta} \phi^\gamma + \tfrac{1}{N-1}\, \P_6 \, \phi^\alpha\phi^\alpha \phi^\beta
+ \tfrac{2}{N(N-1)} \, \P_3  \, \phi^\alpha\phi^\alpha \phi^\alpha - \tfrac{1}{N(N-1)} \, \P_3  \, \phi^2 \phi^\alpha \nn \\
\noalign{\vskip -3pt}
&{} -   \tfrac{2}{(N+1)N(N-1)} \,  \phi^3 \, , \ \alpha\ne \beta \ne \gamma \, ,\qquad \dim \ \tfrac16(N+1)N(N-4)\, , \nn \\
V_S{\!}^{\alpha\beta} ={}& \P_2\,  \phi^\alpha\phi^\alpha \phi^{\beta} + \tfrac{2}{N-1}\, \P_2 \, \phi^\alpha\phi^\alpha \phi^\alpha
\tfrac{1}{N-1} \, \P_2\, \phi^2 \phi^\alpha - \tfrac{2}{N(N-1)} \, \phi^3 \, , \ \alpha\ne \beta\, ,
\   \dim \ \tfrac12(N+1)(N-2)\, , \nn \\
V_A{\!\!}^{\alpha\beta} ={}& \phi^\alpha \phi^\beta \phi^\beta - \phi^\alpha \phi^\alpha \phi^\beta- \tfrac{1}{N+1} \, \phi^2(\phi^\alpha - \phi^\beta) \, , \ \alpha\ne \beta\, ,
\quad \dim \ \tfrac12N(N-1)\, , \nn \\
V_1{\!}^\alpha ={}& \phi^\alpha \phi^\alpha \phi^\alpha - \tfrac{1}{N+1} \, \phi^3 \, , \ \ V_2{\!}^\alpha = \phi^2 \phi^\alpha \, ,
\quad \dim \ N \, , \qquad
  \phi^3 \, , \quad \dim \ 1 \, ,
\end{align}
where
\begin{align}
\D \, V^{\alpha\beta\gamma} = {}&  \vep \, \tfrac{2}{N+3} \, V^{\alpha\beta\gamma} \, , \quad
\D \, V_S{\!}^{\alpha\beta} =  \vep \, \tfrac{1}{3} \, V_S{\!}^{\alpha\beta}  \, ,  \quad \D \, V_A{\!}^{\alpha\beta} =
 \, \tfrac{N+7}{3(N+3)} \, V_A{\!}^{\alpha\beta}  \, ,
\nn \\
\D \begin{pmatrix} V_1{\!}^\alpha  \\ V_2{\!}^\alpha \end{pmatrix} =  {}& \vep \, \frac{1}{3(N+3)}\begin{pmatrix} 3(N+1) & 3 \\ 2(N+1) & 2(N+4) \end{pmatrix}
\begin{pmatrix} V_1{\!}^\alpha \\ V_2{\!}^\alpha  \end{pmatrix} \, , \quad \D \, \phi^3 =   \vep\, \tfrac{N+1}{N+3} \, \phi^3\, .
\end{align}
The eigenvalues of the matrix here are $2(N+1), \, 3(N+3)$.

\section{\texorpdfstring{$\boldsymbol{C_T}$}{CT} for Scalar Theories}
\label{centralCharge}

For a CFT obtained at an RG fixed point a crucial observable is $C_T$, the coefficient of
the energy momentum tensor two point function. In even dimensions this is related to
particular contributions to the anomaly under Weyl recalling of the metric for a
curved space background. For interacting theories the leading contributions to the anomaly
have been calculated in both four and six dimensions as well as $C_T$ directly in both cases.
Higher order terms in the perturbation expansion have not so far been determined. Here
we consider some constraints obtained by matching the perturbation expansion with
$O(N)$ results for large $N$ in any dimension. A similar argument is described in \cite{Sinha3}.

In four dimensions for a potential as in \eqref{Vphi4} then
the general form for $C_T$ to four loops is given by, after rescaling the coupling $\lambda_{ijkl} \to
16\pi^2 \, \lambda_{ijkl}$, has the form
\be
\frac{C_T}{C_{T,{\rm scalar}}} = N -  \tfrac{5}{36} \, \lambda_{ijkl}\lambda_{ijkl}
+ \alpha\, \lambda_{ijkl}\lambda_{klmn}\lambda_{mnij} \, ,
\label{CT4}
\ee
where $C_{T,{\rm scalar}}$ is the result for a single free scalar and the four loop
coefficient $\alpha$ has not been obtained directly.
For the $O(N)$ symmetric theory given by \eqref{VfourN} with $g=0$, so that
$\lambda_{ijkl} = \lambda( \delta_{ij} \delta_{kl} + \delta_{ik} \delta_{jl} +
\delta_{il} \delta_{jk} )$, then \eqref{CT4} becomes
\be
\frac{C_T}{C_{T,{\rm scalar}}} = N\bigg ( 1  -  \tfrac{5}{12}(N+2) \, \lambda^2
+ \alpha  (N+2)(N+8) \, \lambda^3 \bigg ) \, .
\ee
At the fixed point in $4-\vep$ dimensions to ${\rm O}(\vep^2)$,
which follows from $\beta_\lambda$ in \eqref{bgh1} and \eqref{bgh2}.
\be
\lambda_* = \frac{1}{N+8} \, \vep + 3 \, \frac{3N+14}{(N+8)^3} \, \vep^2 \, ,
\ee
which gives
\be
\frac{C_T}{C_{T,{\rm scalar}}} = N\bigg ( 1  -  \frac{5}{12} \frac{N+2}{(N+8)^2} \, \vep^2
- \frac{5}{2} \frac{(N+2)(3N+14)}{(N+8)^4} \, \vep^3
+ \alpha  \frac{N+2}{(N+8)^2} \, \vep ^3 \bigg ) \, .
\ee
The leading ${\rm O}(N^0)$ contributions for large $N$ in \cite{Petkou1,Petkou2,Diab}
${C_T}/{C_{T,{\rm scalar}}} = N - \frac{5}{12}\vep^2 -\frac{7}{36} \vep^3$, determine
\be
\alpha = - \frac{7}{36} \, .
\ee
The result for $N=1$ then gives
\be
\frac{C_T}{C_{T,{\rm scalar}}} =  1  -  \frac{5}{324}  \, \vep^2
- \frac{233}{81\times 108}  \, \vep^3 \, ,
\ee
in accord with \cite{Sinha1,Sinha2}. In an ancillary file we include
results for the central charge for the theories of section
\ref{redSymFour}.

A similar approach may be adopted in six dimensions.
In this case  a renormalisable  scalar theory, for $\N$ component $\phi_i$, with interaction
$\tfrac{1}{6} \, \lambda_{ijk} \phi_i \phi_j \phi_k $
the general form for $C_T$ to three loops in perturbation theory is given by, after rescaling $\lambda_{ijk} \to
(4 \pi)^{\frac32} \, \lambda_{ijk}$,
\be
\frac{C_T}{C_{T,{\rm scalar}}} = \N -  \tfrac{7}{72} \, \lambda_{ijk}\lambda_{ijk}
+ \alpha\, \lambda_{ijk}\lambda_{ilm}\lambda_{jln}\lambda_{kmn}
+ \beta \, \lambda_{ijk}\lambda_{ijl}\lambda_{mnk}\lambda_{mnl} \, ,
\ee
where two three loop contributions are possible.
For the $O(N)$ model with potential as in \eqref{VsixN} with $h=0$
$\phi_i \to (\sigma,\vphi_i), \, \N = N+1$ and the coupling is determined by
$\lambda_{000} = \lambda, \, \lambda_{ij0} = g\, \delta_{ij}$. This gives
\begin{align}
\frac{C_T}{C_{T,{\rm scalar}}} = {}&  N+1  -  \tfrac{7}{72} \, ( 3N g^2 + \lambda^2) \nn \\
&{} + \alpha\, \big ( Ng^3 ( 3 g +  4 \lambda) + \lambda^4 \big )
+ \beta \, \big ( ( Ng^2+\lambda^2)^2 + 4 N g^4 \big ) \, .
\end{align}
This theory has a fixed point for large $N$ with $g_*{\!}^2 \approx\frac{6}{N} ( 1+ \frac{44}{N} - \tfrac{155}{3N} \vep )\vep $
and $
\lambda_*{\!}^2 \approx \frac{6^3}{N} \, \vep $. In the large $N$ limit to leading order in $1/N$ corrections
\be
\frac{C_T}{C_{T,{\rm scalar}}} = N +1 -  \tfrac{7}{4} \, \vep + \tfrac{23}{288} \, \vep^2
+ {\rm O}(\vep^3) \, .
\ee
This would require $6^4 \beta = \frac{23}{8}$ but $\alpha$ is undetermined.

\end{appendices}

\bibliographystyle{utphys}
\bibliography{Phi}

\end{document}